\documentclass[final,authoryear,3p,times,10pt]{elsarticle}

\usepackage{multirow,setspace,times,amssymb,amsmath,graphicx,color,rotating,subfigure,url}
\usepackage{lineno}
\usepackage{natbib}
\usepackage{booktabs}%
\usepackage{longtable}%
\graphicspath{{Figures/}}
\usepackage[table]{xcolor}
\usepackage{tabularx}
\usepackage{graphicx} 
\usepackage{epstopdf}
\usepackage{mathrsfs}
\usepackage{makecell}
\usepackage[bookmarks=true,colorlinks,linkcolor=blue,anchorcolor=blue,citecolor=blue,unicode]{hyperref}
\usepackage{bookmark}

\usepackage{pythonhighlight} 

\hypersetup{CJKbookmarks=true}%
\journal{}

\begin{document}

\begin{frontmatter}
\newpage
\title{Economy importance and structural robustness of the international pesticide trade networks}

\author[SB]{Jian-An Li}
\author[SP,RCE]{Li Wang}
\author[SB,RCE]{Wen-Jie Xie\corref{CorrAuth}}
\ead{wjxie@ecust.edu.cn}
\author[SB,RCE,DM]{Wei-Xing Zhou\corref{CorrAuth}}
\ead{wxzhou@ecust.edu.cn}
\cortext[CorrAuth]{Corresponding authors.} 

\address[SB]{School of Business, East China University of Science and Technology, Shanghai 200237, China}
\address[SP]{School of Physics, East China University of Science and Technology, Shanghai 200237, China}
\address[RCE]{Research Center for Econophysics, East China University of Science and Technology, Shanghai 200237, China}
\address[DM]{School of Mathematics, East China University of Science and Technology, Shanghai 200237, China}

\begin{abstract}
  Pesticides are a kind of agricultural input, whose use can greatly reduce yield loss, regulate plant growth, effectively liberate agricultural productivity, and improve food security. The availability of pesticides in economies all over the world is ensured by pesticide redistribution through international trade and economies play different roles in this process. In this work, we measure and rank the importance of economies using nine node metrics in an evolutionary way. It is found that the clustering coefficient is correlated negatively with the other eight node metrics, while the other eight node metrics are positively correlated with each other and can be grouped into three communities (betweenness; in-degree, PageRank, authority, and in-closeness; out-degree, hub, and out-closeness). We further investigate the structural robustness of the international pesticide trade networks proxied by the giant component size under three types of shocks to economies (node removal in descending order, randomly, and in ascending order). The results show that, except for the clustering coefficient, the international pesticide trade networks are relatively robust under shocks to economies in ascending orders and randomly, but fragile under shocks to economies in descending order. In contrast, removing nodes with the clustering coefficient in ascending and descending orders gives similar robustness curves. Moreover, the structural robustness related to the giant component size evolves over time and exhibits an inverse U-shaped pattern.
\end{abstract}


\end{frontmatter}

\section{Introduction}
\label{intro} 

Pesticides, including insecticides, fungicides, herbicides, disinfectants, and rodenticides and other similar products, are an important class of agricultural inputs, which are used to control pests, diseases, and weeds and regulate plant growth. Hence, the role of pesticides is not only to reduce yield loss and increase food security, but also to effectively liberate agricultural productivity. The application of pesticides reduces food losses remarkably during production (about 78\% loss of fruits, 54\% loss of vegetables, and 32\% loss of cereals) and storage. The availability of pesticides in the economies all over the world is maintained by pesticide redistribution through international trade, which forms international pesticide trade networks, in which nodes represent economies and links stand for import/export trade relationships. 

The basic structure and evolution of the international pesticide trade networks have been investigated \citep{Li-Xie-Zhou-2021-FrontPhysics}. What is more important is to understand the structural robustness of the international pesticide trade networks. Many complex networks are robust to random failures but fragile to intentional attacks \citep{Albert-Jeong-Barabasi-2000-Nature,Cohen-Erez-benAvraham-Havlin-2000-PhysRevLett,Callaway-Newman-Strogatz-Watts-2000-PhysRevLett,Cohen-Erez-benAvraham-Havlin-2001-PhysRevLett}. Both random failures and intentional attacks can be internal or external shocks to the networks. The main aim of this work is to quantify the structural robustness of the international pesticide trade networks under different types of shocks to economies. The robustness and fragility of complex networks have been extensively studied in diverse fields, such as international oil trade networks \citep{Foti-Pauls-Rockmore-2013-JEconDynControl,Fair-Bauch-Anand-2017-SciRep,Xie-Wei-Zhou-2021-JStatMech,Wei-Xie-Zhou-2022-JComplexNetw,Chen-Ding-Zhang-Zhang-Nie-2022-Energy,Wei-Xie-Zhou-2022-Energy} and other infrastructure networks \citep{Albert-Albert-Nakarado-2004-PhysRevE,Latora-Marchiori-2005-PhysRevE,Pagani-Aiello-2013-PhysicaA,Wang-Guan-Lai-2009-NewJPhys,Zanin-Lillo-2013-EurPhysJ-SpecTop,Diao-Sweetapple-Farmani-Fu-Ward-Butler-2016-WaterRes}.

In order to investigate the structural robustness of complex networks, one needs to determine the attack strategies (or shock types) and measures for structural robustness. The most adopted strategies are random removal of nodes and intentional removal of nodes in descending order of node metrics \citep{Albert-Jeong-Barabasi-2000-Nature}. We also consider the intentional node removal strategy in ascending order of node metrics for comparison, which is rarely considered in the literature. The two intentional node removal strategies are executed based on node metrics to rank economies. There are many measures of node importance \citep{Lu-Chen-Ren-Zhang-Zhang-Zhou-2016-PhysRep}. We adopt nine node metrics for the removal strategies. Most studies in the literature utilized fewer node metrics. There are also a lot of network metrics for structural robustness \citep{Oehlers-Fabian-2021-Mathematics,Schaeffer-Valdes-Figols-Bachmann-Morales-BustosJimenez-2021-JComplexNetw}. We consider in this work the size of the giant component \citep{Albert-Jeong-Barabasi-2000-Nature}, which quantifies the degree of pesticide trade globalization.

The remainder of this work is organized as follows. We briefly describe the data in Section~\ref{S1:Data}. In Section~\ref{S1:NodeRank}, we quantify and rank the importance or influence of economies in the international pesticide trade networks based on nine node metrics and anatomize their cross-correlation structure. We are able to classify the node metrics into three groups: the first group related to incoming links, the second group related to outgoing links, and the third group unrelated to link directions. The structural robustness is investigated in Section~\ref{S1:GiantComponent} by considering the giant component size. In addition to the random attack (failures) strategy and the targeted attack strategy based on descending node metrics, we also consider the intentional attack strategy based on ascending node metrics. Section~\ref{S1:Summary} summarizes our results.

\section{Data description}
\label{S1:Data}

\begin{figure}[!ht]
  \centering
  \includegraphics[width=0.5\linewidth]{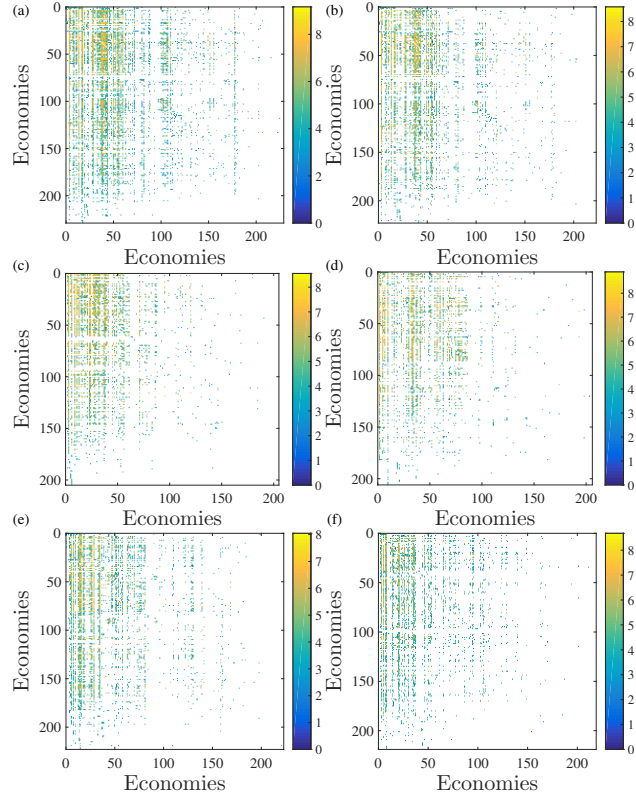}
  \caption{Heat maps of the international pesticide trade networks $V$ of all pesticides (a), insecticides (b), fungicides (c), herbicides (d), disinfectants (e), and rodenticides and other similar products (f) in 2018. The color bars show the magnitude order of the trading value quantified by $\log_{10}v_{ij}$.}
  \label{Fig:Pesticide:vij:value}
\end{figure}

The data sets of the international pesticide trade from 2007 to 2018 are retrieved from the UN Comtrade database, which contain the import and export economies and trading values of five categories, including insecticides (380891), fungicides (380892), herbicides (380893), disinfectants (380894), and rodenticides and other similar products (380899). The six network matrices $\mathbf{V}(t)=\{v_{ij}: i,j=1, \cdots, N(t)\}$ of the international pesticide trade networks $V$ of all pesticides, insecticides, fungicides, herbicides, disinfectants, and rodenticides and other similar products in 2018 are illustrated in Fig.~\ref{Fig:Pesticide:vij:value}, where the element $v_{ij}(t)$ is the trading value from economy $i$ to economy $j$.

Fig.~\ref{Fig:Pesticide:vij:value}(a) shows the trading value matrix of the aggregated international trade network of all the five networks for individual pesticide categories, while Fig.~\ref{Fig:Pesticide:vij:value}(b-f) show the trading value matrices for the five pesticide categories. It is trivial that all the matrices are asymmetric ($v_{ij}\neq v_{ji}$). Although the six matrices are not identical, they share similar patterns to a certain extent. We also observe that the trading values span about nine orders of magnitude, showing the high heterogeneity of international pesticide trade.

\section{Quantifying economy importance}
\label{S1:NodeRank}

There are many node metrics that capture the local or global characteristics of nodes that can be used to measure the importance of nodes in different aspects \citep{Chen-Lu-Shang-Zhang-Zhou-2012-PhysicaA,Lu-Chen-Ren-Zhang-Zhang-Zhou-2016-PhysRep}. We consider nine node metrics in this work, including clustering coefficient, betweenness, PageRank, in-degree, out-degree, in-closeness, out-closeness, authority score, and hub score. These node metrics have different traits. The clustering coefficient is defined for undirected networks, while the other eight metrics are extracted from directed networks. Concerning the other eight node metrics, betweenness does not distinguish importing and exporting economies; in-degree, PageRank, authority and in-closeness are calculated from exporting (thus the node under consideration is a target node) economies; and out-degree, out-closeness, and hub are obtained from importing economies (thus the node under consideration is a source node). In this section, we quantify economy importance in the international pesticide trade based on these nine node metrics, and further investigate their mutual correlations.

\subsection{In-degree and out-degree}
\label{S2:In:Out:Degree}

\begin{figure}[!ht]
\centering
    \includegraphics[width=0.33\linewidth]{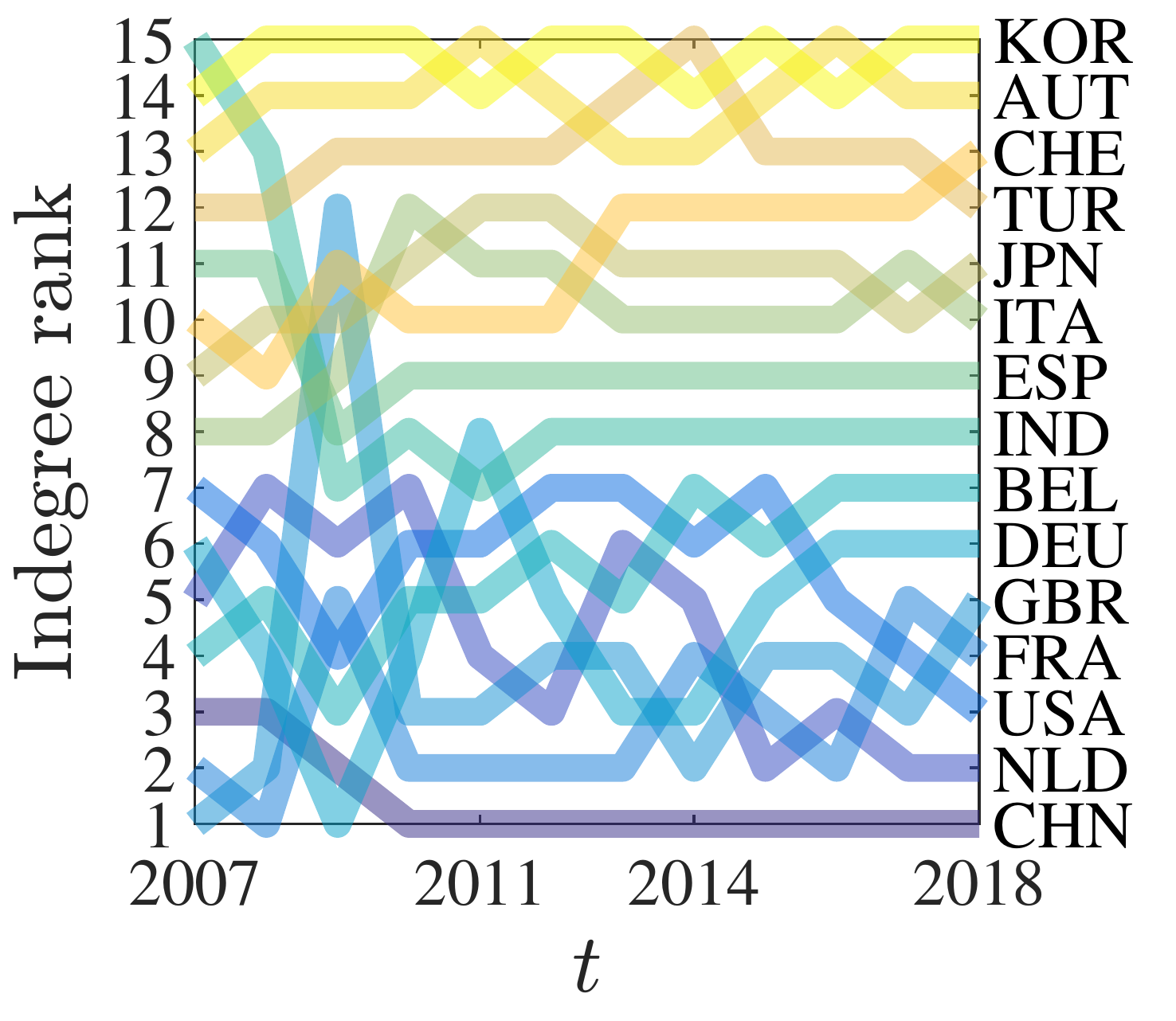}
    \includegraphics[width=0.33\linewidth]{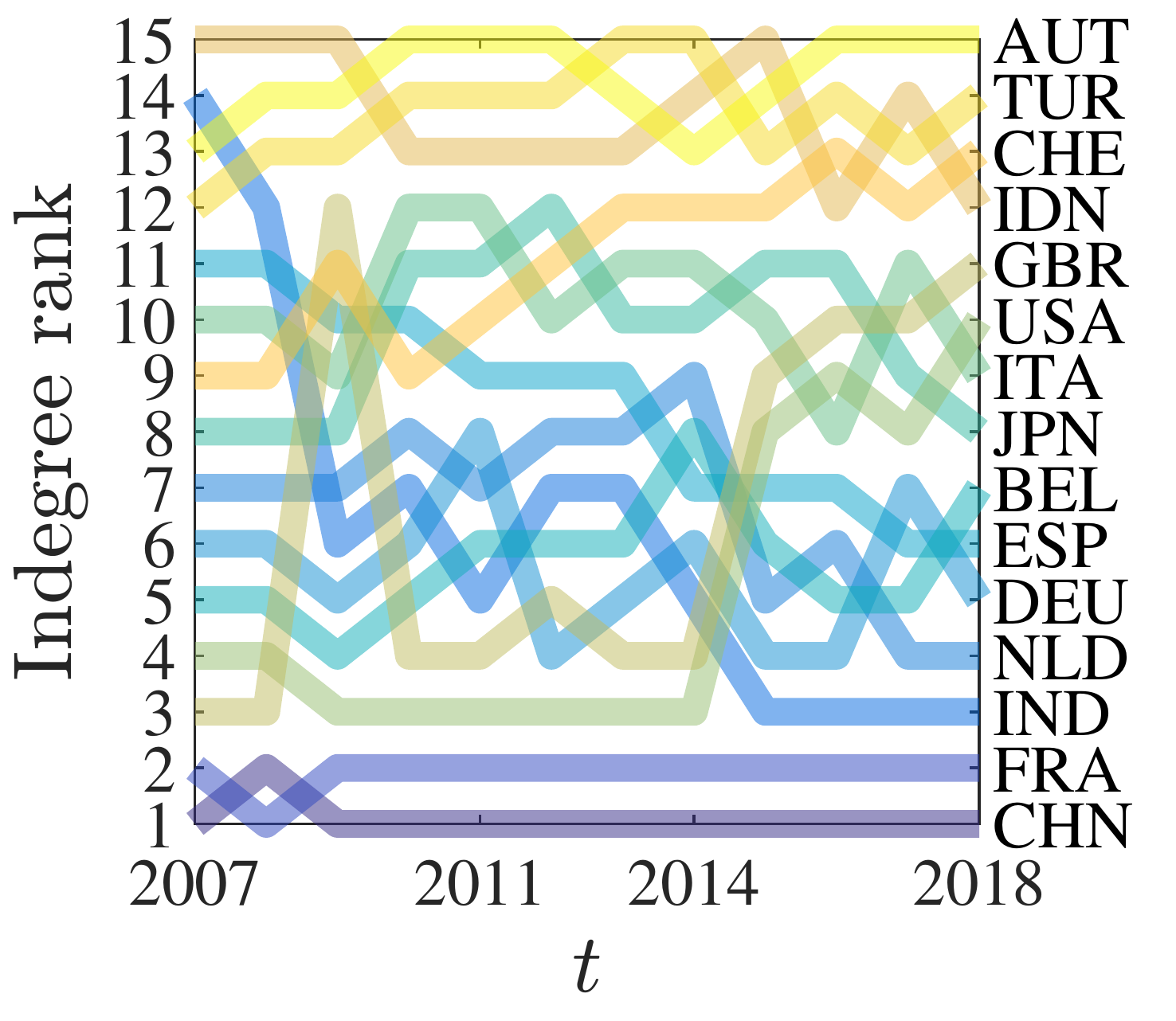}
    \includegraphics[width=0.33\linewidth]{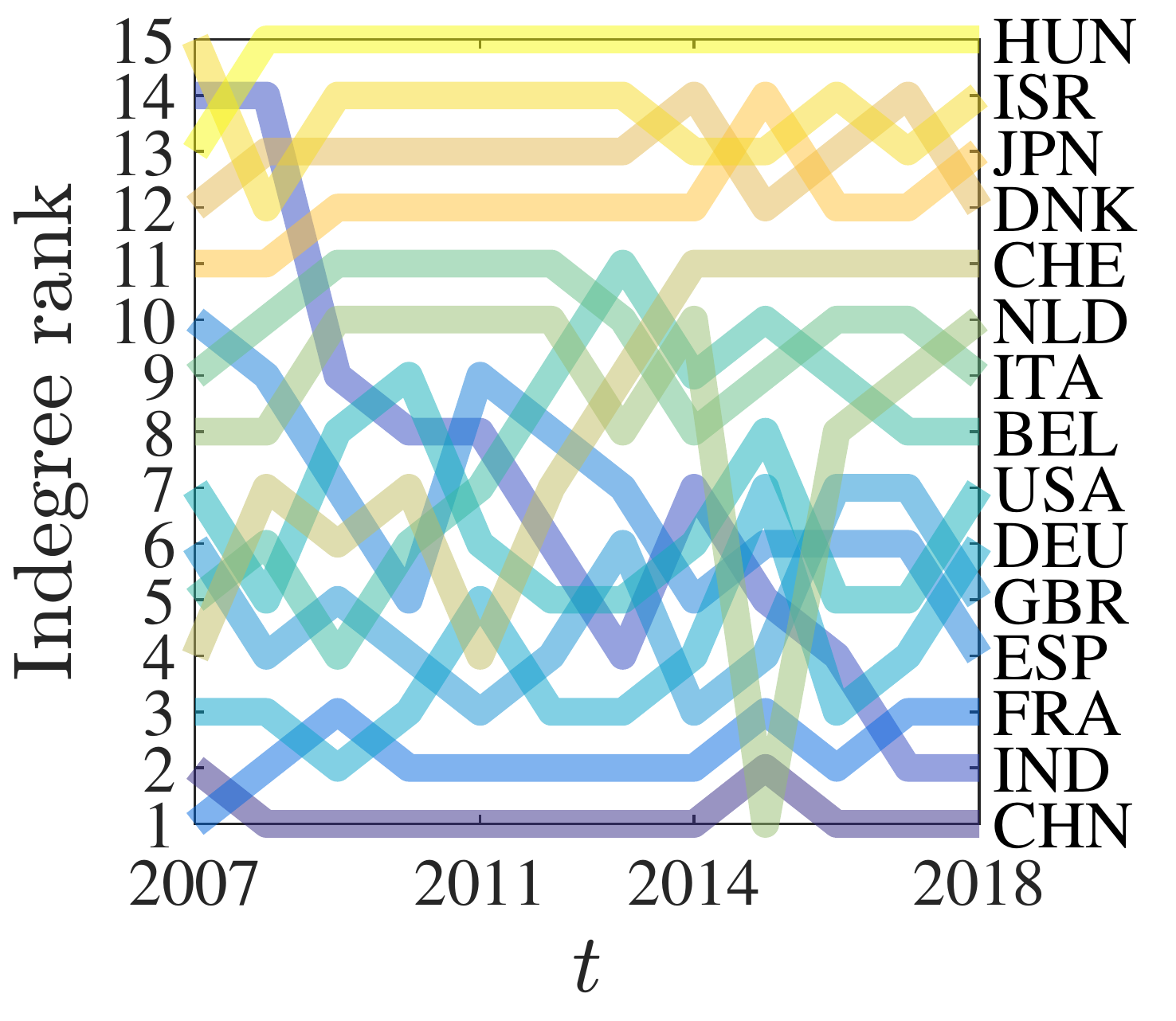}
    \includegraphics[width=0.33\linewidth]{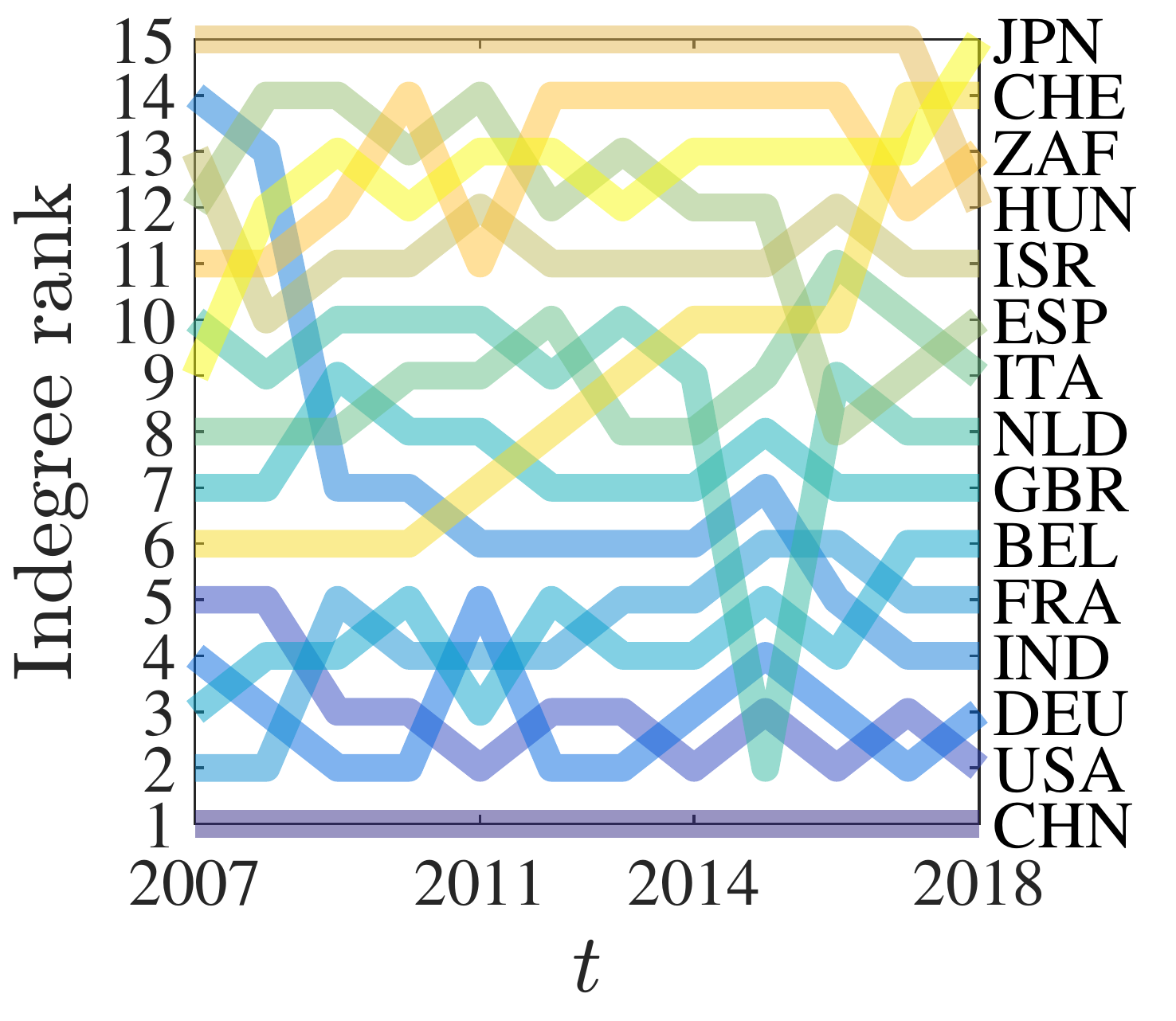}
    \includegraphics[width=0.33\linewidth]{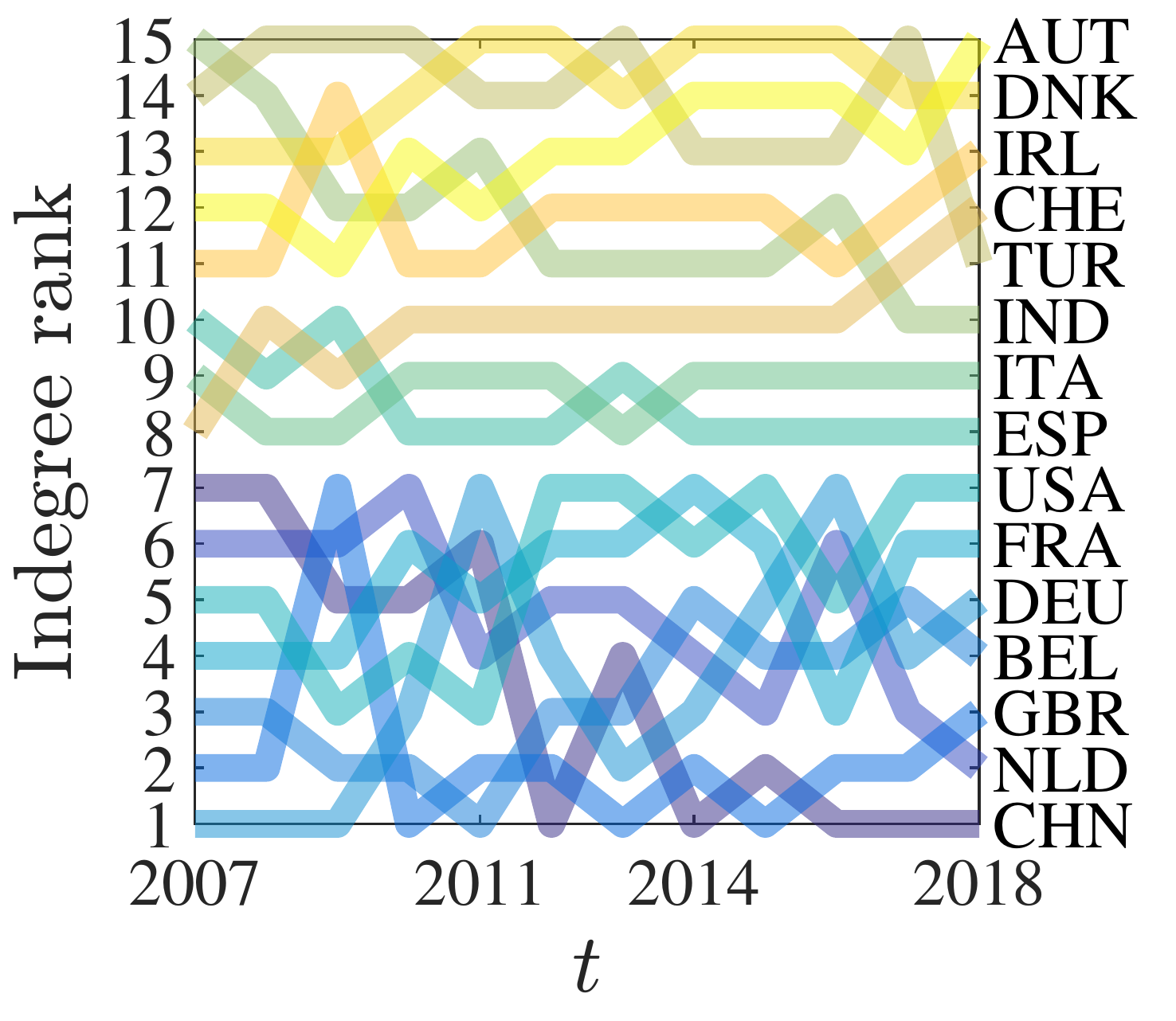}
    \includegraphics[width=0.33\linewidth]{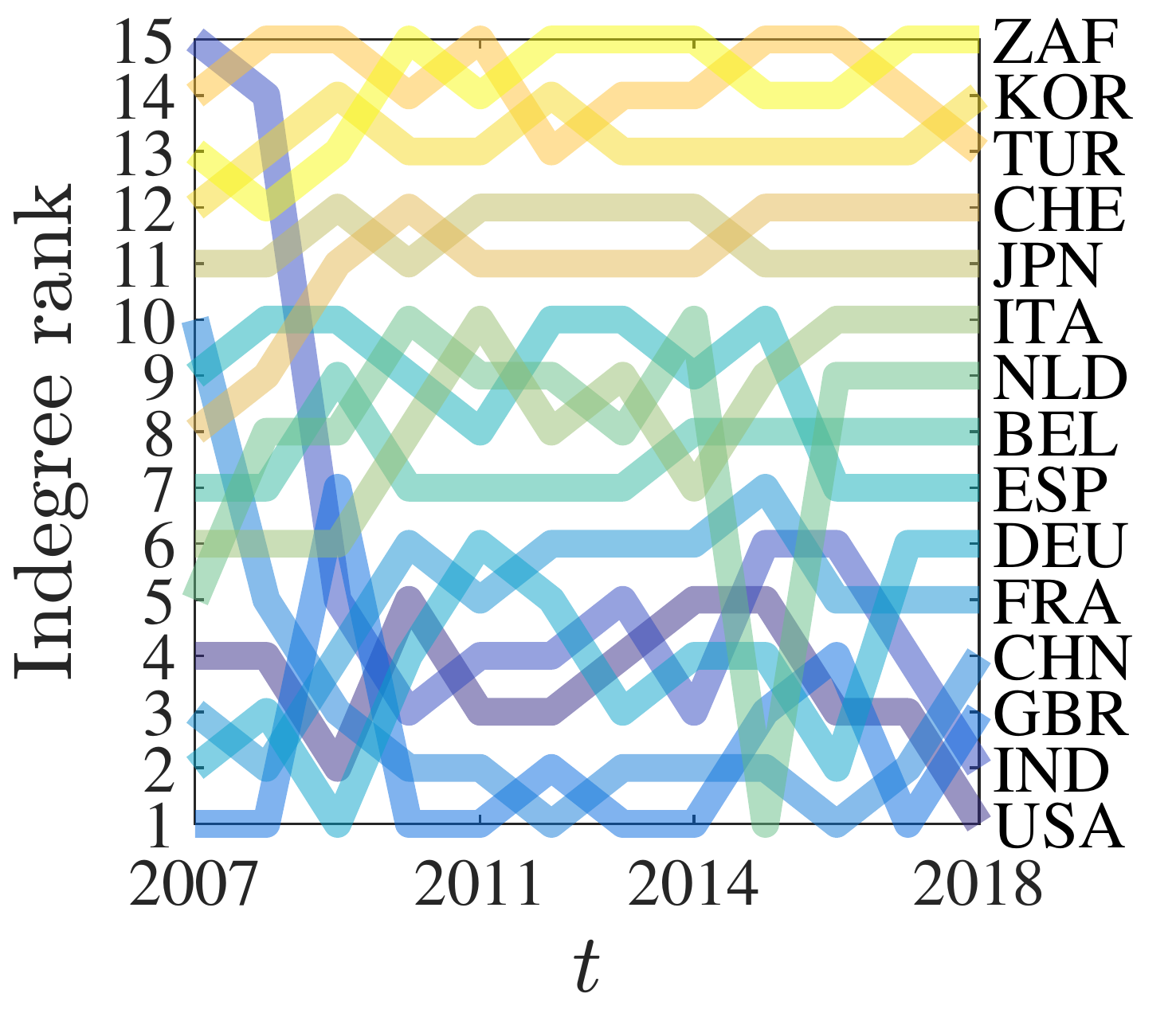}
\vskip    -9.55cm   \hskip   -16.0cm {(a)}
\vskip    -0.43cm   \hskip    -4.9cm {(b)}
\vskip    -0.43cm   \hskip     6.0cm {(c)}
\vskip     4.33cm   \hskip   -16.0cm {(d)}
\vskip    -0.43cm   \hskip    -4.9cm {(e)}
\vskip    -0.43cm   \hskip     6.0cm {(f)}
\vskip 4.05cm
  \caption{Rank evolution of the top-15 economies in the international trade networks for all pesticides (a), insecticides (b), fungicides (c), herbicides (d), disinfectants (e), and rodenticides and other similar products (f) from 2007 to 2018. The ranking is based on the in-degrees of nodes.}
  \label{Fig:Pesticide:Rank:Evo:in-degree}
\end{figure}

The adjacent matrix $\mathbf{A}=[a_{ij}]$ of an international pesticide trade network can be extracted from its trading value matrix $\mathbf{V}$, where $a_{ij}=1$ if $v_{ij}>0$ and $a_{ij}=0$ otherwise. In other words, $a_{ij}=1$ if economy $i$ exports a category of pesticides to economy $j$. The in-degree of target node $j$, $k_{j}^{\rm{\mathrm{in}}}$, is the count of exporting partners of economy $j$, expressed as follows:
\begin{equation}
  k_{j}^{\mathrm{in}} = \sum_{i=1}^{N}a_{ij},
  \label{Eq:k_in}
\end{equation}
where $N$ is the number of nodes in the trade network. We rank the economies in the international trade networks for all pesticides, insecticides, fungicides, herbicides, disinfectants, and rodenticides and other similar products in each year from 2007 to 2018. The rank evolution of the top-15 economies is shown in Fig.~\ref{Fig:Pesticide:Rank:Evo:in-degree}.


For the aggregated network in 2018, as shown in Fig.~\ref{Fig:Pesticide:Rank:Evo:in-degree}(a), the top-15 economies are 
China, the Netherlands, the USA, France, the United Kingdom, Germany, Belgium, India, Spain, Italy, Japan, Switzerland, Turkey, Austria, and Korea,
which are developed economies or the largest developing economies. Except for the rodenticide trade network, China ranks No. 1 and has the largest number of importing partners. The USA has the largest in-degree in 2018 for importing rodenticides.

\begin{figure}[!ht]
\centering
    \includegraphics[width=0.33\linewidth]{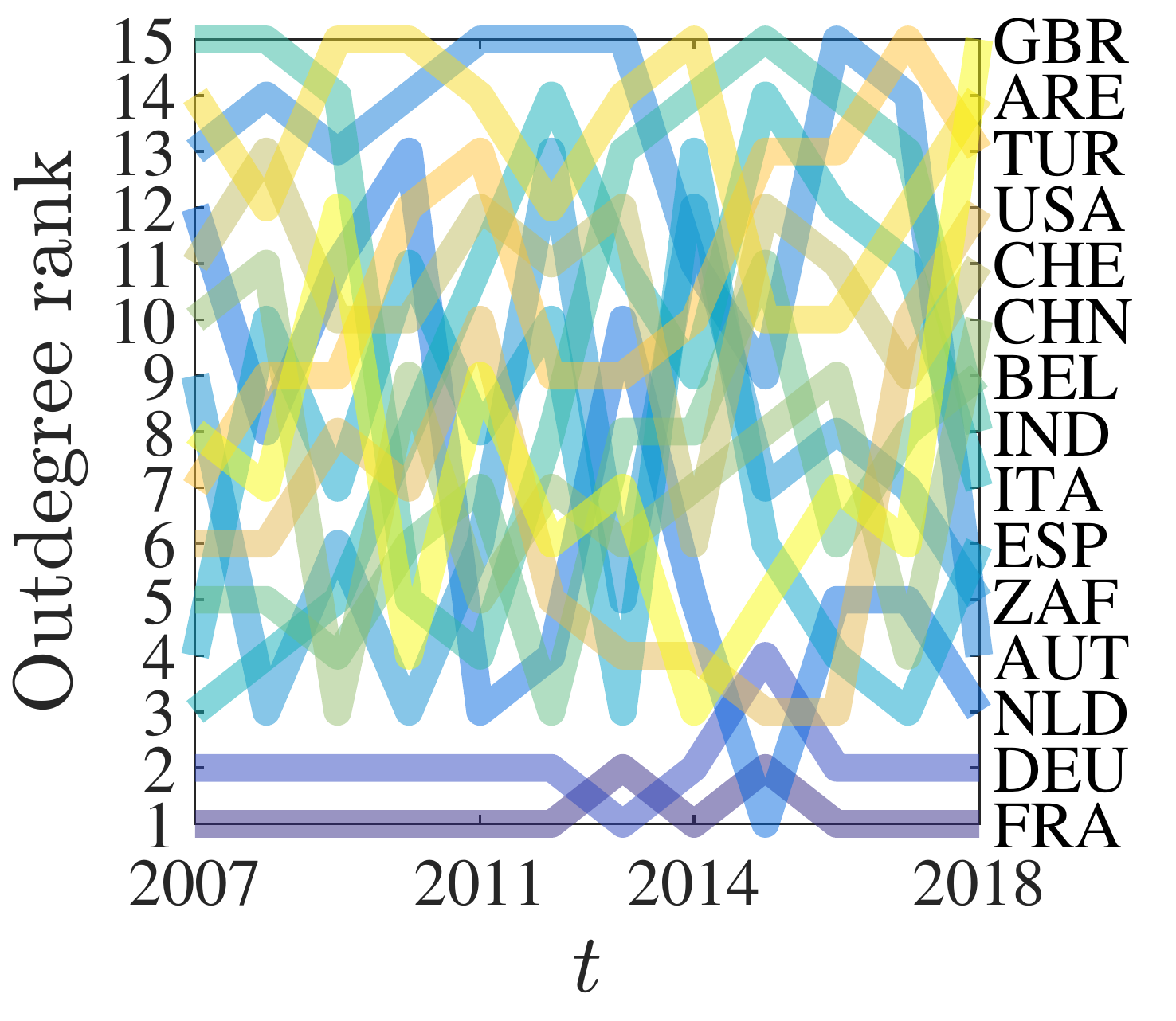}
    \includegraphics[width=0.33\linewidth]{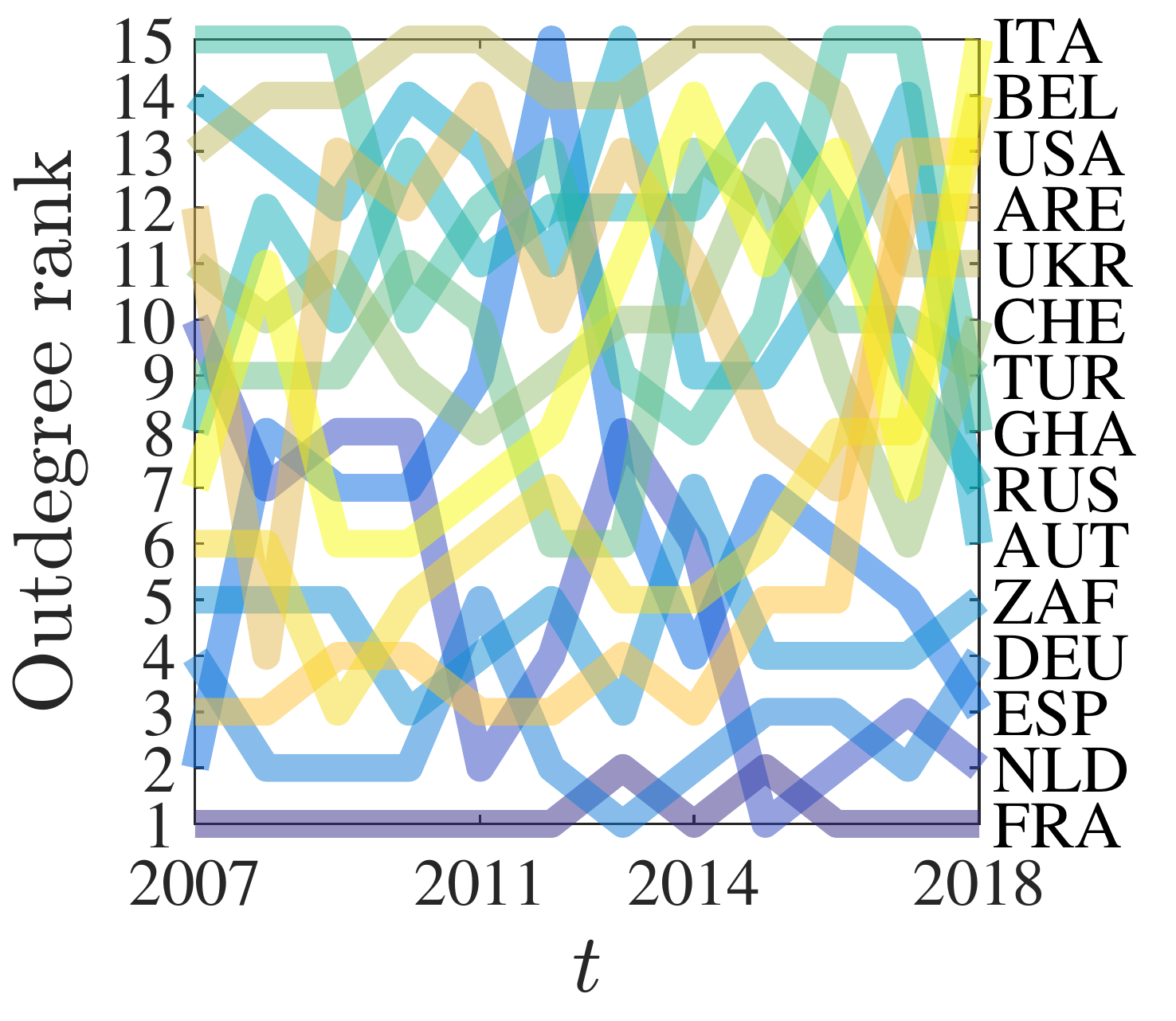}
    \includegraphics[width=0.33\linewidth]{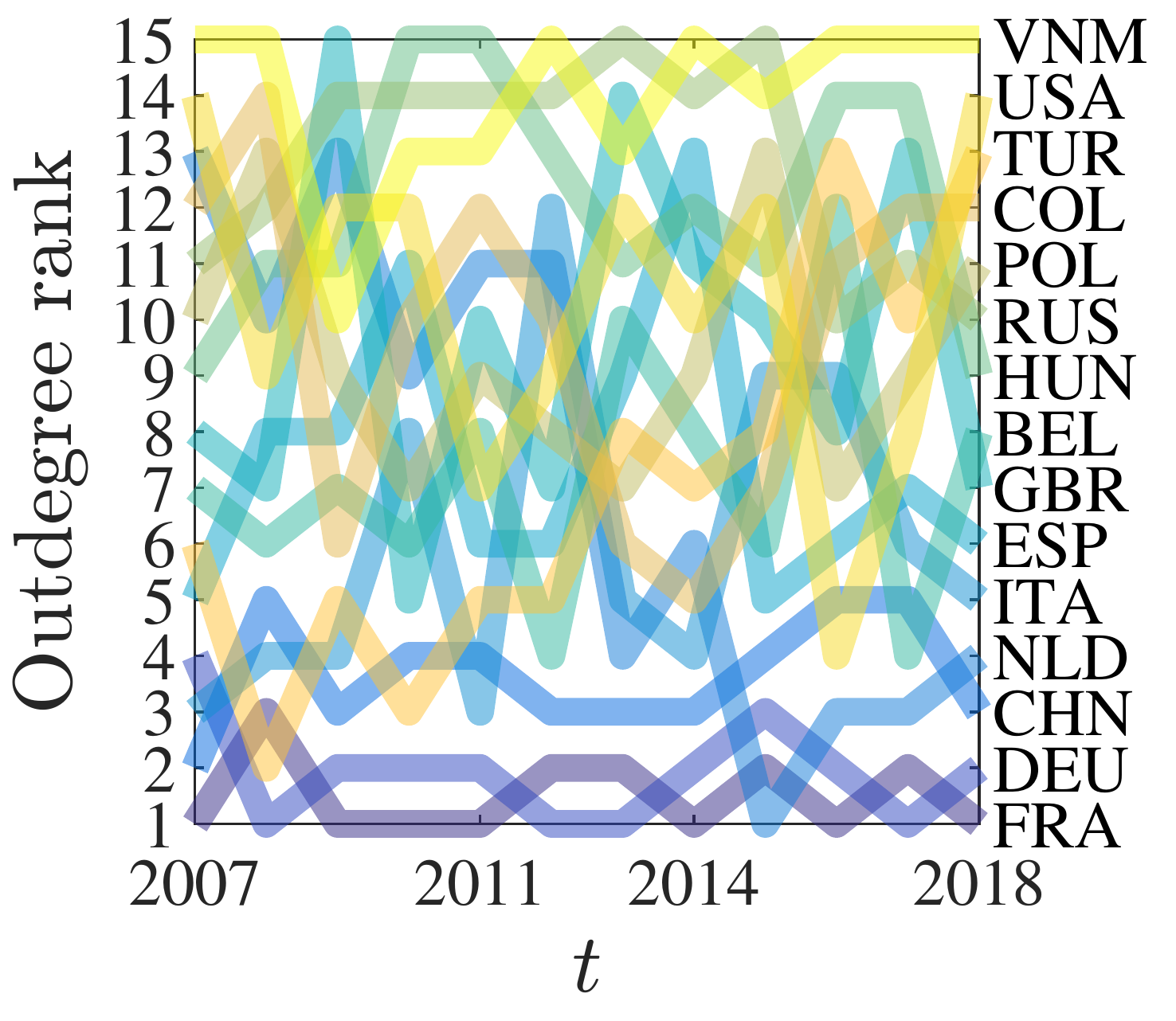}
    \includegraphics[width=0.33\linewidth]{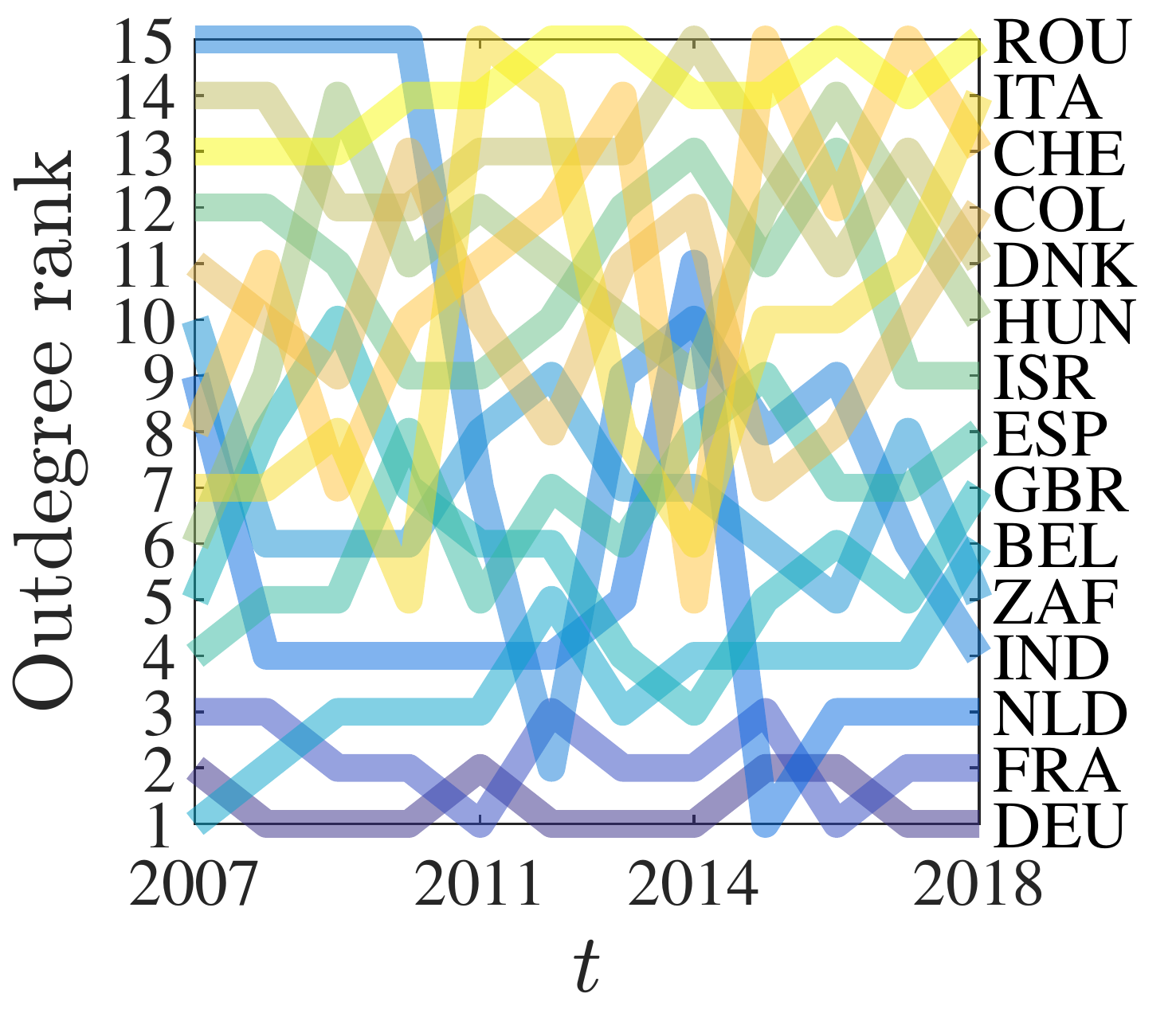}
    \includegraphics[width=0.33\linewidth]{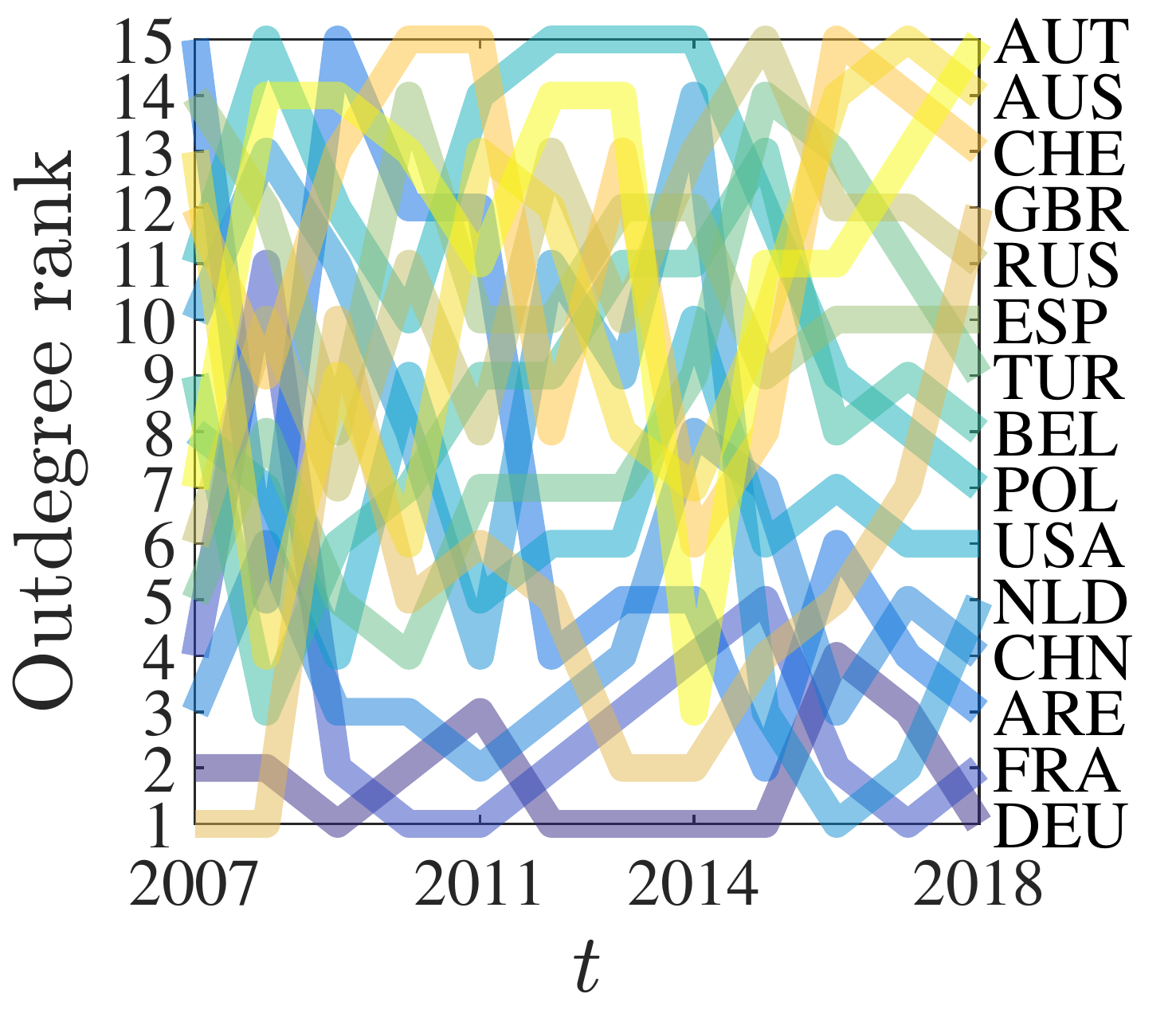}
    \includegraphics[width=0.33\linewidth]{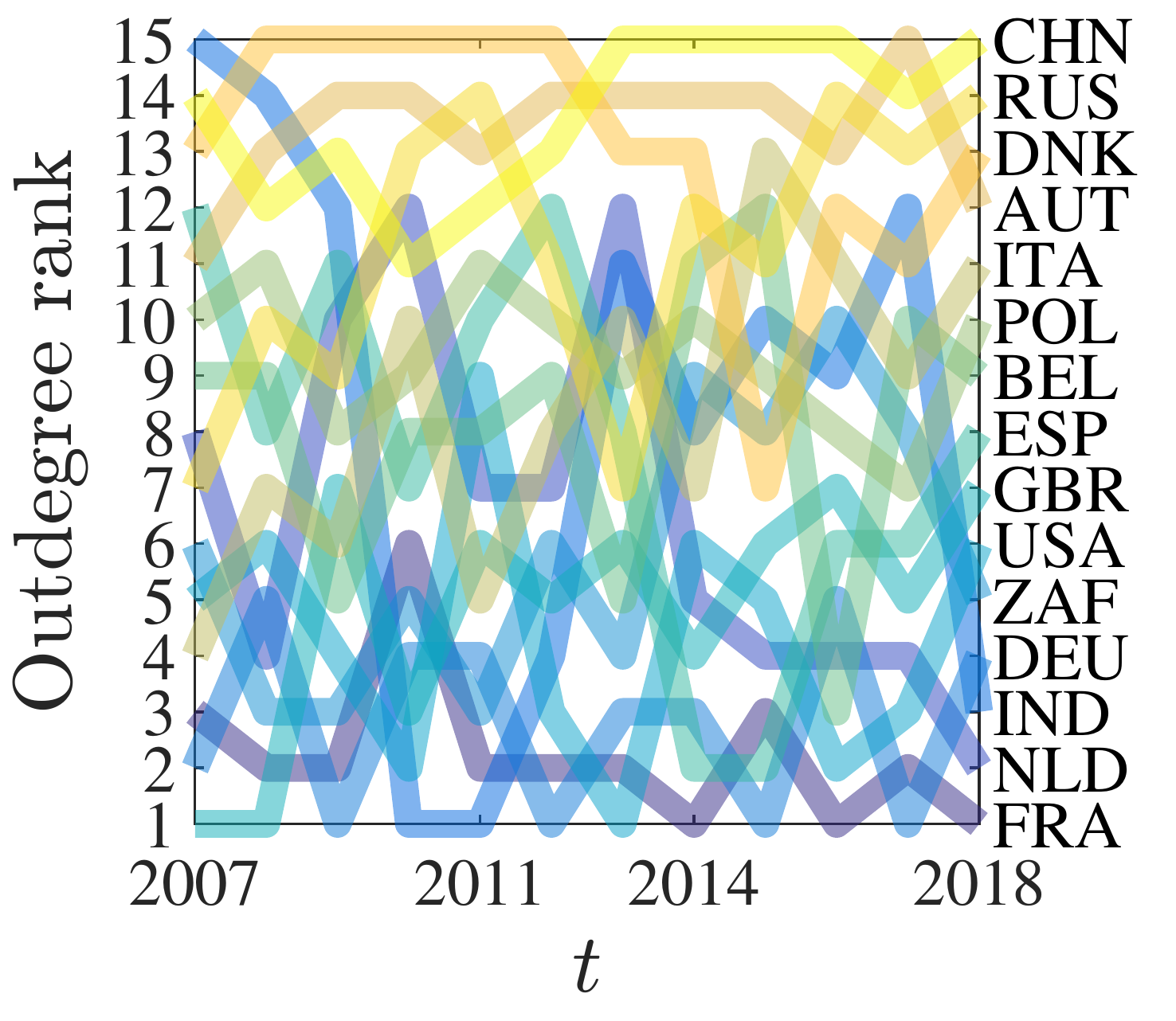}
\vskip    -9.55cm   \hskip   -16.0cm {(a)}
\vskip    -0.43cm   \hskip    -4.9cm {(b)}
\vskip    -0.43cm   \hskip     6.0cm {(c)}
\vskip     4.33cm   \hskip   -16.0cm {(d)}
\vskip    -0.43cm   \hskip    -4.9cm {(e)}
\vskip    -0.43cm   \hskip     6.0cm {(f)}
\vskip 4.05cm
  \caption{Rank evolution of the top-15 economies in the international trade networks for all pesticides (a), insecticides (b), fungicides (c), herbicides (d), disinfectants (e), and rodenticides and other similar products (f) from 2007 to 2018. The ranking is based on the out-degrees of nodes.}
  \label{Fig:Pesticide:Rank:Evo:out-degree}
\end{figure}

The out-degree of source node $i$, $k_{i}^{\rm{\mathrm{out}}}$, is the count of importing partners of economy $i$, expressed as follows:
\begin{equation}
  k_{i}^{\rm{\mathrm{out}}} = \sum_{j=1}^{N}a_{ij}.
  \label{Eq:k_out}
\end{equation}
We rank the economies in the international trade networks based on node out-degrees in each year from 2007 to 2018 and illustrate the rank evolution of the top-15 economies in Fig.~\ref{Fig:Pesticide:Rank:Evo:out-degree}.


For the aggregated network in 2018, as shown in Fig.~\ref{Fig:Pesticide:Rank:Evo:out-degree}(a), the top-15 economies are France, Germany, the Netherlands, Austria, South Africa, Spain, Italy, India, Belgium, China, Switzerland, the USA, Turkey, United Arab Emirates, and the United Kingdom, 
which are also developed economies or large developing economies. France ranks No. 1 in the networks for insecticides, fungicides\textbf{,} and rodenticides and No. 2 in the networks for herbicides and disinfectants, while Germany ranks No. 1 in the networks for herbicides and disinfectants and No. 2 in the networks for fungicides and rodenticides. The Netherlands also ranks highly in all the networks. It is clear that economies in North Europe play the most important role in exporting pesticides.

\subsection{In-closeness and out-closeness}
\label{S2:In:Out:Closeness}

The closeness of node $i$ is to measure the inverse of information propagation length, which is defined as the reciprocal of the mean distance between $i$ and all other nodes. Since the network might not be connected, the closeness can be calculated as follows
\begin{equation}
    C_i =\frac{N-1}{\sum_{j\in{{\mathbf{n}}_i}}l_{ij}},
    \label{Eq:Centrality:Closeness}
\end{equation}
where ${\mathbf{n}}_i$ is the set of nodes that $i$ can reach and $l_{ij}$ is the length of the shortest path between $i$ and $j$. 

For connected networks, we can define the in-closeness as follows,
\begin{equation}
    C_i^{\mathrm{in}} =\frac{N-1}{\sum_{j\in{{\mathbf{n}}_i^{\mathrm{in}}}}l_{ij}},
    \label{Eq:Centrality:inCloseness}
\end{equation}
where ${\mathbf{n}}_i^{\mathrm{in}}$ is the set of nodes that have at least one directed path pointing to $i$. 

\begin{figure}[!ht]
\centering
    \includegraphics[width=0.33\linewidth]{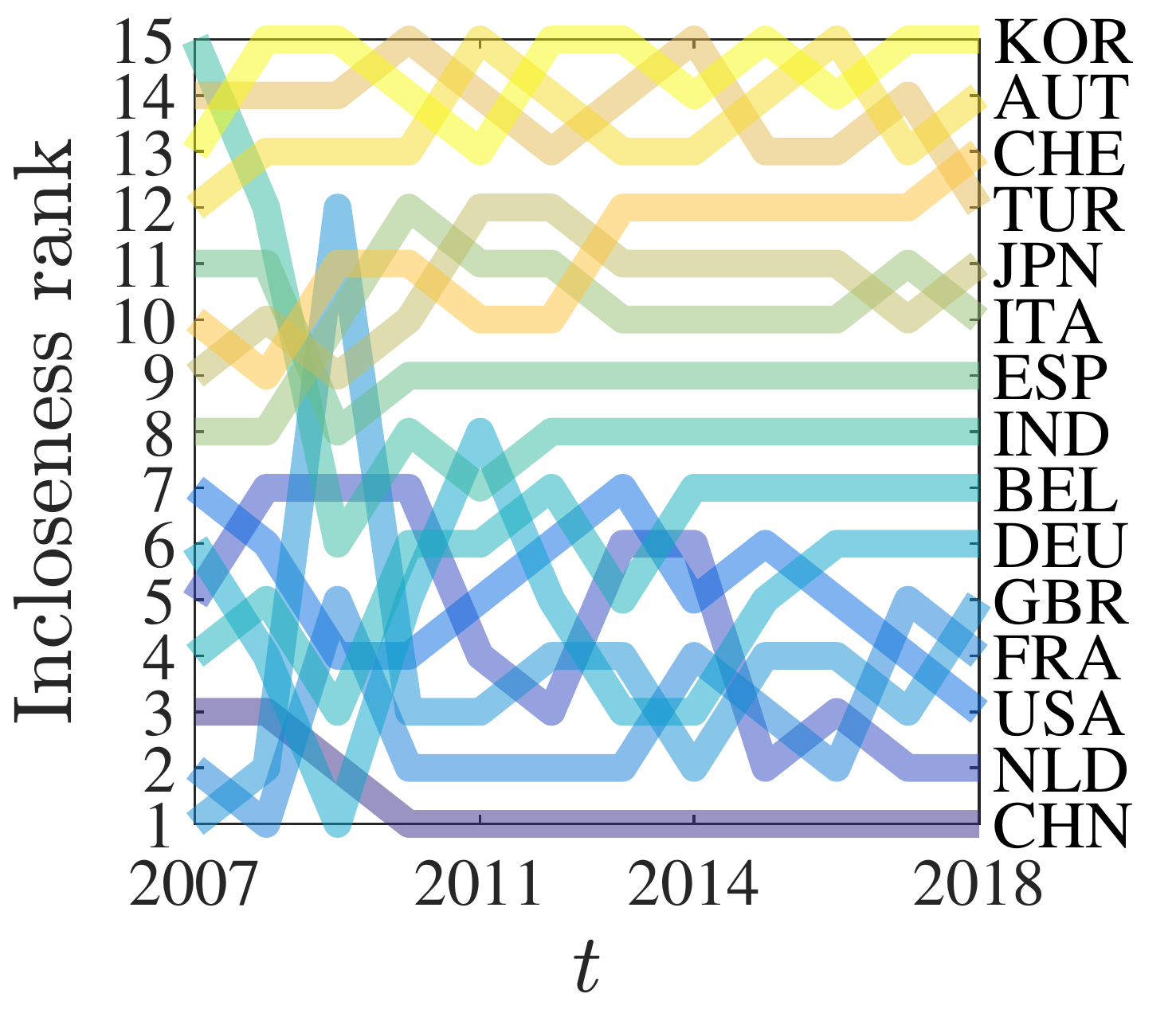}
    \includegraphics[width=0.33\linewidth]{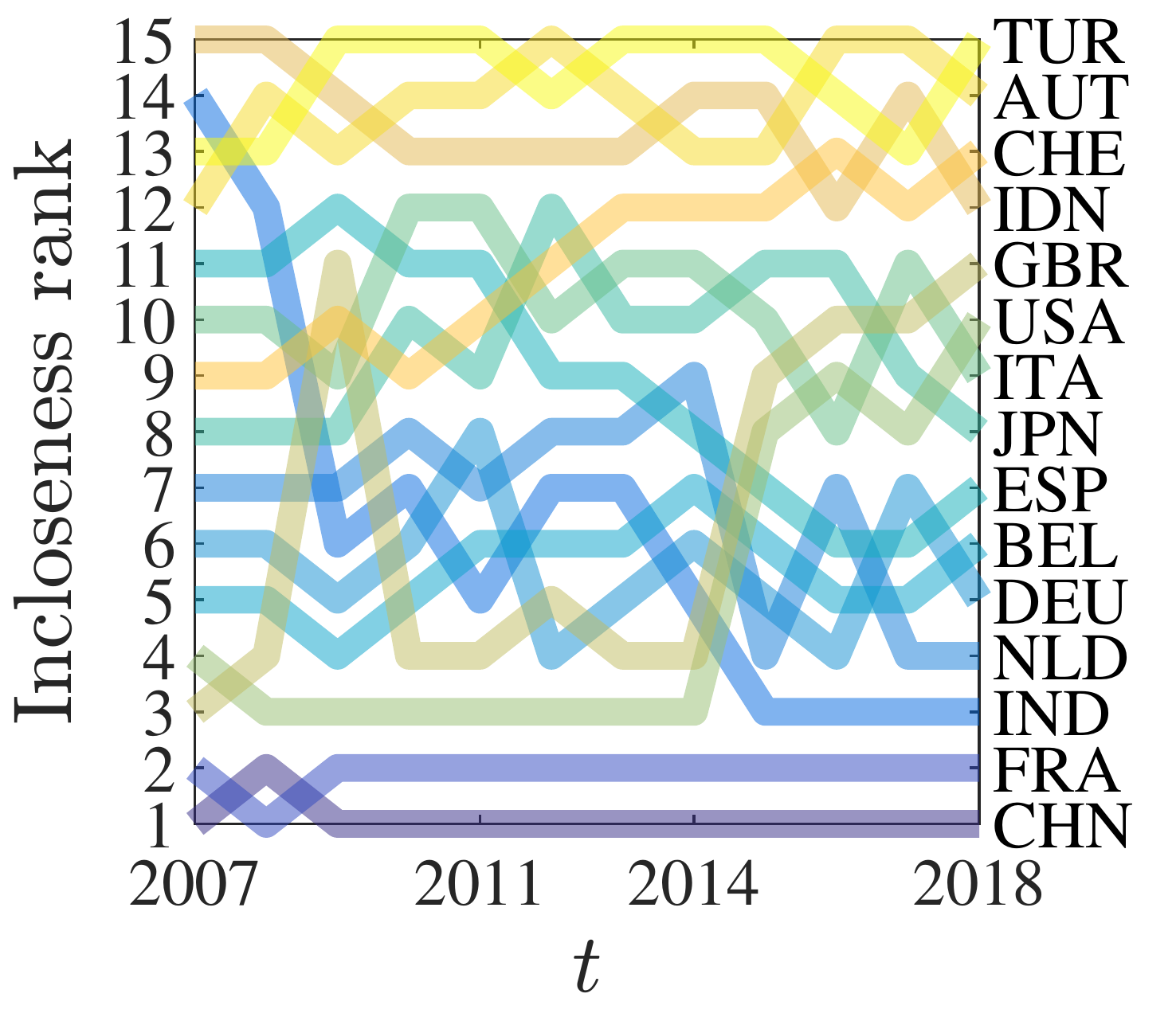}
    \includegraphics[width=0.33\linewidth]{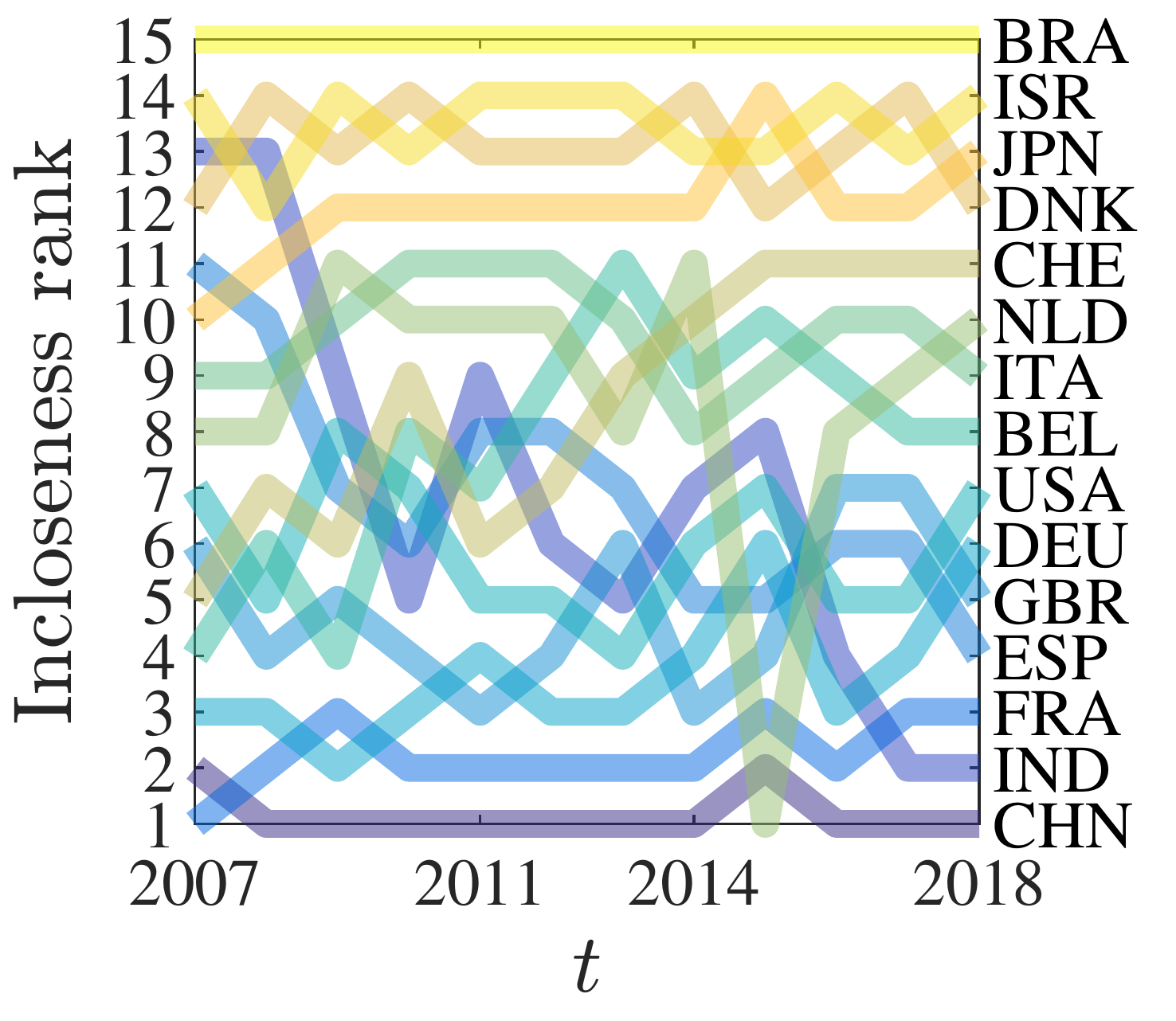}
    \includegraphics[width=0.33\linewidth]{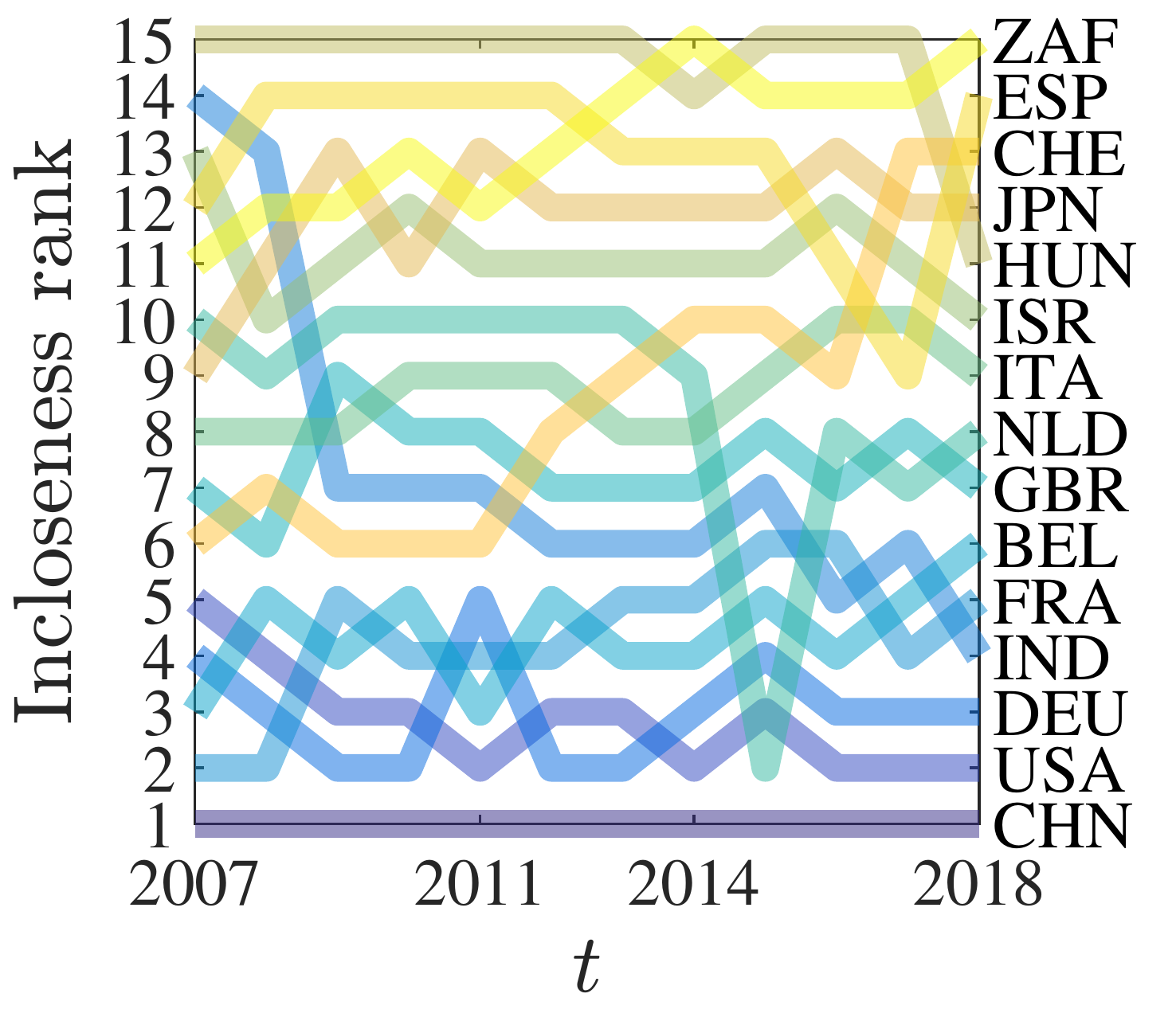}
    \includegraphics[width=0.33\linewidth]{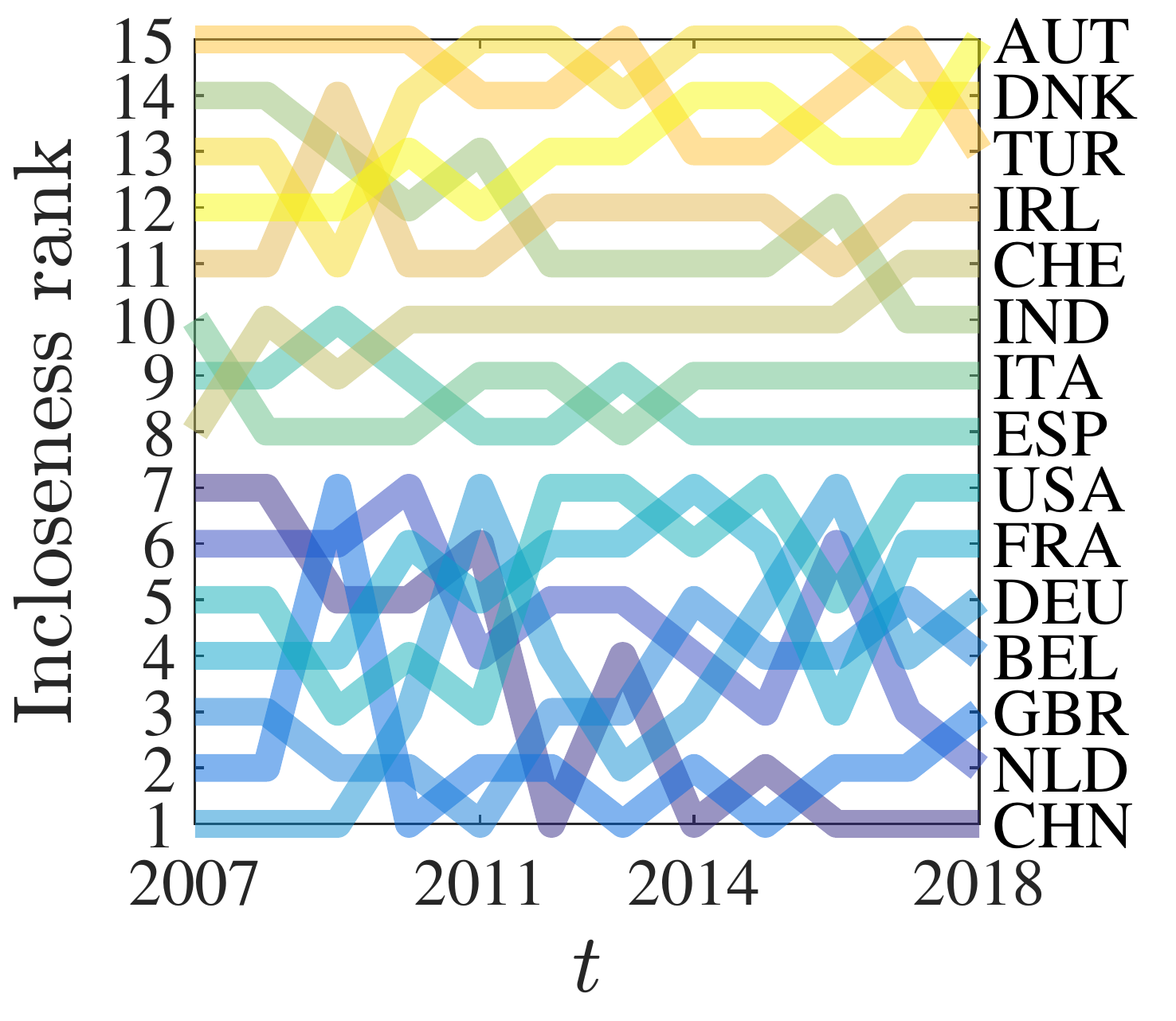}
    \includegraphics[width=0.33\linewidth]{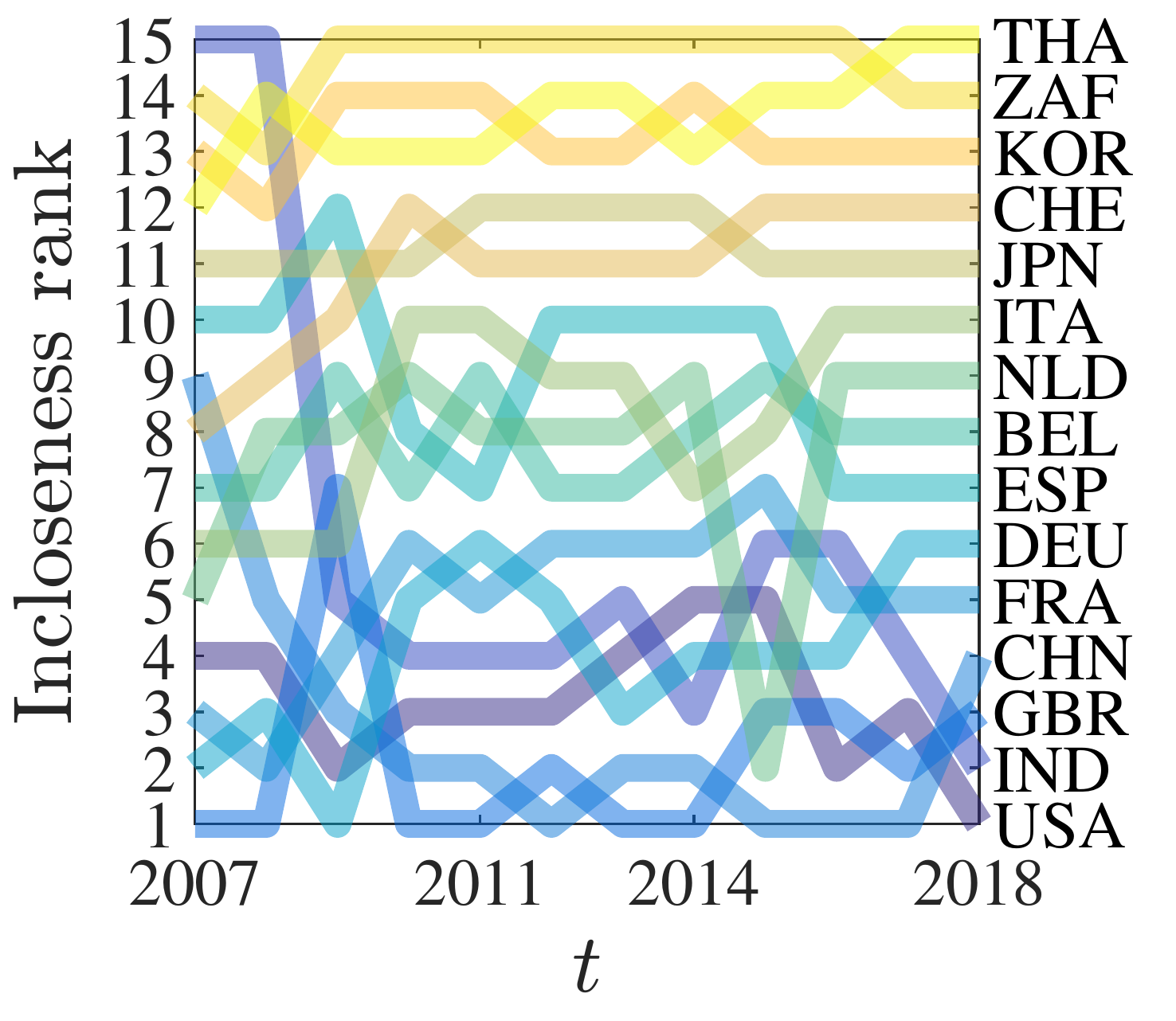}
\vskip    -9.55cm   \hskip   -16.0cm {(a)}
\vskip    -0.43cm   \hskip    -4.9cm {(b)}
\vskip    -0.43cm   \hskip     6.0cm {(c)}
\vskip     4.33cm   \hskip   -16.0cm {(d)}
\vskip    -0.43cm   \hskip    -4.9cm {(e)}
\vskip    -0.43cm   \hskip     6.0cm {(f)}
\vskip 4.05cm
  \caption{Rank evolution of the top-15 economies in the international trade networks for all pesticides (a), insecticides (b), fungicides (c), herbicides (d), disinfectants (e), and rodenticides and other similar products (f) from 2007 to 2018. The ranking is based on node in-closeness.}
  \label{Fig:Pesticide:Rank:Evo:in-closeness}
\end{figure}

We rank the economies in the international trade networks based on node in-closeness in each year from 2007 to 2018 and illustrate the rank evolution of the top-15 economies in Fig.~\ref{Fig:Pesticide:Rank:Evo:in-closeness}. For the aggregated network in 2018, as shown in Fig.~\ref{Fig:Pesticide:Rank:Evo:in-closeness}(a), the top-15 economies are China, the Netherlands, the USA, France, the United Kingdom, Germany, Belgium, India, Spain, Italy, Japan, Turkey, Switzerland, Austria, and Korea, which are large economies. For the categorized networks, the list of the top-15 economies may change slightly. We observe that, in 2018, China ranks No. 1 based on in-closeness in the international trade networks for insecticides, fungicides, herbicides, and disinfectants, while the USA ranks No. 1 in the rodenticides network.

Similarly, we can define the out-closeness as follows,
\begin{equation}
    C_i^{\mathrm{out}} =\frac{N-1}{\sum_{j\in{{\mathbf{n}}_i^{\mathrm{out}}}}l_{ij}},
    \label{Eq:Centrality:outCloseness}
\end{equation}
where ${\mathbf{n}}_i^{\mathrm{out}}$ is the set of nodes that have at least one directed path pointing from $i$ to $j$.

\begin{figure}[!ht]
\centering
    \includegraphics[width=0.33\linewidth]{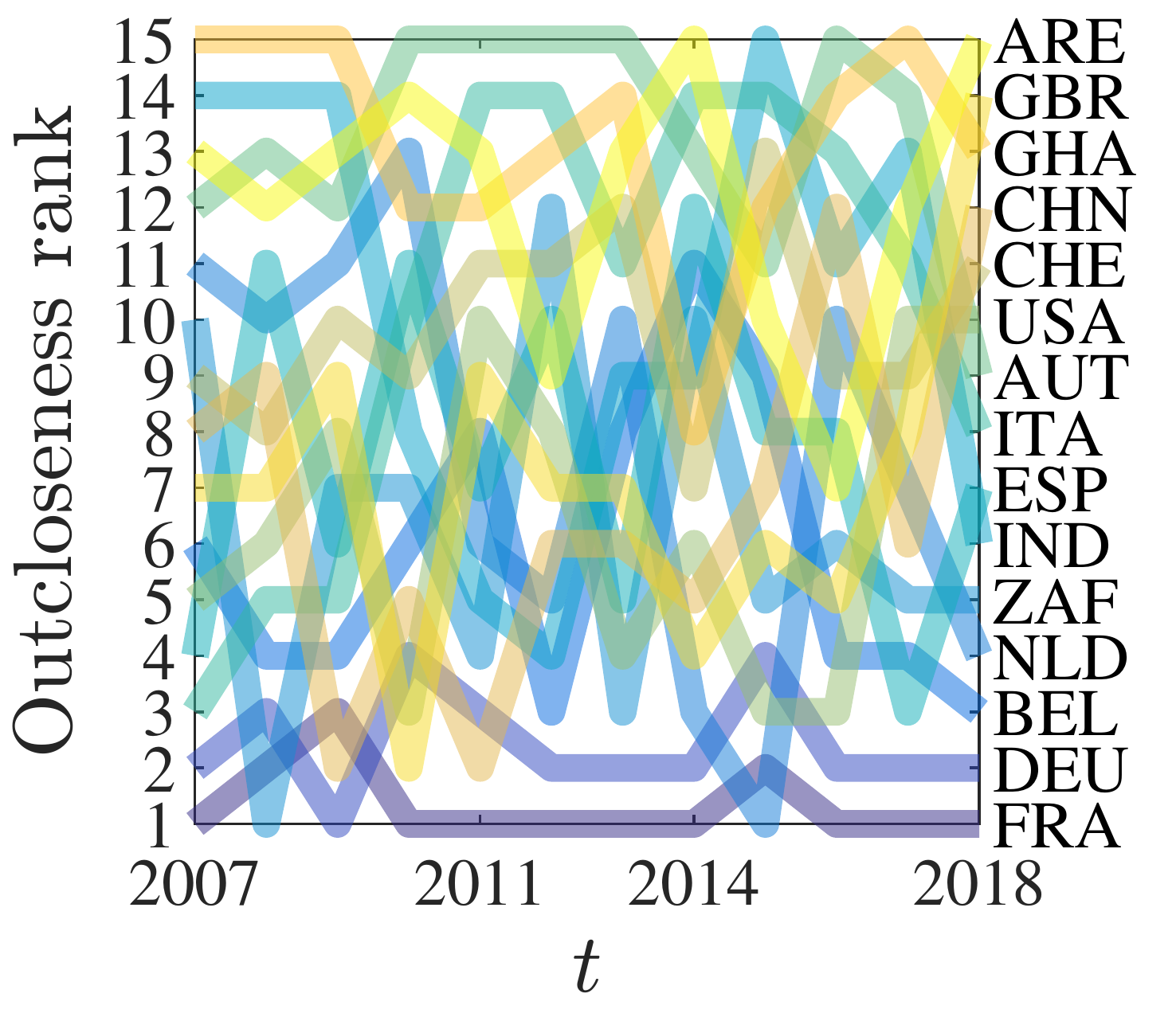}
    \includegraphics[width=0.33\linewidth]{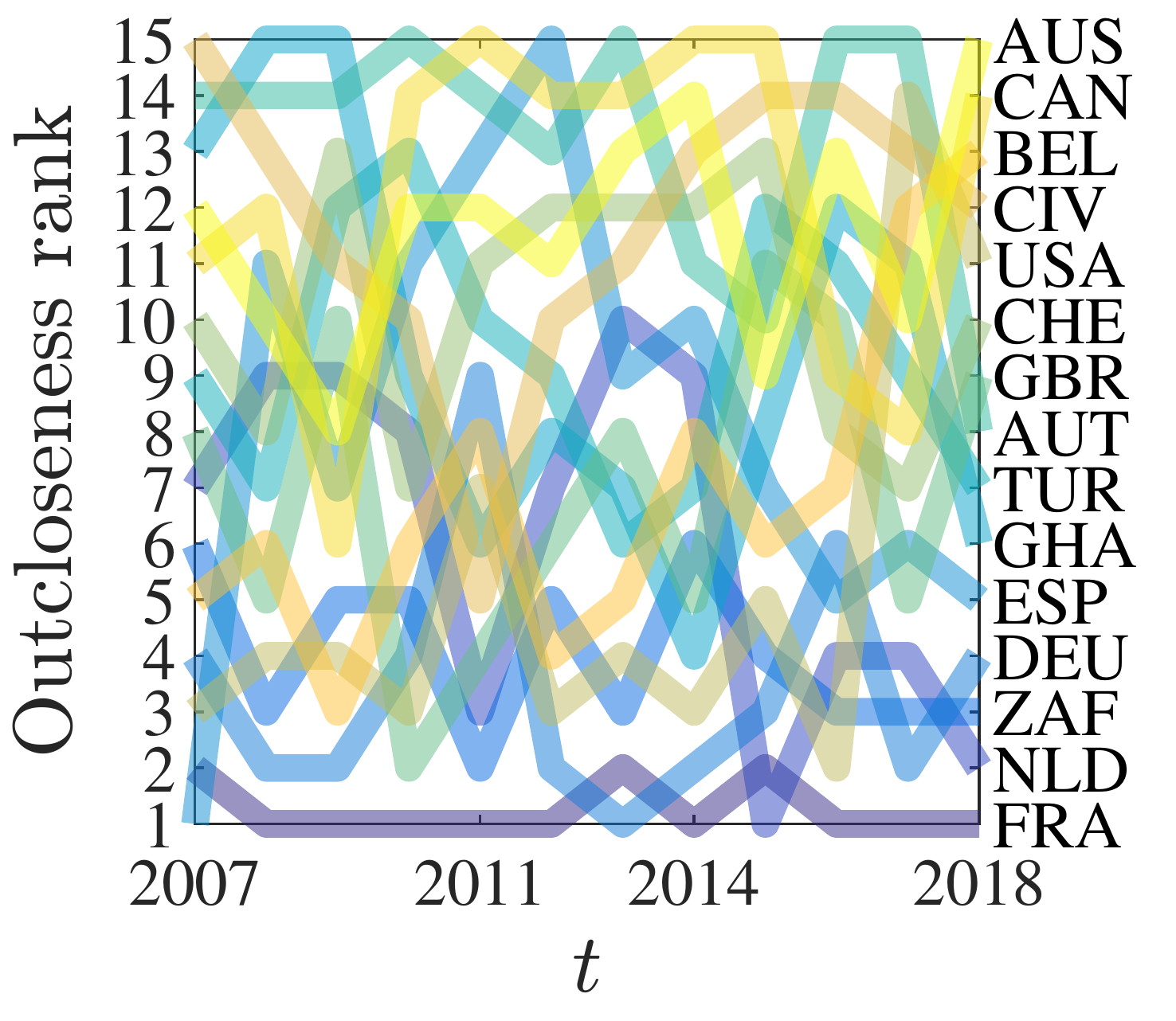}
    \includegraphics[width=0.33\linewidth]{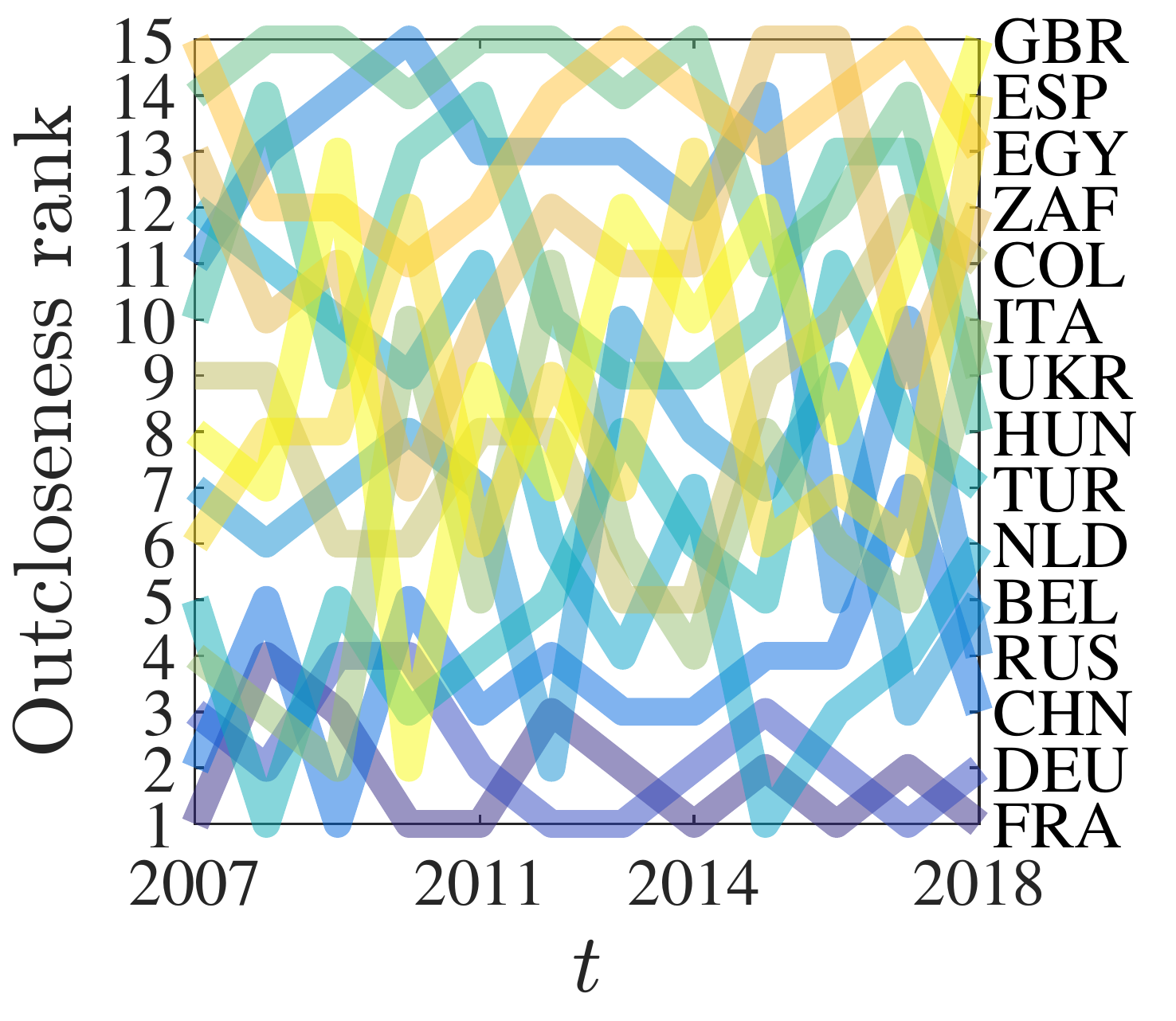}
    \includegraphics[width=0.33\linewidth]{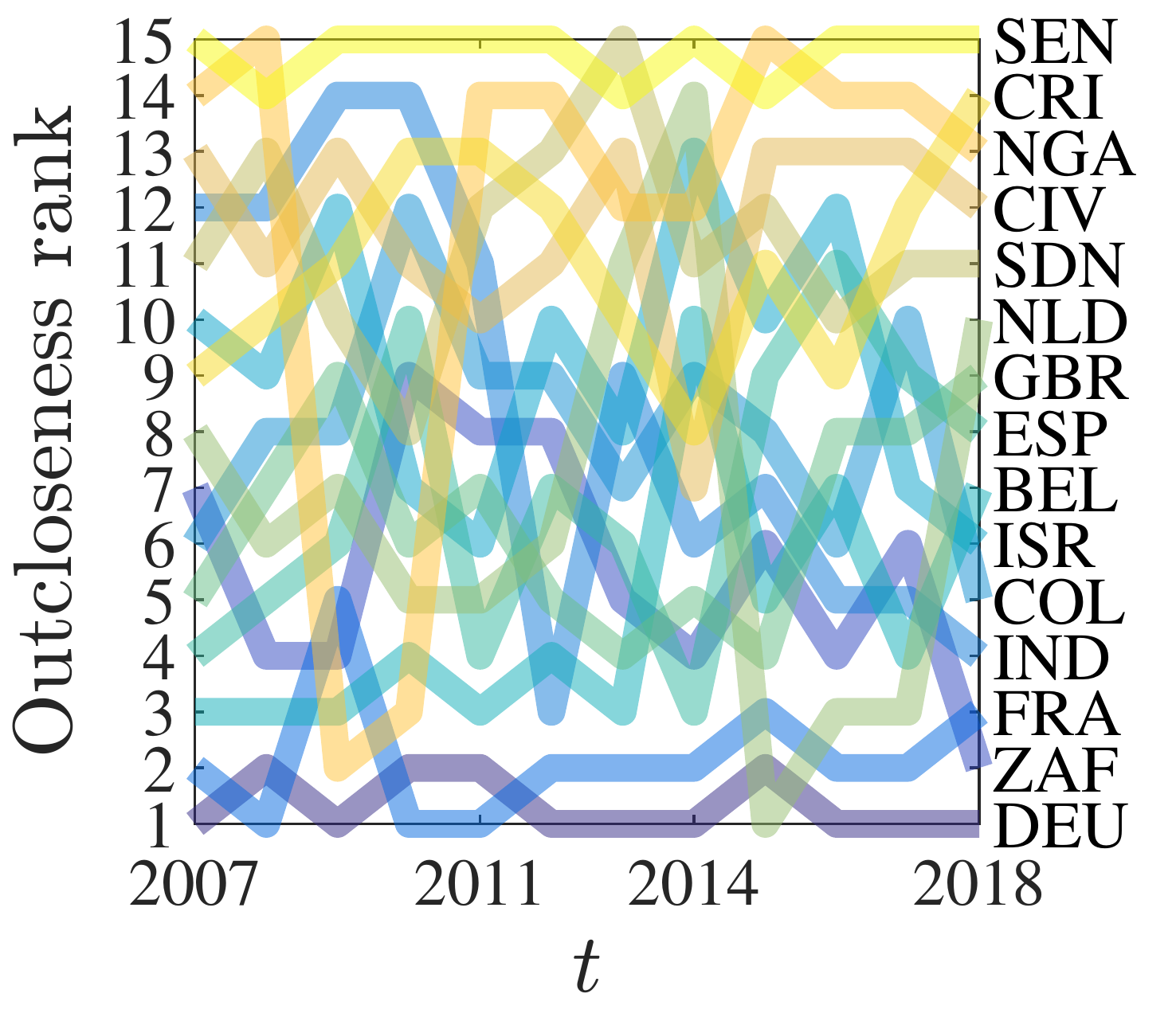}
    \includegraphics[width=0.33\linewidth]{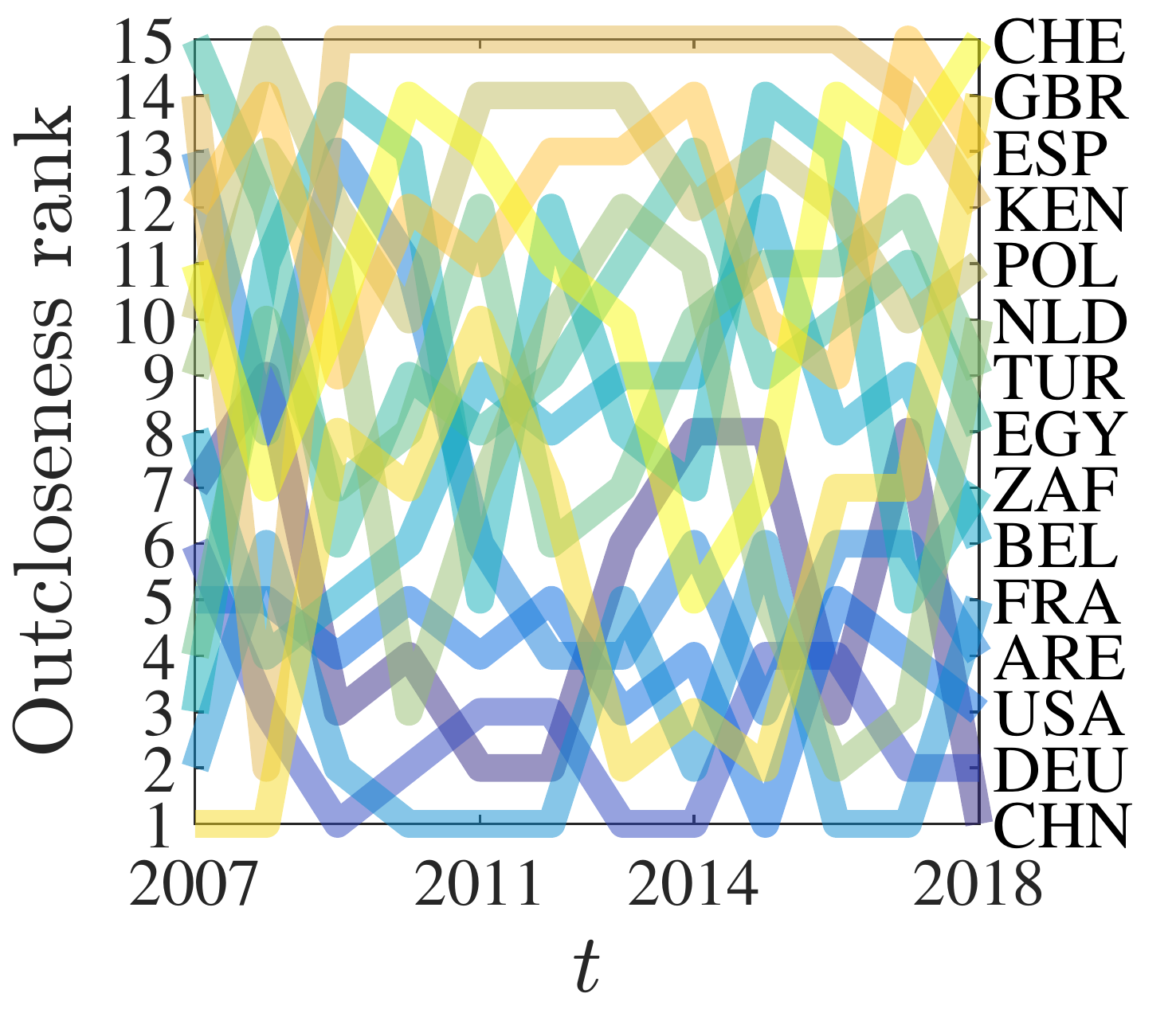}
    \includegraphics[width=0.33\linewidth]{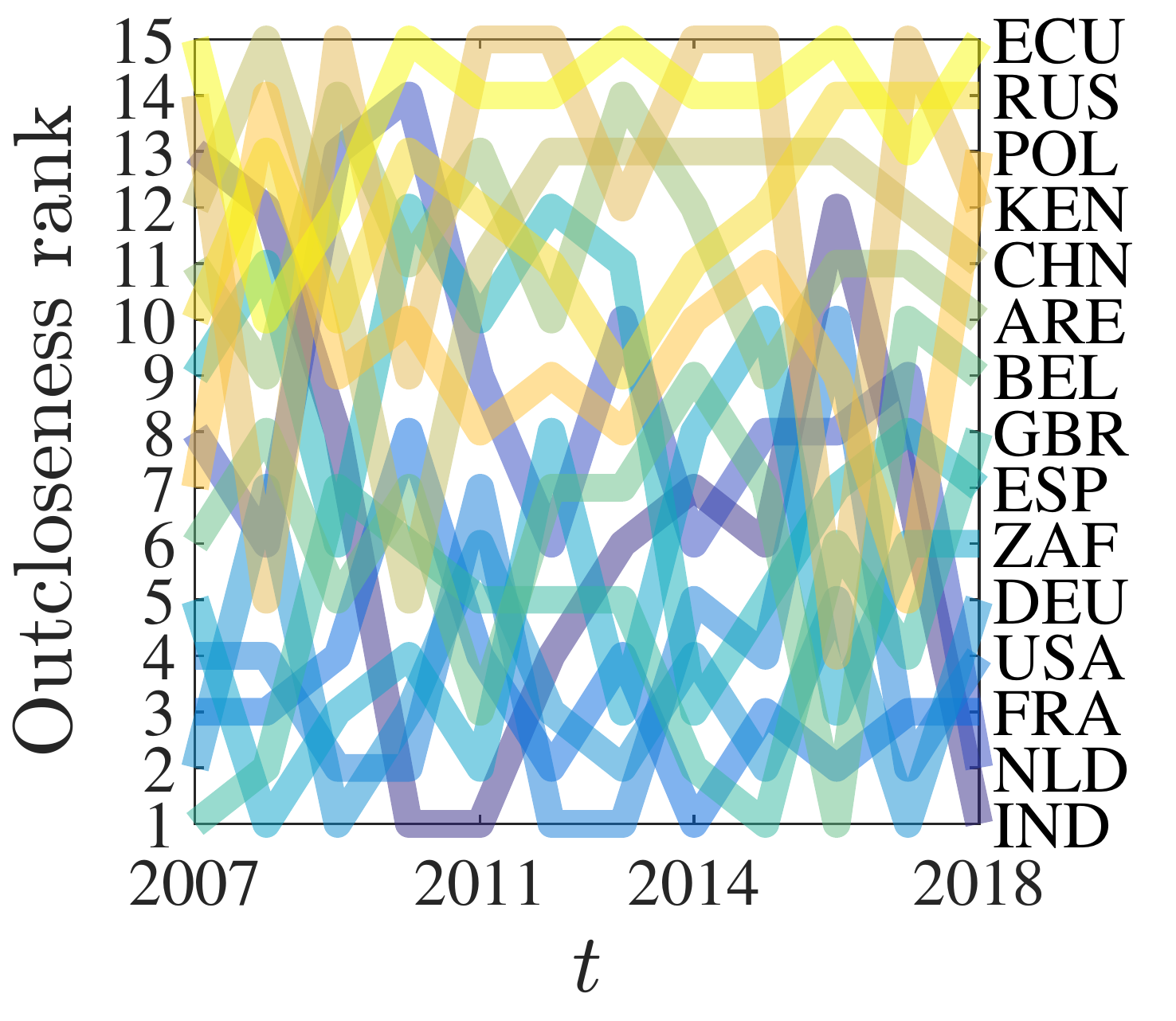}
\vskip    -9.55cm   \hskip   -16.0cm {(a)}
\vskip    -0.43cm   \hskip    -4.9cm {(b)}
\vskip    -0.43cm   \hskip     6.0cm {(c)}
\vskip     4.33cm   \hskip   -16.0cm {(d)}
\vskip    -0.43cm   \hskip    -4.9cm {(e)}
\vskip    -0.43cm   \hskip     6.0cm {(f)}
\vskip 4.05cm
  \caption{Rank evolution of the top-15 economies in the international trade networks for all pesticides (a), insecticides (b), fungicides (c), herbicides (d), disinfectants (e), and rodenticides and other similar products (f) from 2007 to 2018. The ranking is based on node out-closeness.}
  \label{Fig:Pesticide:Rank:Evo:out-closeness}
\end{figure}

We also rank the economies in the international trade networks based on node out-closeness in each year from 2007 to 2018 and illustrate the rank evolution of the top-15 economies in Fig.~\ref{Fig:Pesticide:Rank:Evo:out-closeness}. For the aggregated network in 2018, as shown in Fig.~\ref{Fig:Pesticide:Rank:Evo:out-closeness}(a), the top-15 economies are France, Germany, Belgium, the Netherlands, South Africa, India, Spain, Italy, Austria, the USA, Switzerland, China, Ghana, the United Kingdom, and the United Arab Emirates, which are mainly large economies.

In 2018, France ranks No. 1 in the networks for insecticides and fungicides, Germany ranks No. 1 in the herbicides network, China ranks No. 1 in the disinfectants network, and India ranks No. 1 in the rodenticides network. The historic ranks of France and Germany are more stable than China and India. Although Belgium ranks fifth or lower in the categorized trade networks, it ranks third in the aggregated network in 2018.

\begin{figure}[!ht]
\centering
    \includegraphics[width=0.33\linewidth]{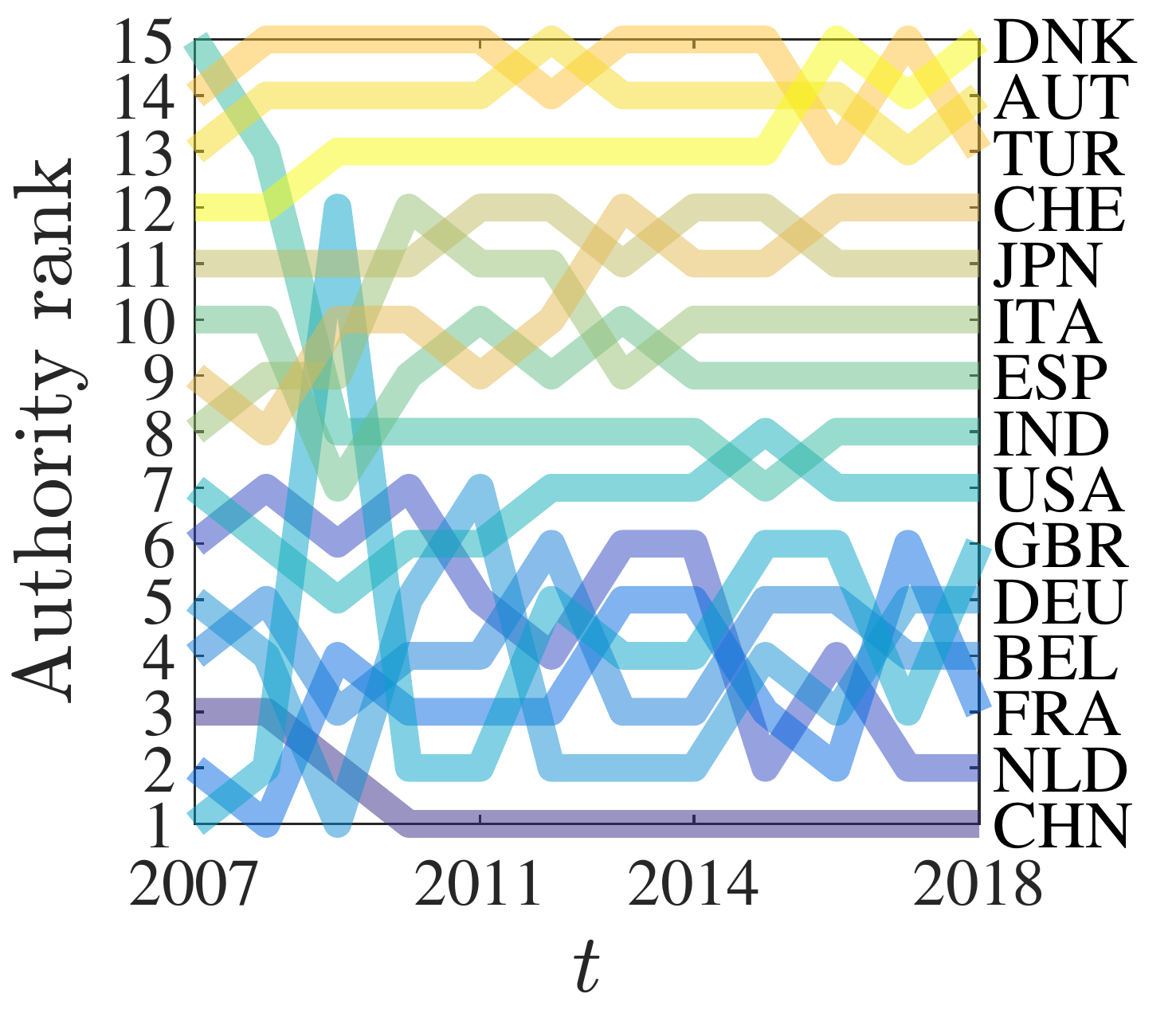}
    \includegraphics[width=0.33\linewidth]{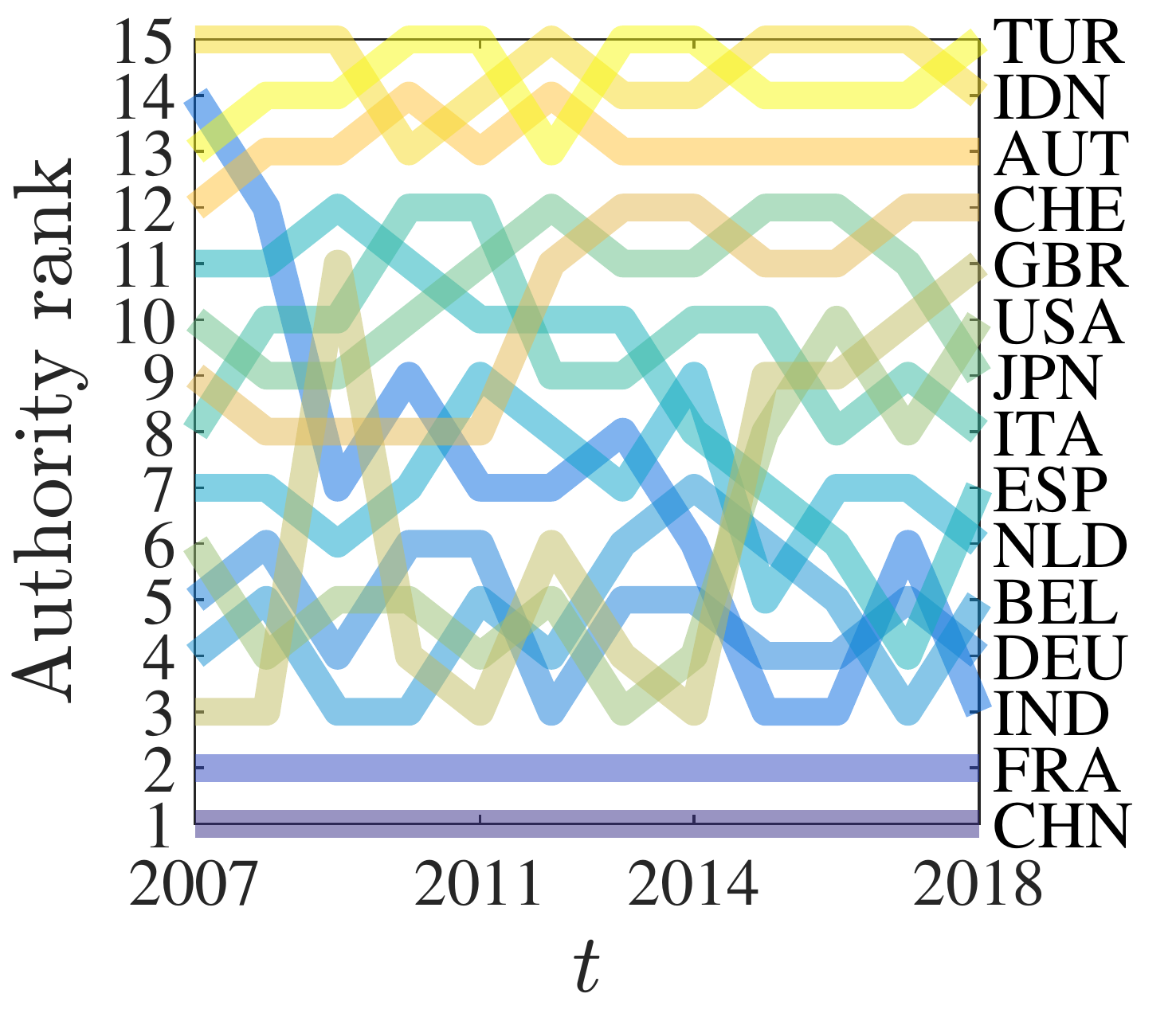}
    \includegraphics[width=0.33\linewidth]{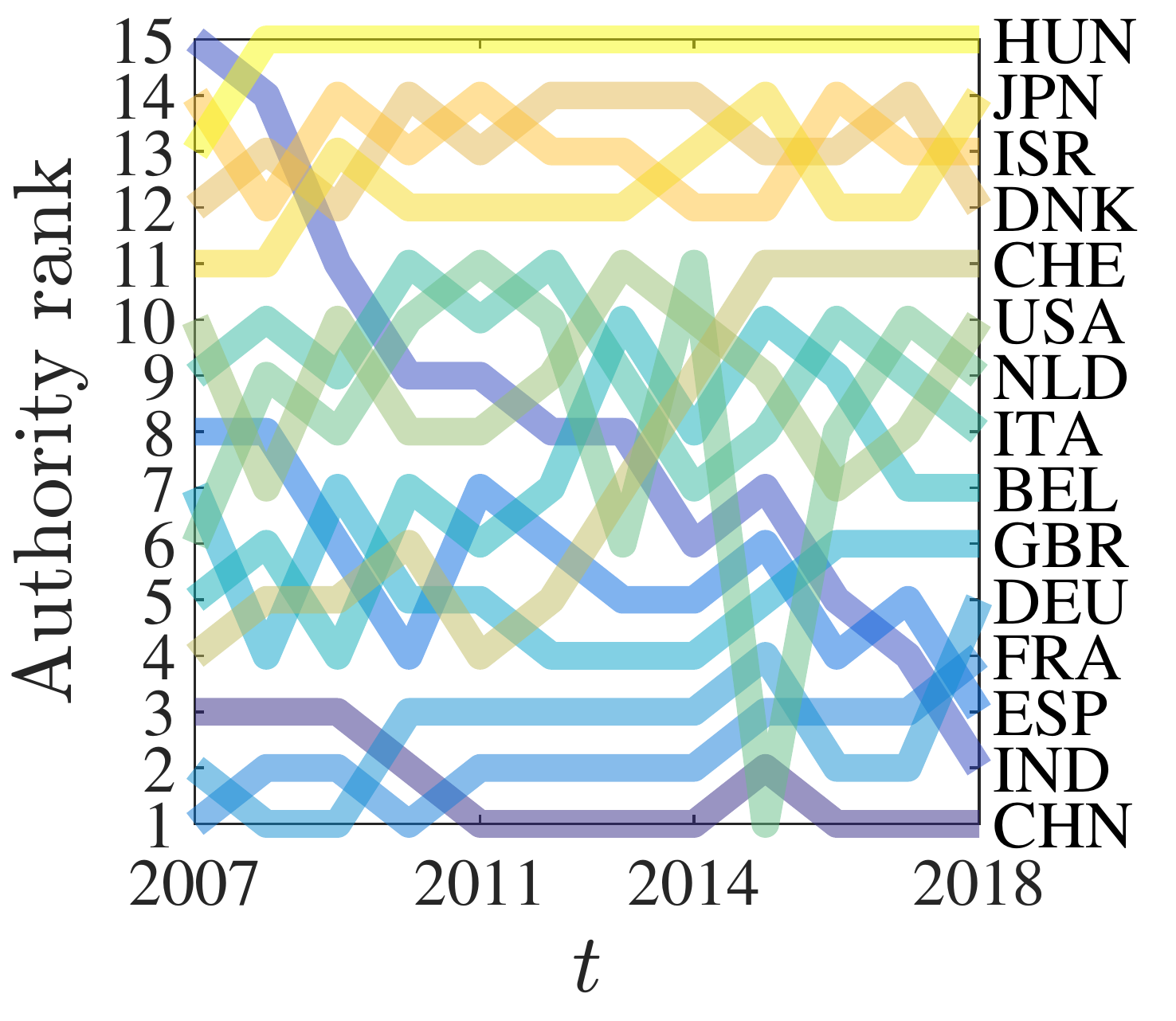}
    \includegraphics[width=0.33\linewidth]{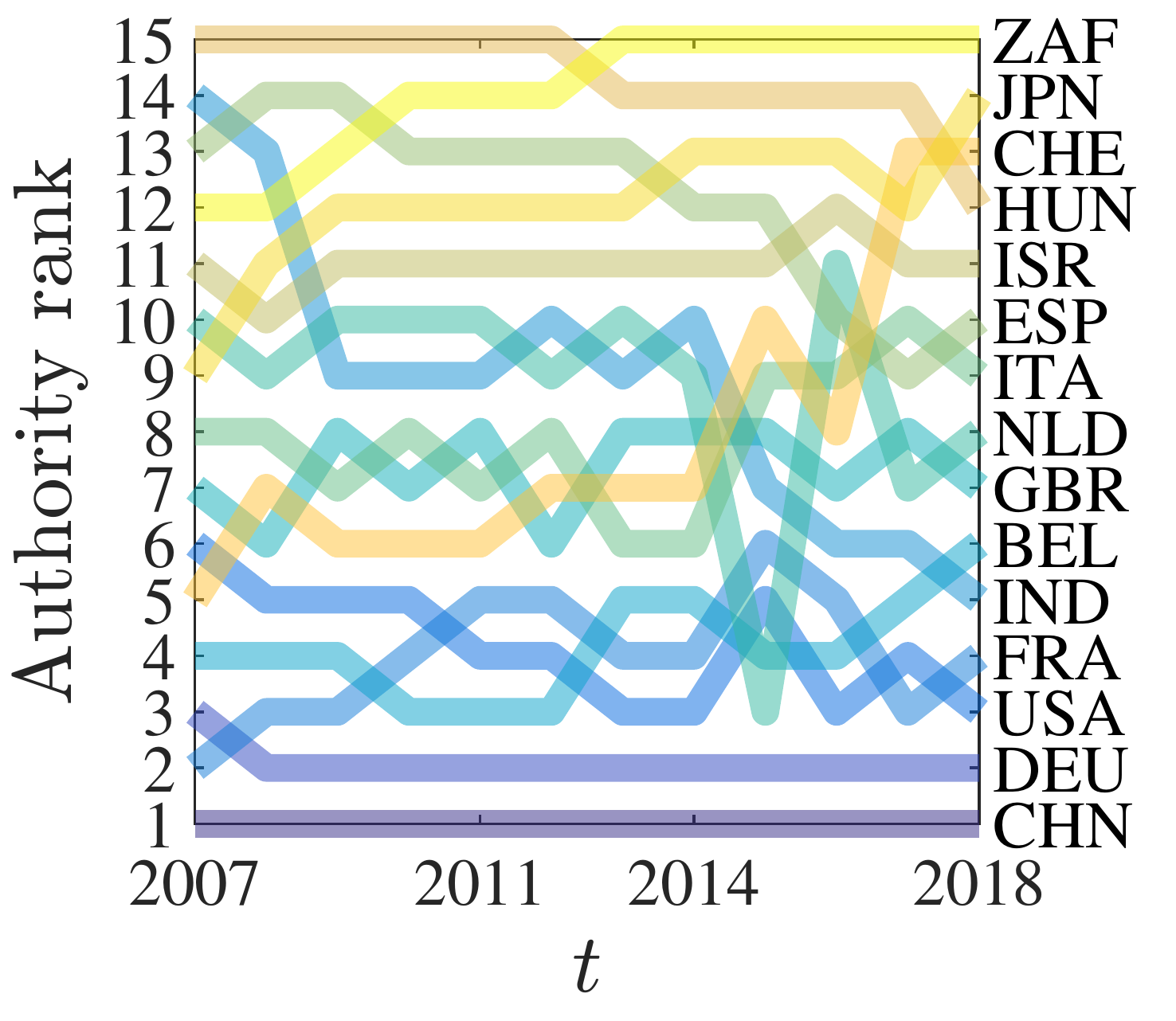}
    \includegraphics[width=0.33\linewidth]{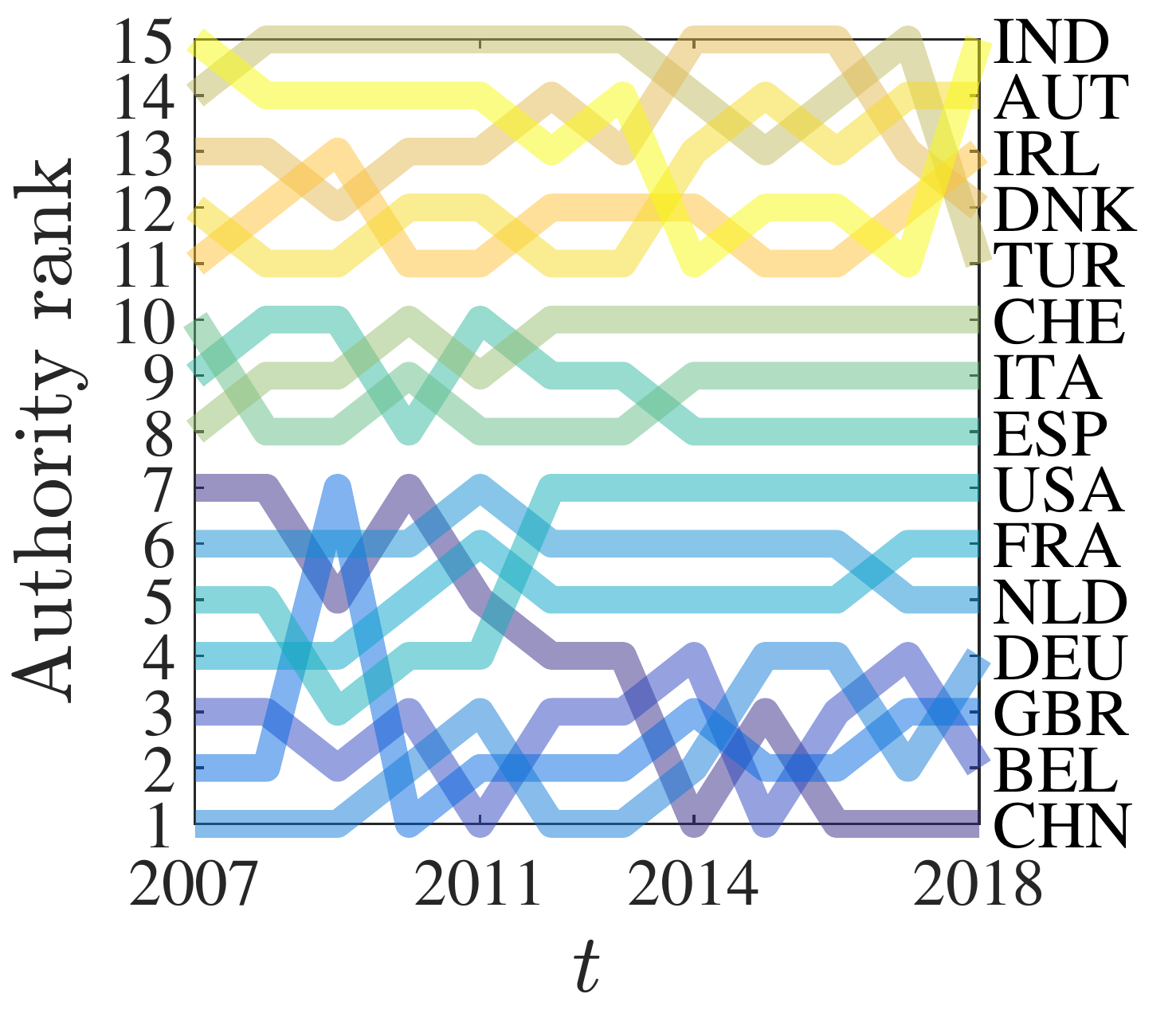}
    \includegraphics[width=0.33\linewidth]{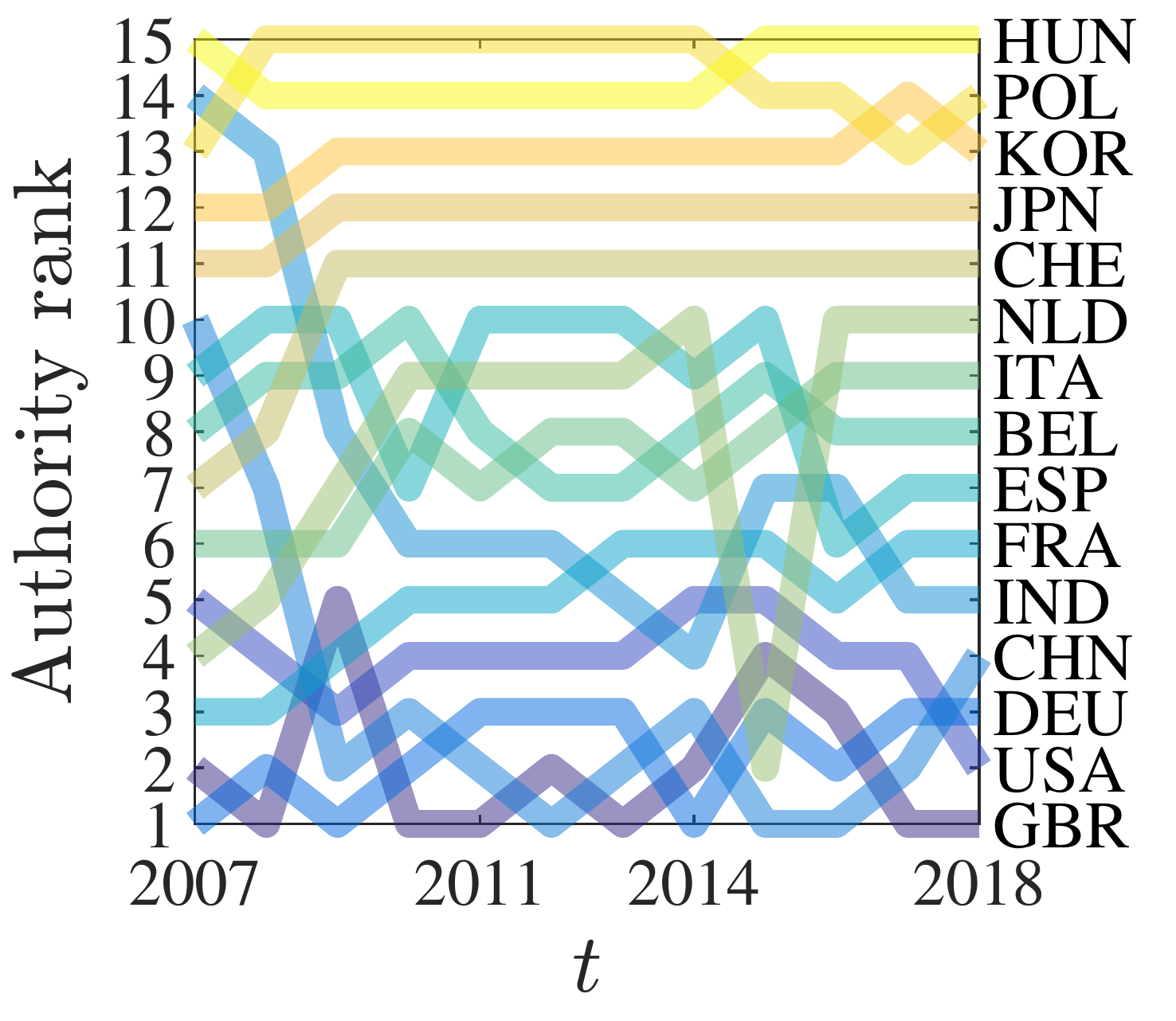}
\vskip    -9.55cm   \hskip   -16.0cm {(a)}
\vskip    -0.43cm   \hskip    -4.9cm {(b)}
\vskip    -0.43cm   \hskip     6.0cm {(c)}
\vskip     4.33cm   \hskip   -16.0cm {(d)}
\vskip    -0.43cm   \hskip    -4.9cm {(e)}
\vskip    -0.43cm   \hskip     6.0cm {(f)}
\vskip 4.05cm
  \caption{Rank evolution of the top-15 economies in the international trade networks for all pesticides (a), insecticides (b), fungicides (c), herbicides (d), disinfectants (e), and rodenticides and other similar products (f) from 2007 to 2018. The ranking is based on nodes' authority scores.}
  \label{Fig:Pesticide:Rank:Evo:Auth}
\end{figure}

\subsection{Authority score and hub score}
\label{S2:HITS}

Authorities and hubs are two kinds of vital nodes in directed networks, which were introduced by Kleinberg in his HITS algorithm \citep{Kleinberg-1999-JACM}. Authorities are target nodes with large in-degrees, while hubs are source nodes with large out-degrees. The authority score $a_i$ and hub score $h_i$ of each node can be defined and calculated in an iterative way, 
\begin{equation}
\left\{
   \begin{aligned}
       a_i = \sum_j a_{ji} h_j \\
       h_i = \sum_j a_{ij} a_j
\end{aligned}\right.
\end{equation}
which are normalized in each iteration.

We determine the authority scores and hub scores of nodes in each international pesticide trade network and rank the economies in each year from 2007 to 2018. The rank evolution of the top-15 economies based on authority scores is illustrated in Fig.~\ref{Fig:Pesticide:Rank:Evo:Auth}. For the aggregated network in 2018, as shown in Fig.~\ref{Fig:Pesticide:Rank:Evo:Auth}(a), the top-15 economies are China, the Netherlands, France, Belgium, Germany, the United Kingdom, the USA, India, Spain, Italy, Japan, Switzerland, Turkey, Austria, and Denmark, which are large economies. Concerning the categorized networks shown in Fig.~\ref{Fig:Pesticide:Rank:Evo:Auth}(b-f), in 2018, China ranks No. 1 in the international trade networks for insecticides, fungicides, herbicides, and disinfectants, while the United Kingdom ranks No. 1 in the rodenticides network.

\begin{figure}[!ht]
\centering
    \includegraphics[width=0.33\linewidth]{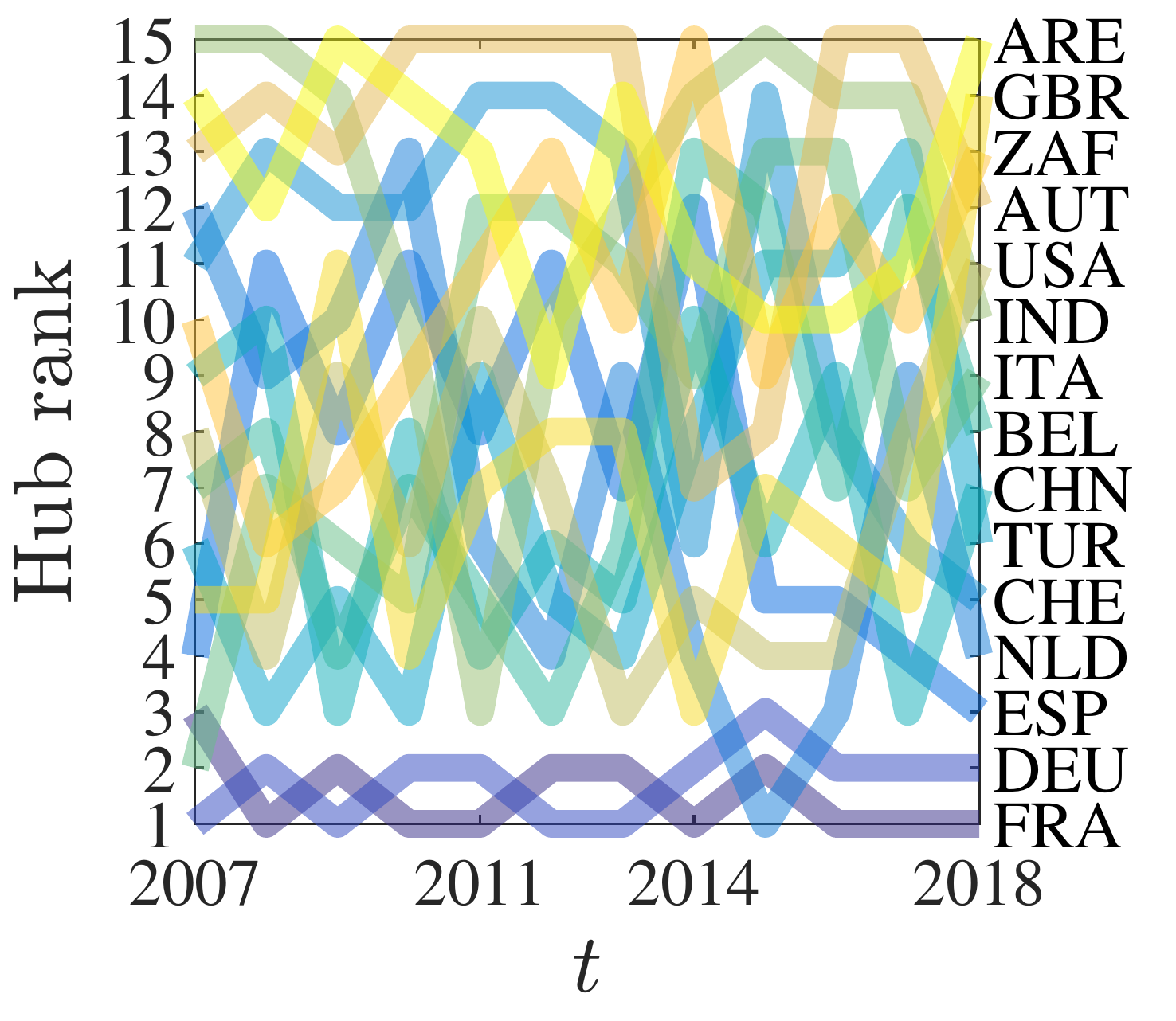}
    \includegraphics[width=0.33\linewidth]{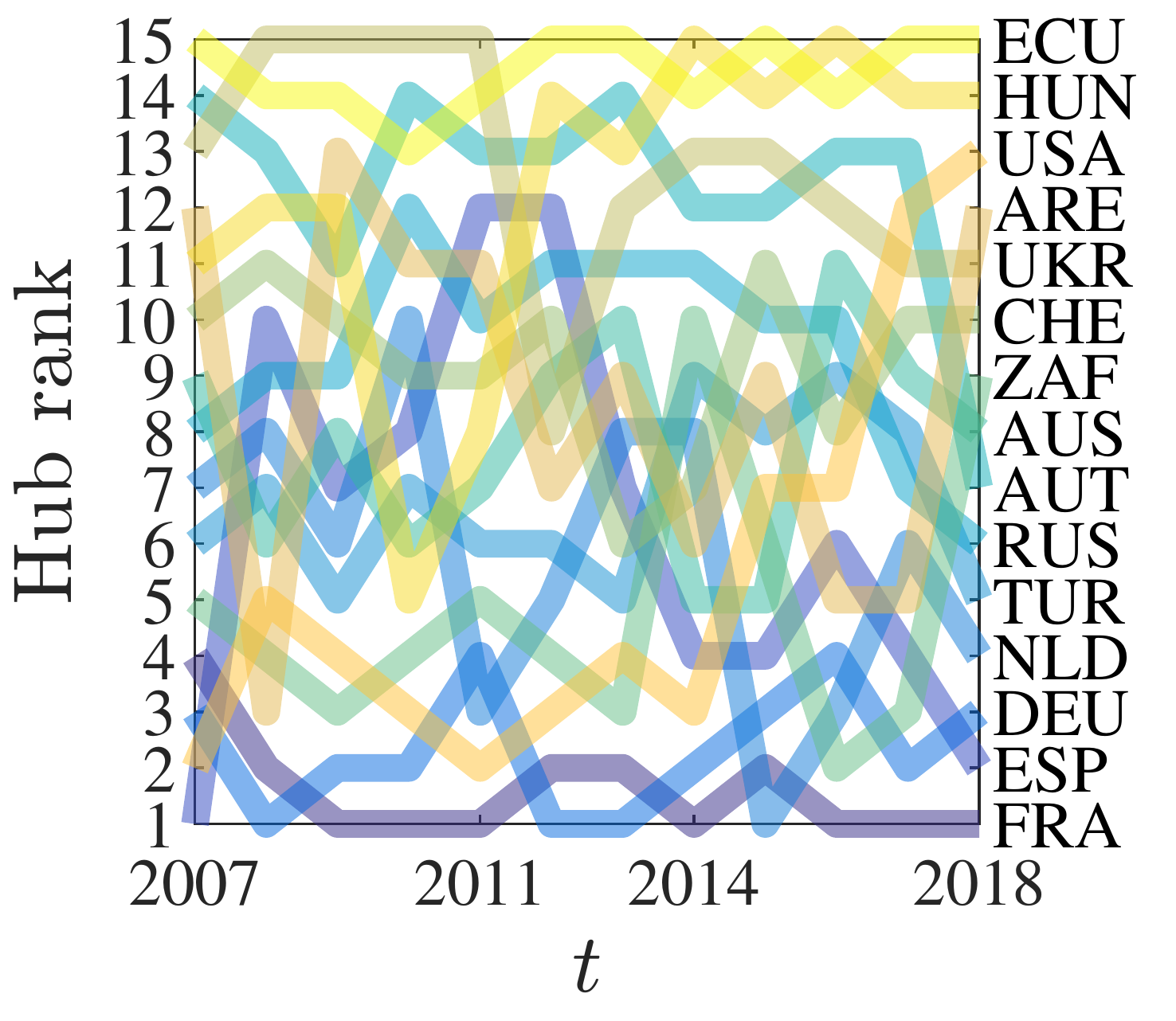}
    \includegraphics[width=0.33\linewidth]{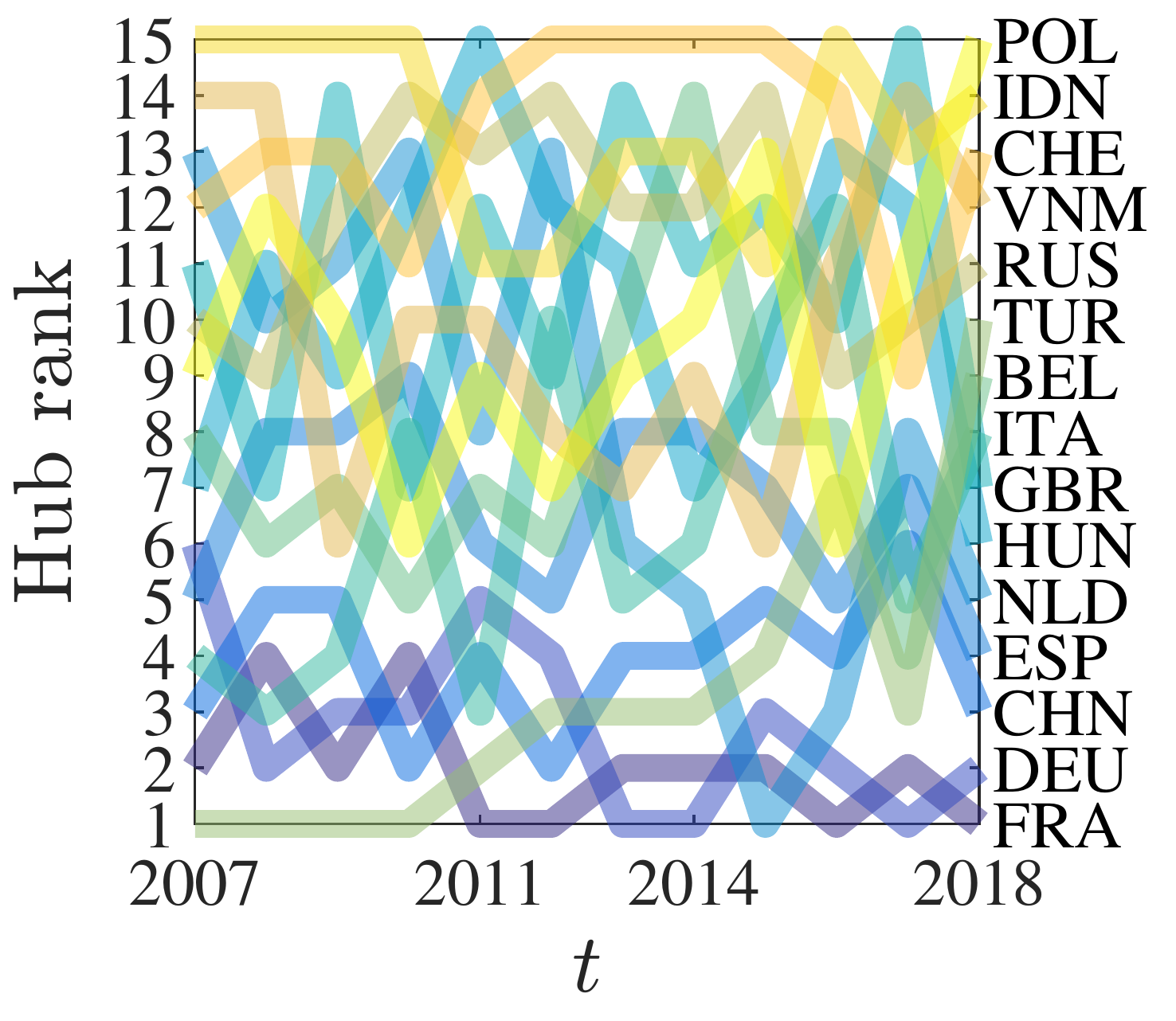}
    \includegraphics[width=0.33\linewidth]{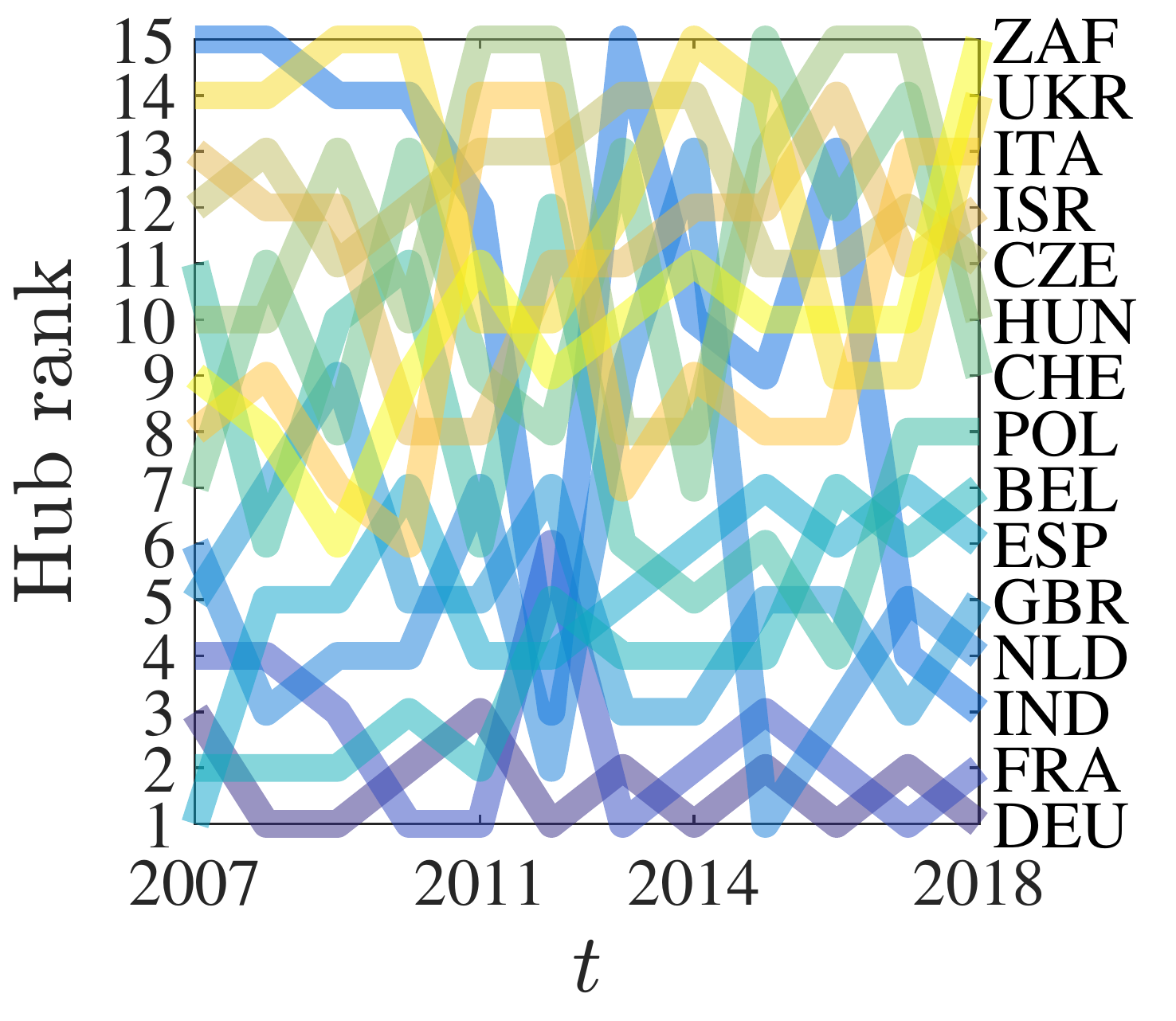}
    \includegraphics[width=0.33\linewidth]{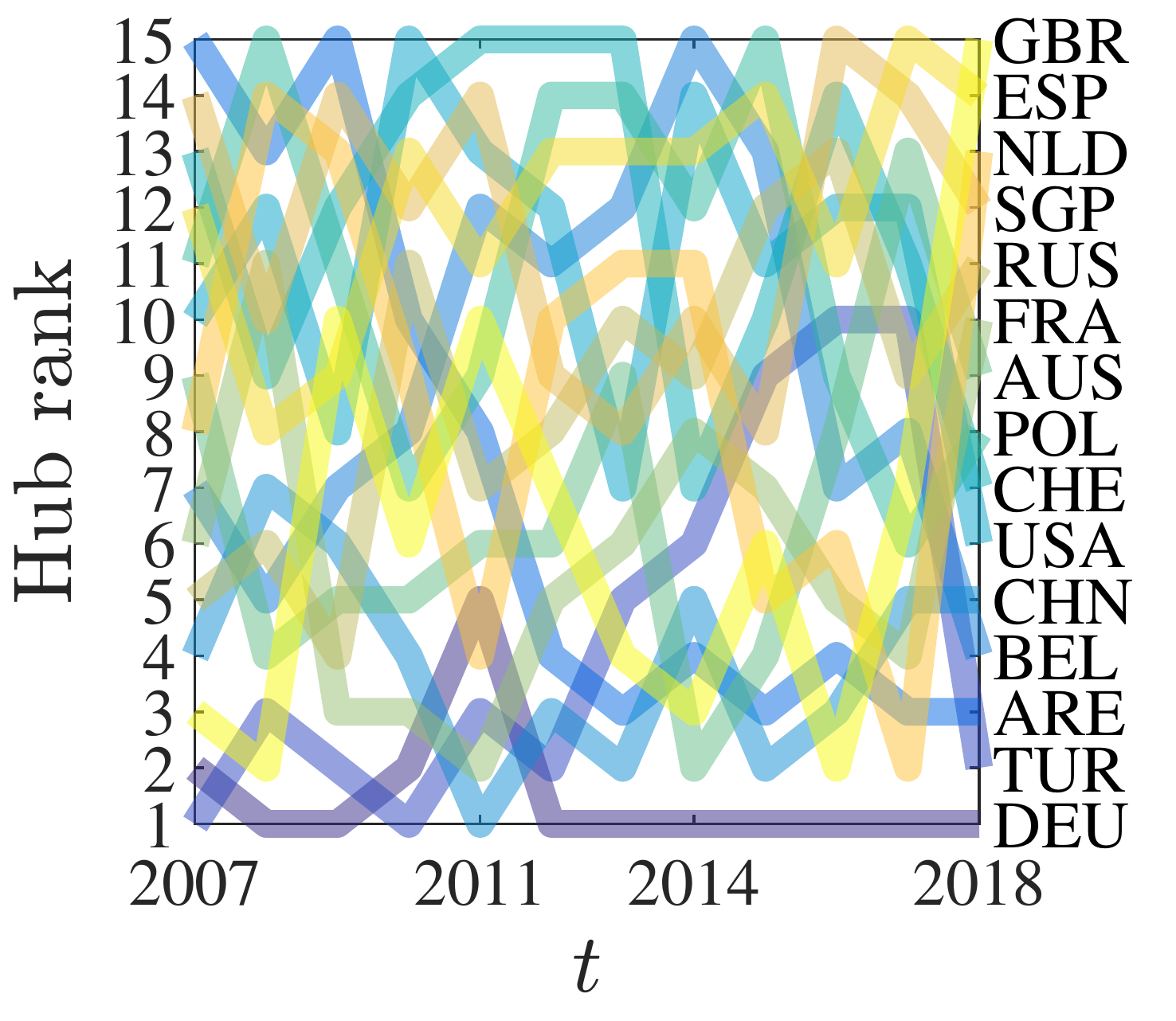}
    \includegraphics[width=0.33\linewidth]{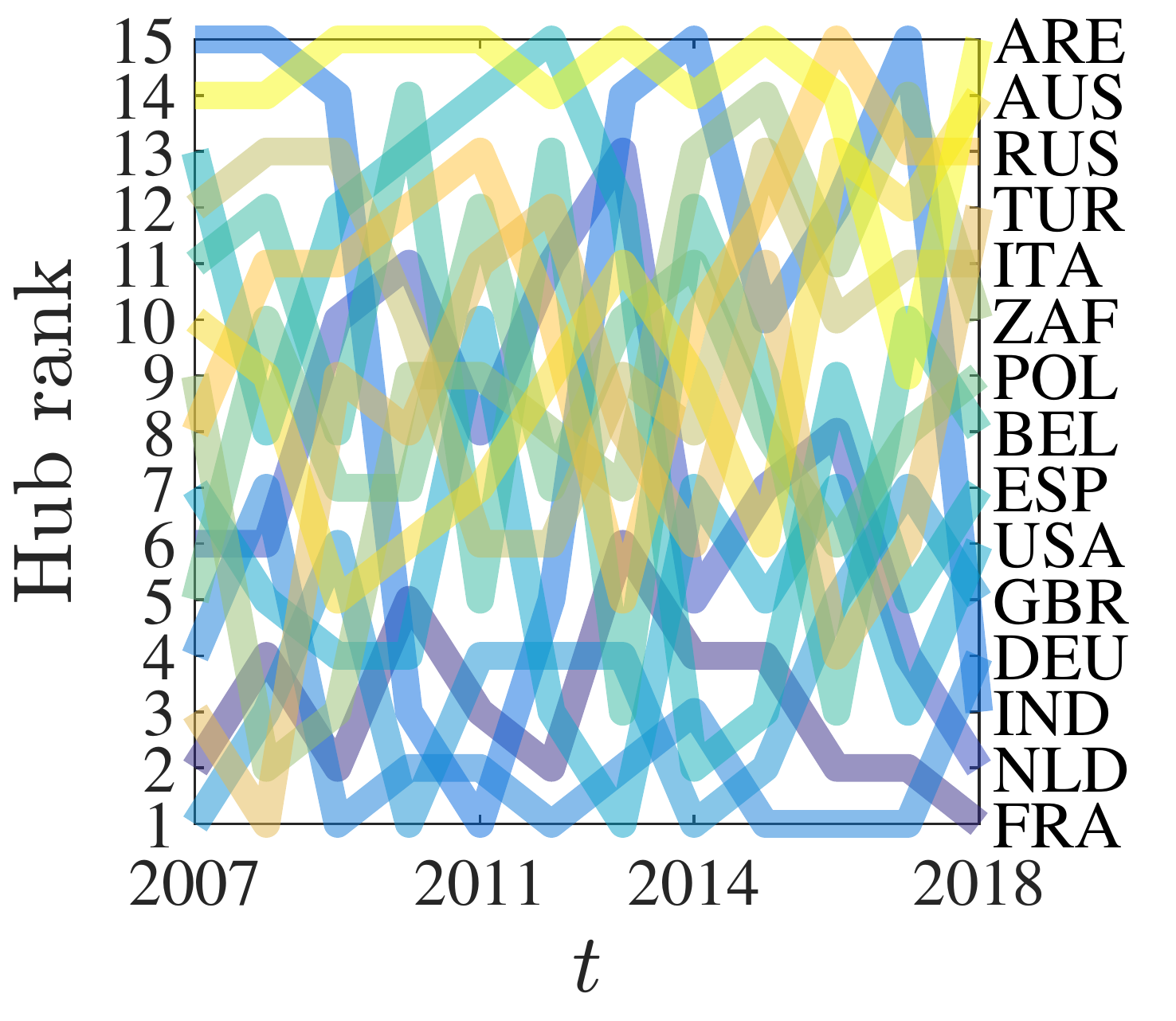}
\vskip    -9.55cm   \hskip   -16.0cm {(a)}
\vskip    -0.43cm   \hskip    -4.9cm {(b)}
\vskip    -0.43cm   \hskip     6.0cm {(c)}
\vskip     4.33cm   \hskip   -16.0cm {(d)}
\vskip    -0.43cm   \hskip    -4.9cm {(e)}
\vskip    -0.43cm   \hskip     6.0cm {(f)}
\vskip 4.05cm
  \caption{Rank evolution of the top-15 economies in the international trade networks for all pesticides (a), insecticides (b), fungicides (c), herbicides (d), disinfectants (e), and rodenticides and other similar products (f) from 2007 to 2018. The ranking is based on nodes' hub scores.}
  \label{Fig:Pesticide:Rank:Evo:Hub}
\end{figure}

The rank evolution of the top-15 economies based on hub scores is illustrated in Fig.~\ref{Fig:Pesticide:Rank:Evo:Hub}. For the aggregated network in 2018, as shown in Fig.~\ref{Fig:Pesticide:Rank:Evo:Hub}(a), the top-15 economies are France, Germany, Spain, the Netherlands, Switzerland, Turkey, China, Belgium, Italy, India, the USA, Austria, South Africa, the United Kingdom, and United Arab Emirates, which are mainly large economies. 
In 2018, France ranks No. 1 in the international trade networks for insecticides, fungicides and rodenticides, while Germany ranks No. 1 in the networks for herbicides and disinfectants.

\begin{figure}[!ht]
\centering
    \includegraphics[width=0.33\linewidth]{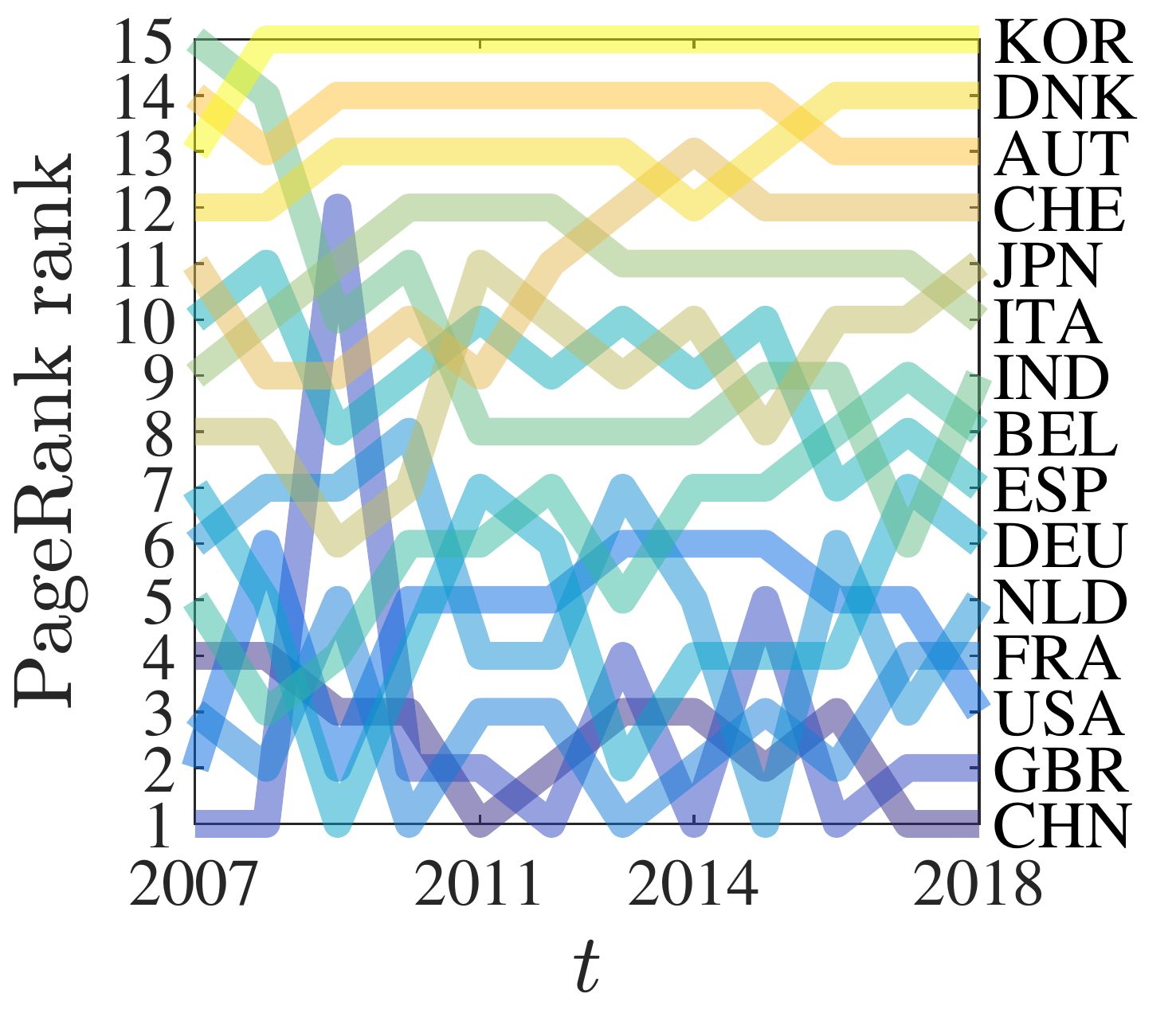}
    \includegraphics[width=0.33\linewidth]{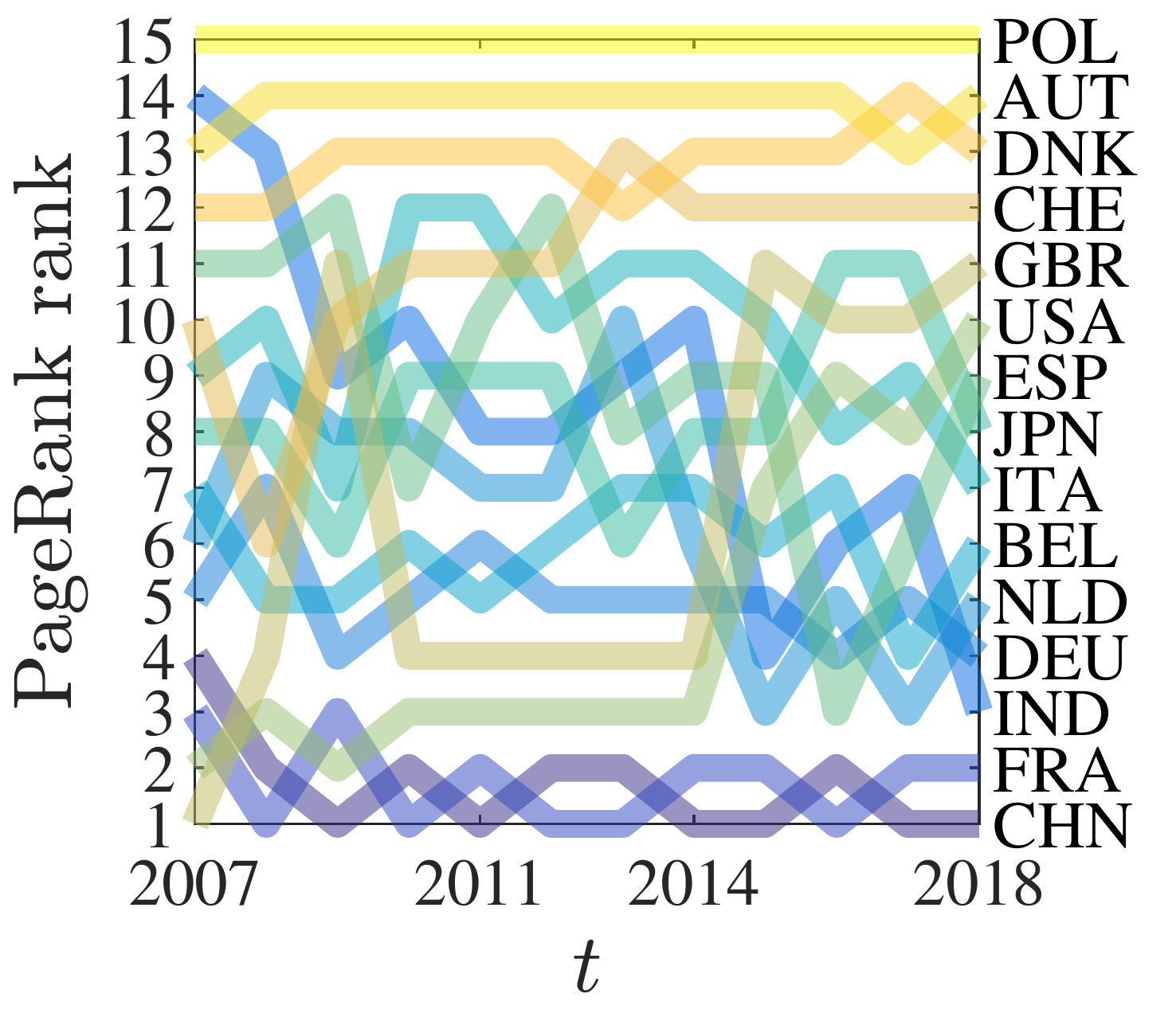}
    \includegraphics[width=0.33\linewidth]{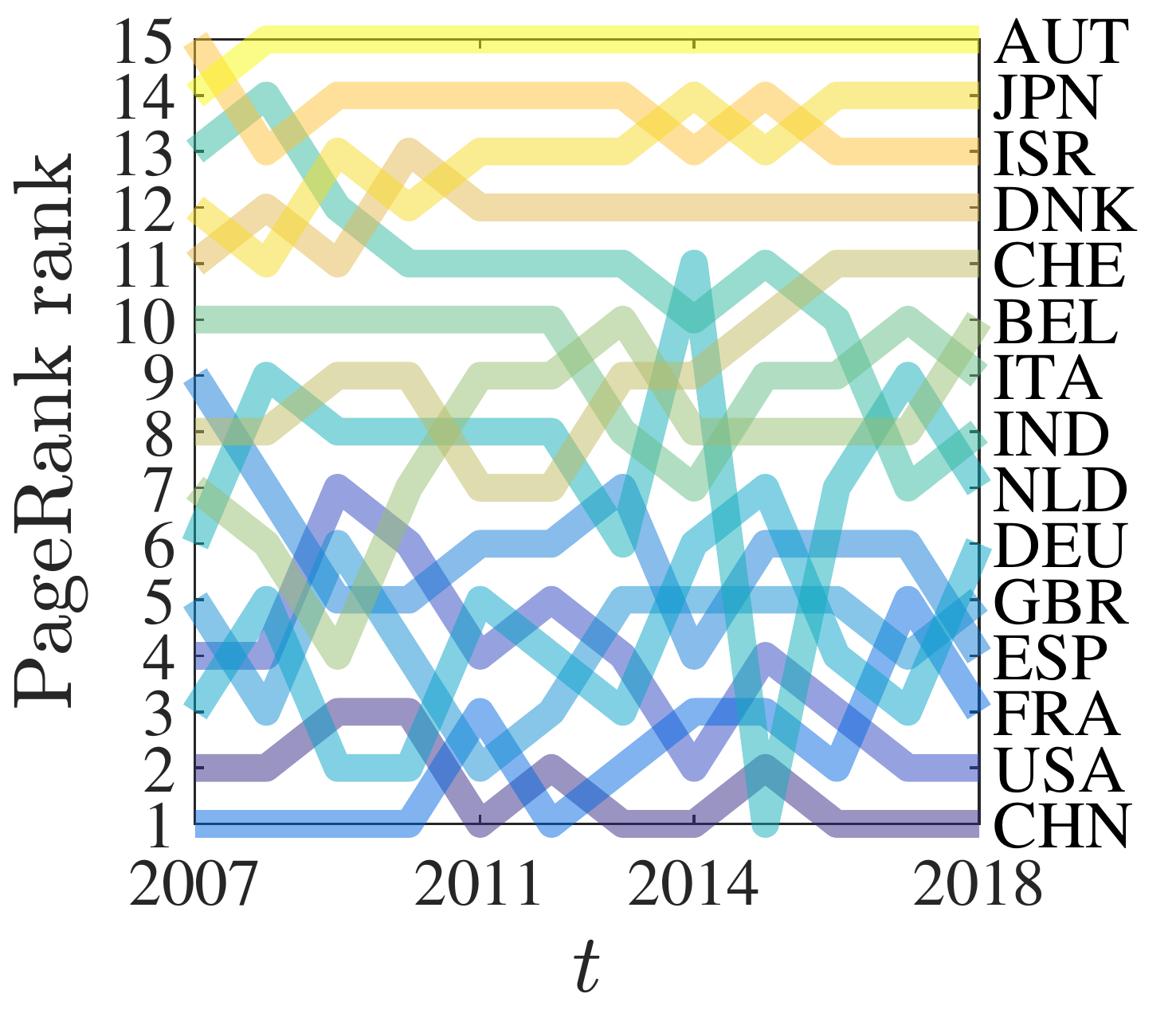}
    \includegraphics[width=0.33\linewidth]{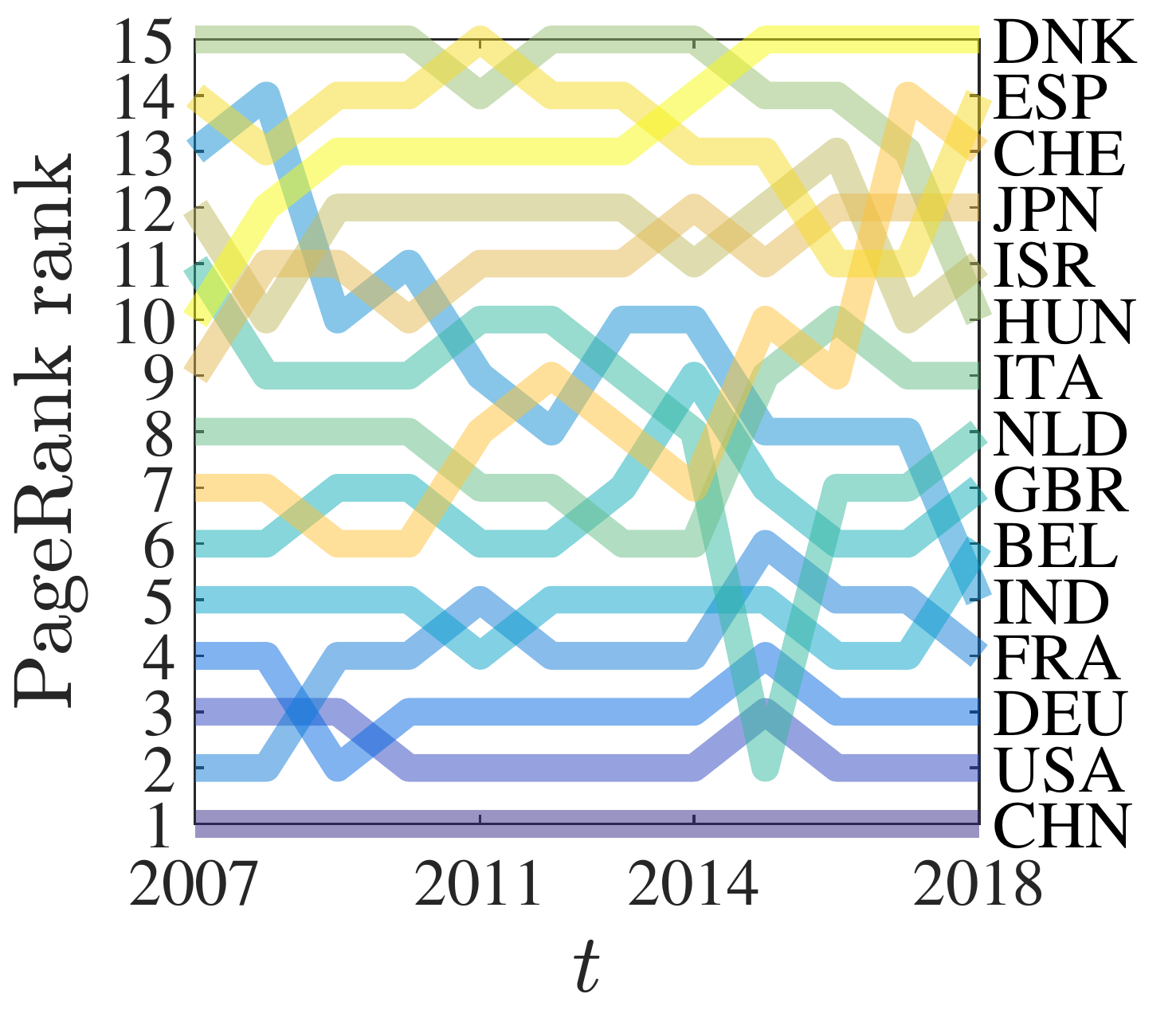}
    \includegraphics[width=0.33\linewidth]{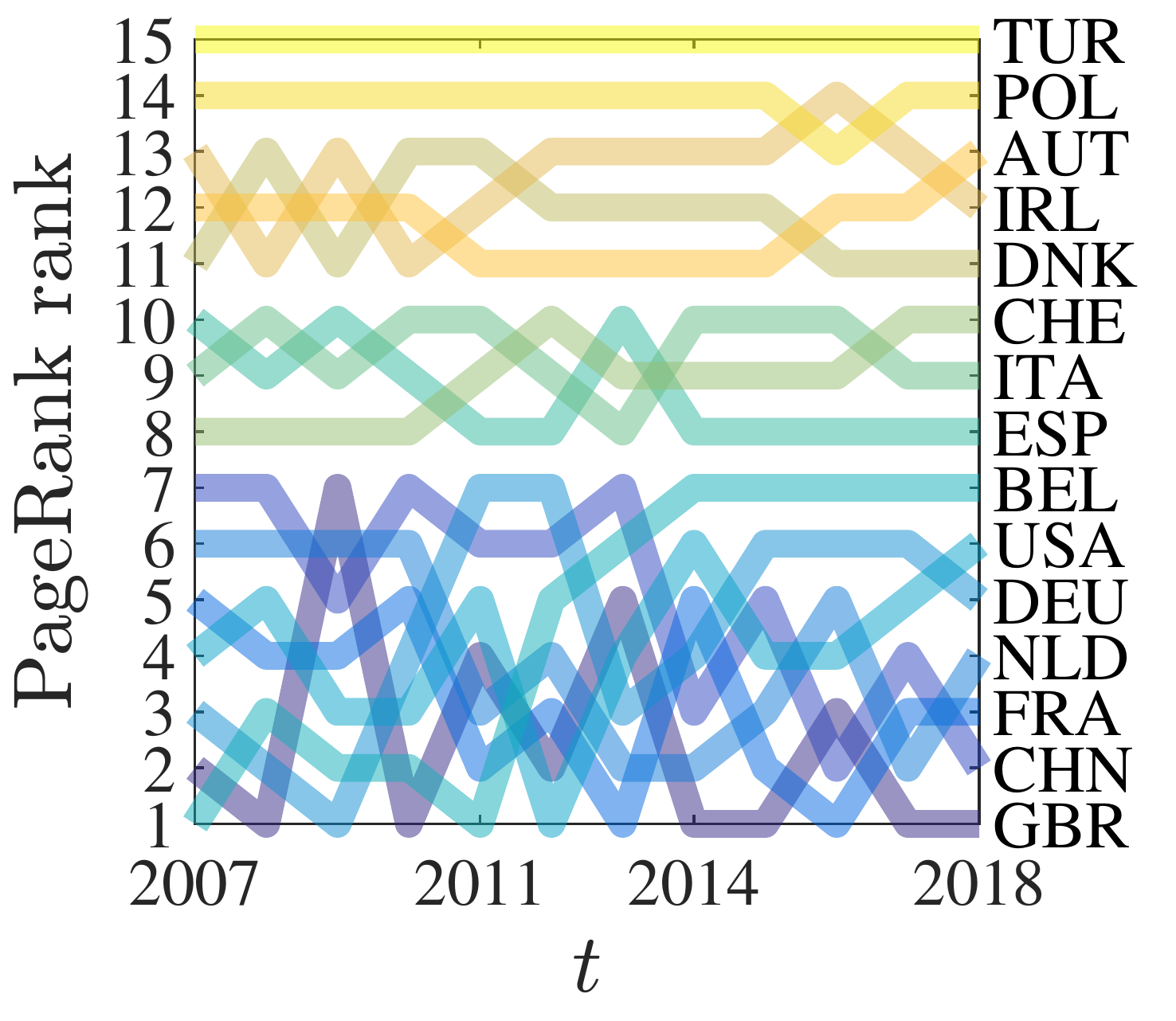}
    \includegraphics[width=0.33\linewidth]{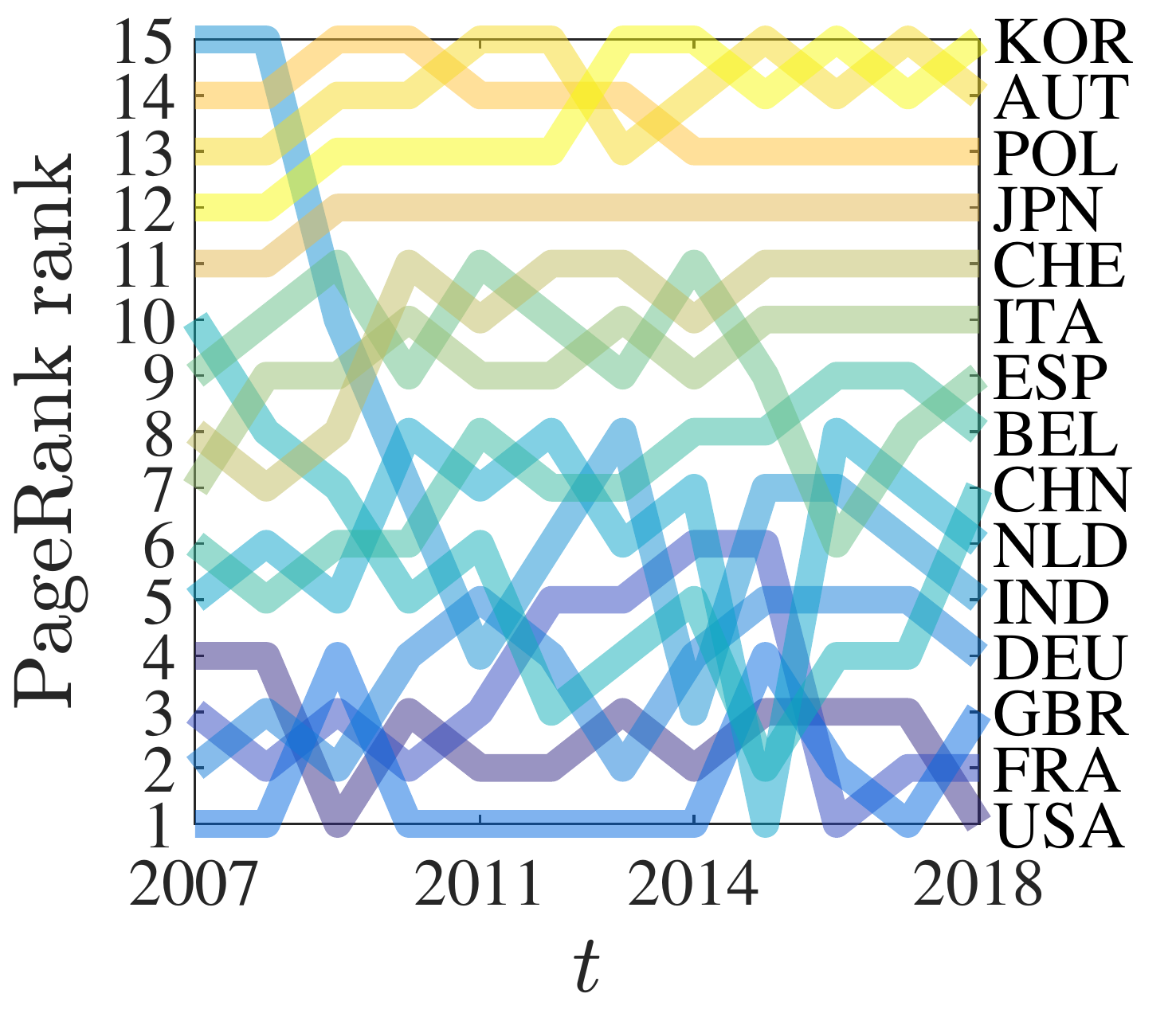}
\vskip    -9.55cm   \hskip   -16.0cm {(a)}
\vskip    -0.43cm   \hskip    -4.9cm {(b)}
\vskip    -0.43cm   \hskip     6.0cm {(c)}
\vskip     4.33cm   \hskip   -16.0cm {(d)}
\vskip    -0.43cm   \hskip    -4.9cm {(e)}
\vskip    -0.43cm   \hskip     6.0cm {(f)}
\vskip 4.05cm
  \caption{Rank evolution of the top-15 economies in the international trade networks for all pesticides (a), insecticides (b), fungicides (c), herbicides (d), disinfectants (e), and rodenticides and other similar products (f) from 2007 to 2018. The ranking is based on node PageRank.}
  \label{Fig:Pesticide:Rank:Evo:Pagerank}
\end{figure}

\subsection{PageRank centrality}
\label{S2:PageRank}

PageRank centrality is an important variant of eigenvector centrality, which is measured by the sum of weighted contributions of nodes pointing to the target node \citep{Brin-Page-1998-ComputNetwISDNSyst}. PageRank is a measure based on incoming links.  
We calculate the PageRank of nodes of each international pesticide trade network with 100 iterations and rank the economies in each year from 2007 to 2018. The rank evolution of the top-15 economies based on PageRank is illustrated in Fig.~\ref{Fig:Pesticide:Rank:Evo:Pagerank}. For the aggregated network in 2018, as shown in Fig.~\ref{Fig:Pesticide:Rank:Evo:Pagerank}(a), the top-15 economies are China, the United Kingdom, the USA, France, the Netherlands, Germany, Spain, Belgium, India, Italy, Japan, Switzerland, Austria, Denmark, and Korea, which are all developed or large developing economies. 

Concerning the categorized networks shown in Fig.~\ref{Fig:Pesticide:Rank:Evo:Pagerank}(b-f), in 2018, China ranks No. 1 in the international trade networks for insecticides, fungicides, and herbicides; the United Kingdom ranks No. 1 in the disinfectants network; and the USA ranks No. 1 in the rodenticides network.

\begin{figure}[!ht]
\centering
    \includegraphics[width=0.33\linewidth]{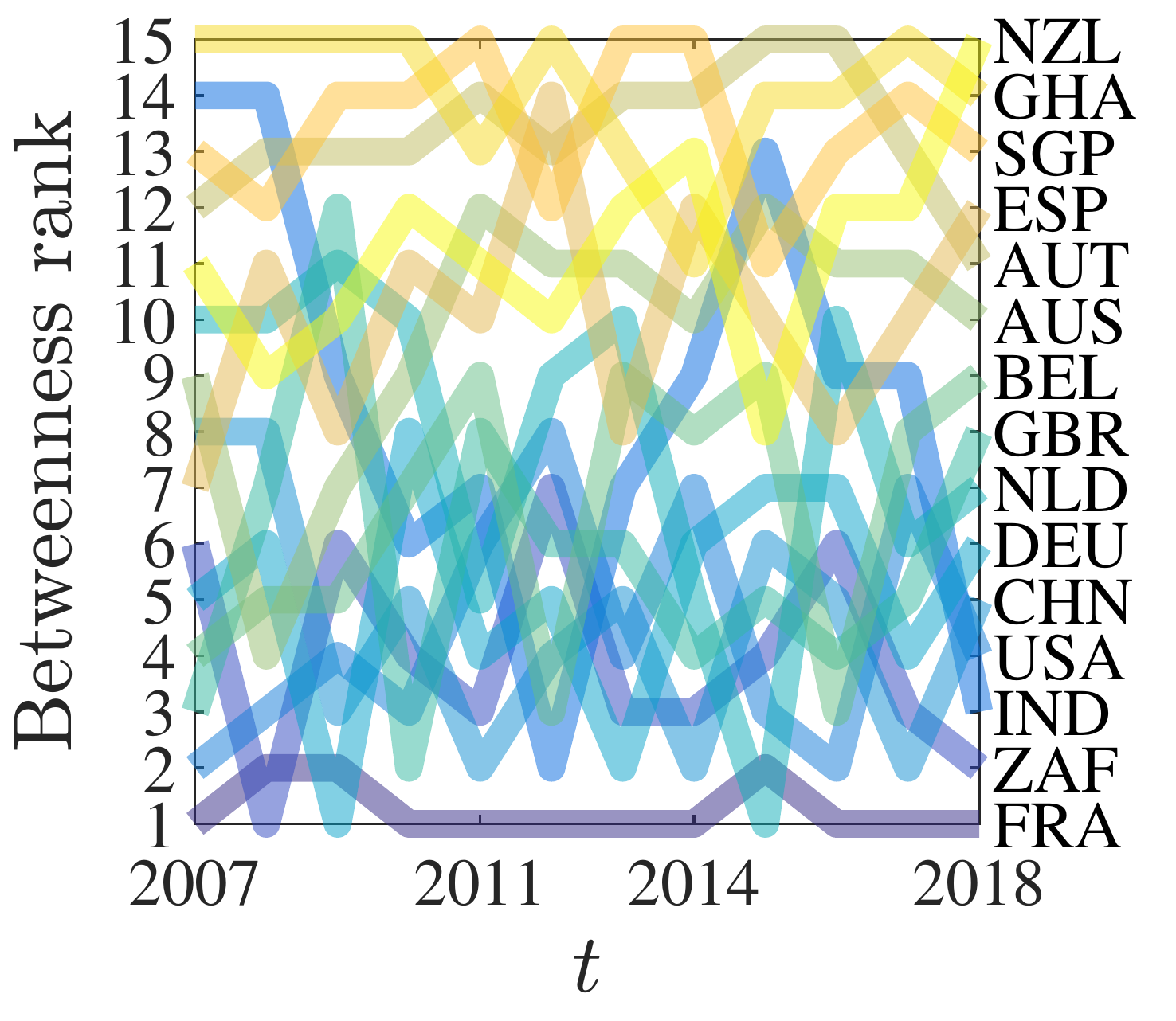}
    \includegraphics[width=0.33\linewidth]{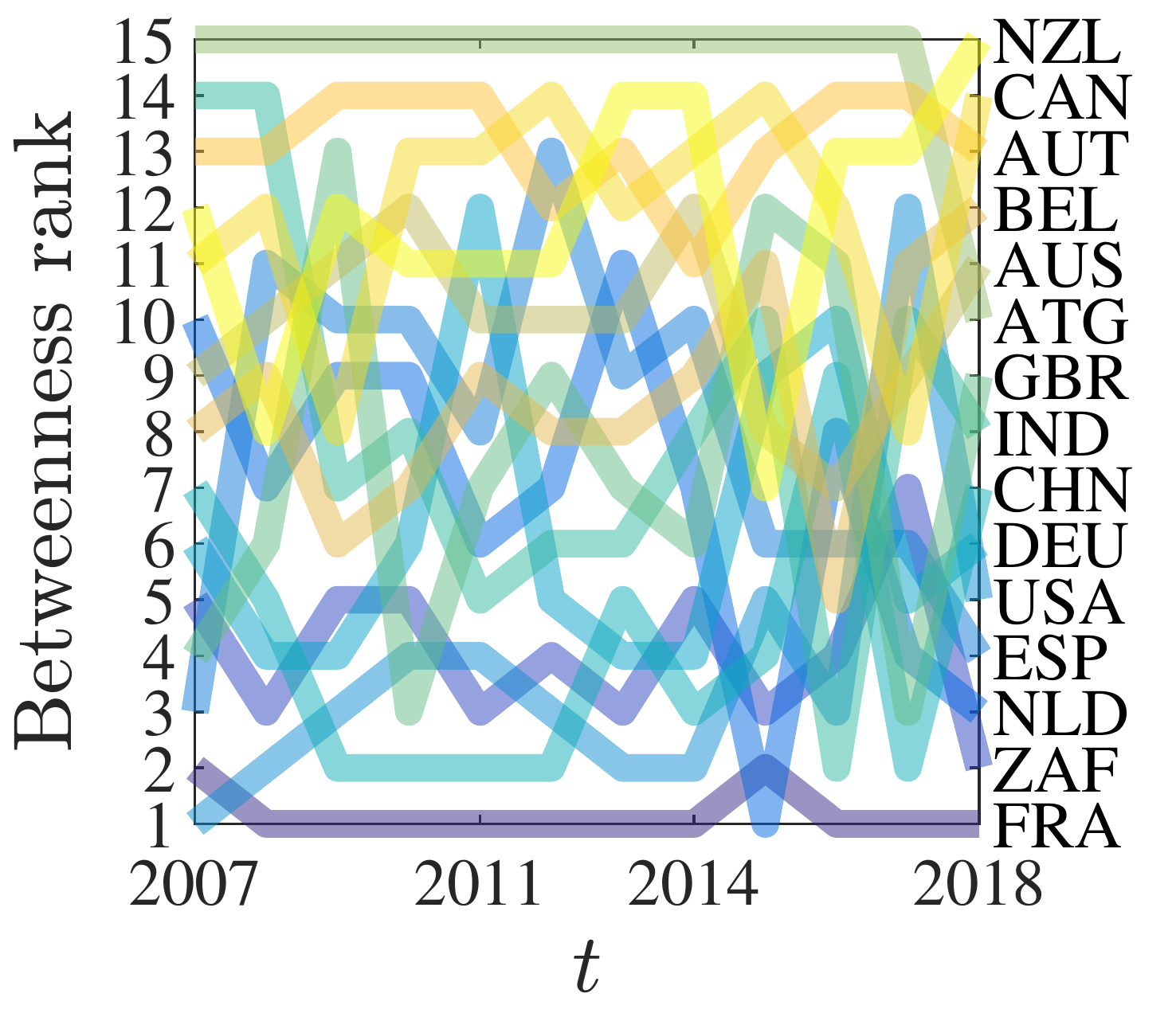}
    \includegraphics[width=0.33\linewidth]{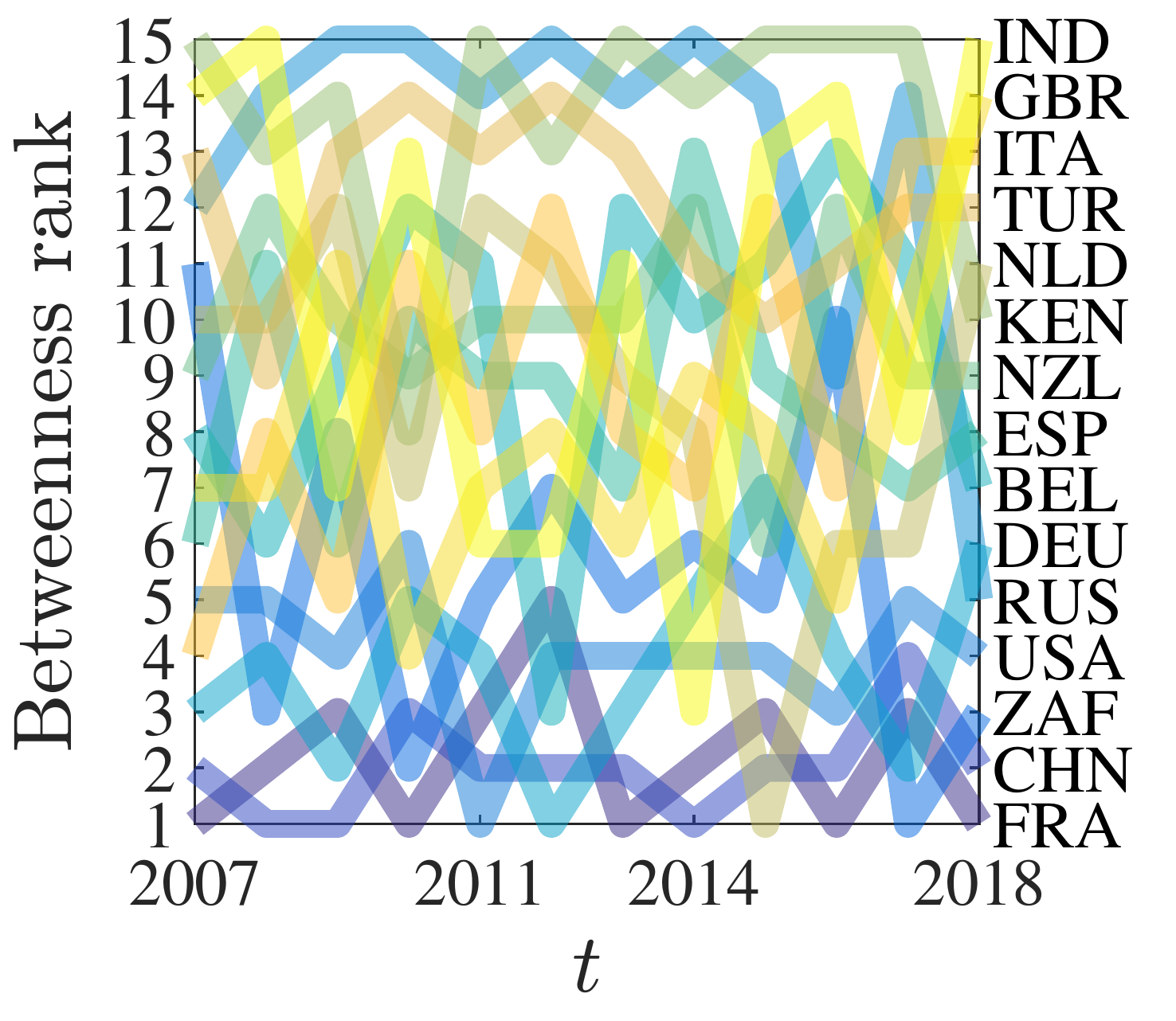}
    \includegraphics[width=0.33\linewidth]{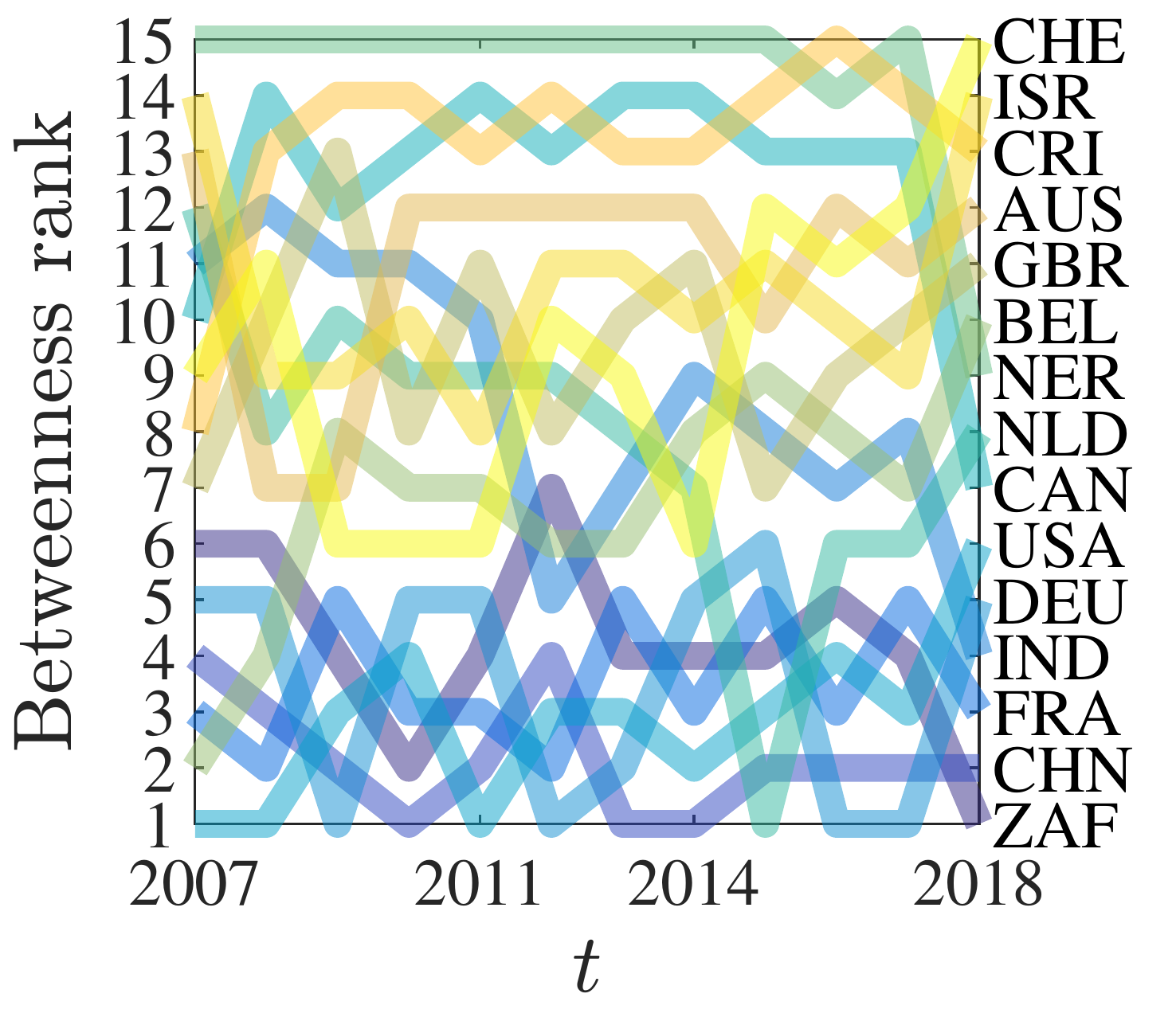}
    \includegraphics[width=0.33\linewidth]{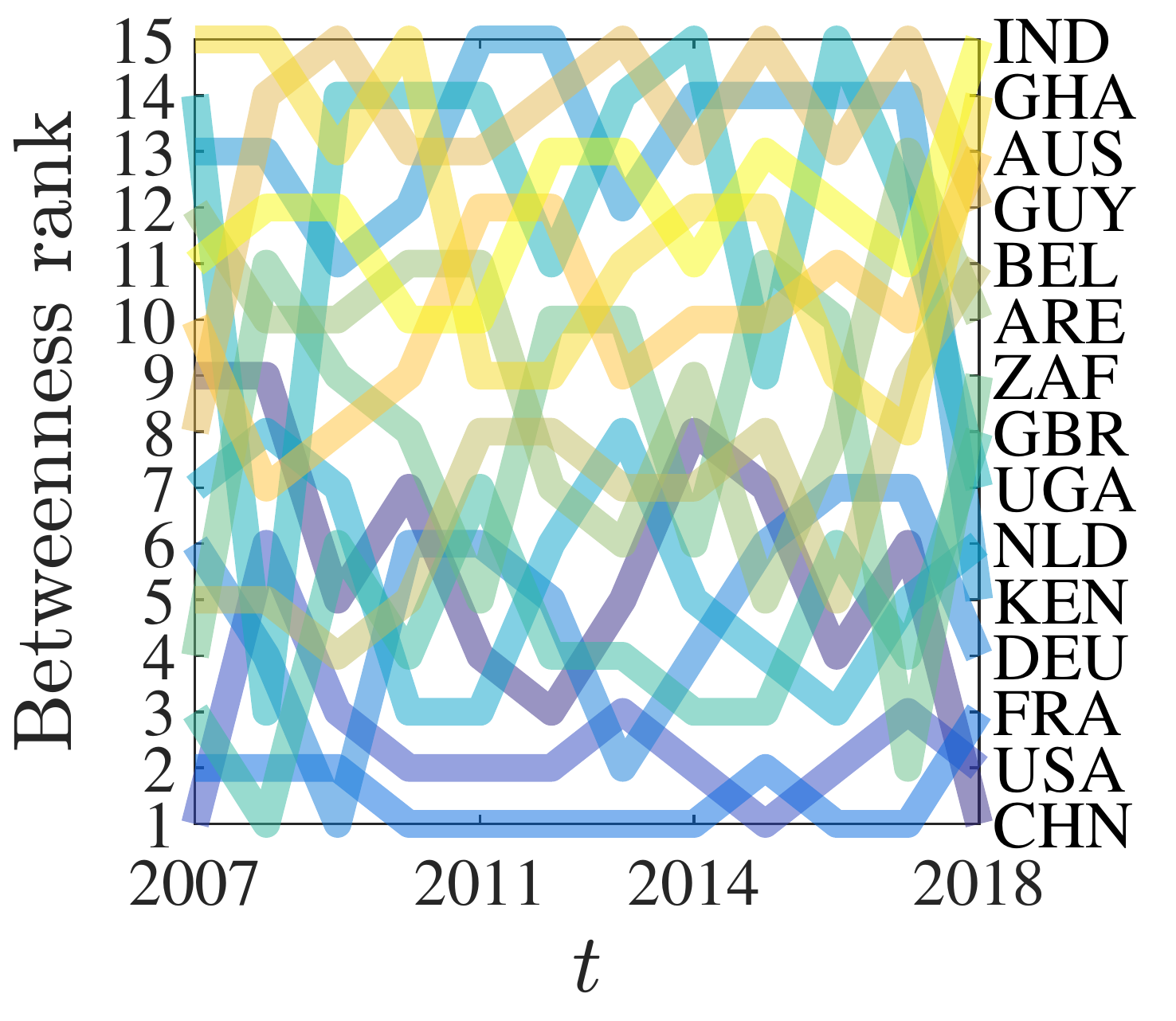}
    \includegraphics[width=0.33\linewidth]{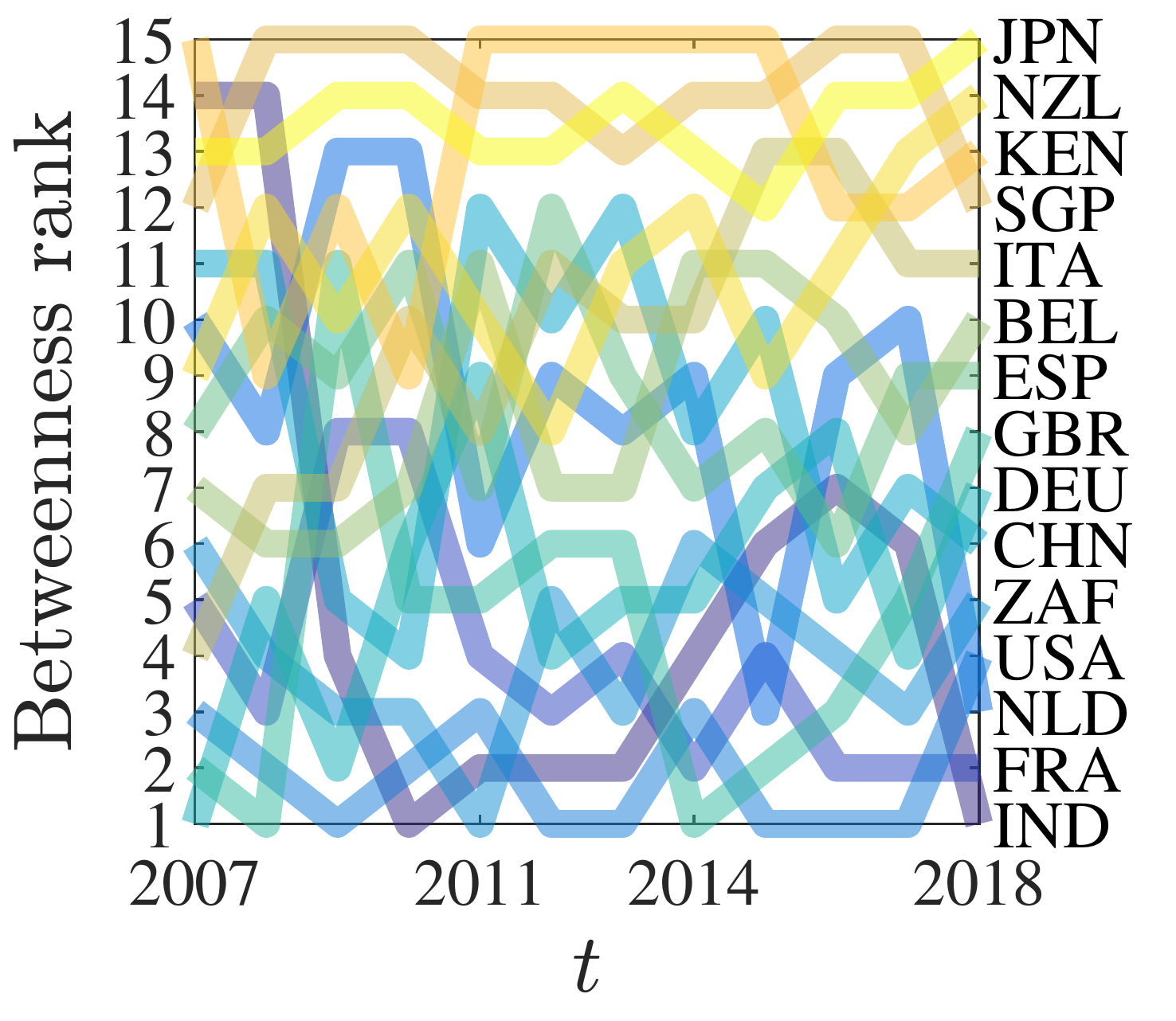}
\vskip    -9.55cm   \hskip   -16.0cm {(a)}
\vskip    -0.43cm   \hskip    -4.9cm {(b)}
\vskip    -0.43cm   \hskip     6.0cm {(c)}
\vskip     4.33cm   \hskip   -16.0cm {(d)}
\vskip    -0.43cm   \hskip    -4.9cm {(e)}
\vskip    -0.43cm   \hskip     6.0cm {(f)}
\vskip 4.05cm
  \caption{Rank evolution of the top-15 economies in the international trade networks for all pesticides (a), insecticides (b), fungicides (c), herbicides (d), disinfectants (e), and rodenticides and other similar products (f) from 2007 to 2018. The ranking is based on node betweenness.}
  \label{Fig:Pesticide:Rank:Evo:Betweenness}
\end{figure}

\subsection{Betweenness centrality}
\label{S2:Betweenness}

Betweenness centrality is a node metric based on shortest paths, which is the ratio of the number of shortest paths going through the investigated node to the number of shortest paths between all pairs of nodes in the network \citep{Freeman-1977-Sociometry}:
\begin{equation}
B_c(i)=\sum_{st}\frac{n_{st}^i}{g_{st}},
\end{equation}
where $n_{st}^i$ is the number of shortest paths between node $s$ and node $t$ passing through node $i$ and $g_{st}$ is the total number of shortest paths between nodes $s$ and $t$. Although the shortest paths are directed, betweenness cannot be decomposed into ``in-'' and ``out-'' parts, unlike the in-degree and out-degree pair in Sec.~\ref{S2:In:Out:Degree}, the in-closeness and out-closeness pair in Sec.~\ref{S2:In:Out:Closeness}, or the authority and hub pair in Sec.~\ref{S2:HITS}.

We calculate the betweenness of nodes of each international pesticide trade network and rank the economies in each year from 2007 to 2018. The rank evolution of the top-15 economies based on betweenness centrality is illustrated in Fig.~\ref{Fig:Pesticide:Rank:Evo:Betweenness}. For the aggregated network in 2018, as shown in Fig.~\ref{Fig:Pesticide:Rank:Evo:Betweenness}(a), the top-15 economies are France, South Africa, India, the USA, China, Germany, the Netherlands, the United Kingdom, Belgium, Austria, Spain, Singapore, Ghana, and New Zealand, which are mainly developed or large developing economies.

Concerning the categorized networks shown in Fig.~\ref{Fig:Pesticide:Rank:Evo:Betweenness}(b-f), in 2018, France ranks No. 1 in the international trade networks for insecticides and fungicides, South Africa ranks No. 1 in the herbicides network, China ranks No. 1 in the disinfectants network, and India ranks No. 1 in the rodenticides network. This figure provides information about the rise of BRICS economies, especially China, India, and South Africa.

\subsection{Clustering coefficient}

An economy's clustering coefficient characterizes the connectance of its trade partners, which can be calculated as follow \citep{Watts-Strogatz-1998-Nature}:
\begin{equation}
  C_{\rm{c}}\left(i\right)=\frac{|\{a_{jk}:i,j \in {\mathbf{N}}_{i},a_{ij} \in A\}|}{k_i(k_i-1)}.
\end{equation}
where ${\mathbf{N}}_i$ is the set of neighbors of $i$.

\begin{figure}[!ht]
\centering
    \includegraphics[width=0.33\linewidth]{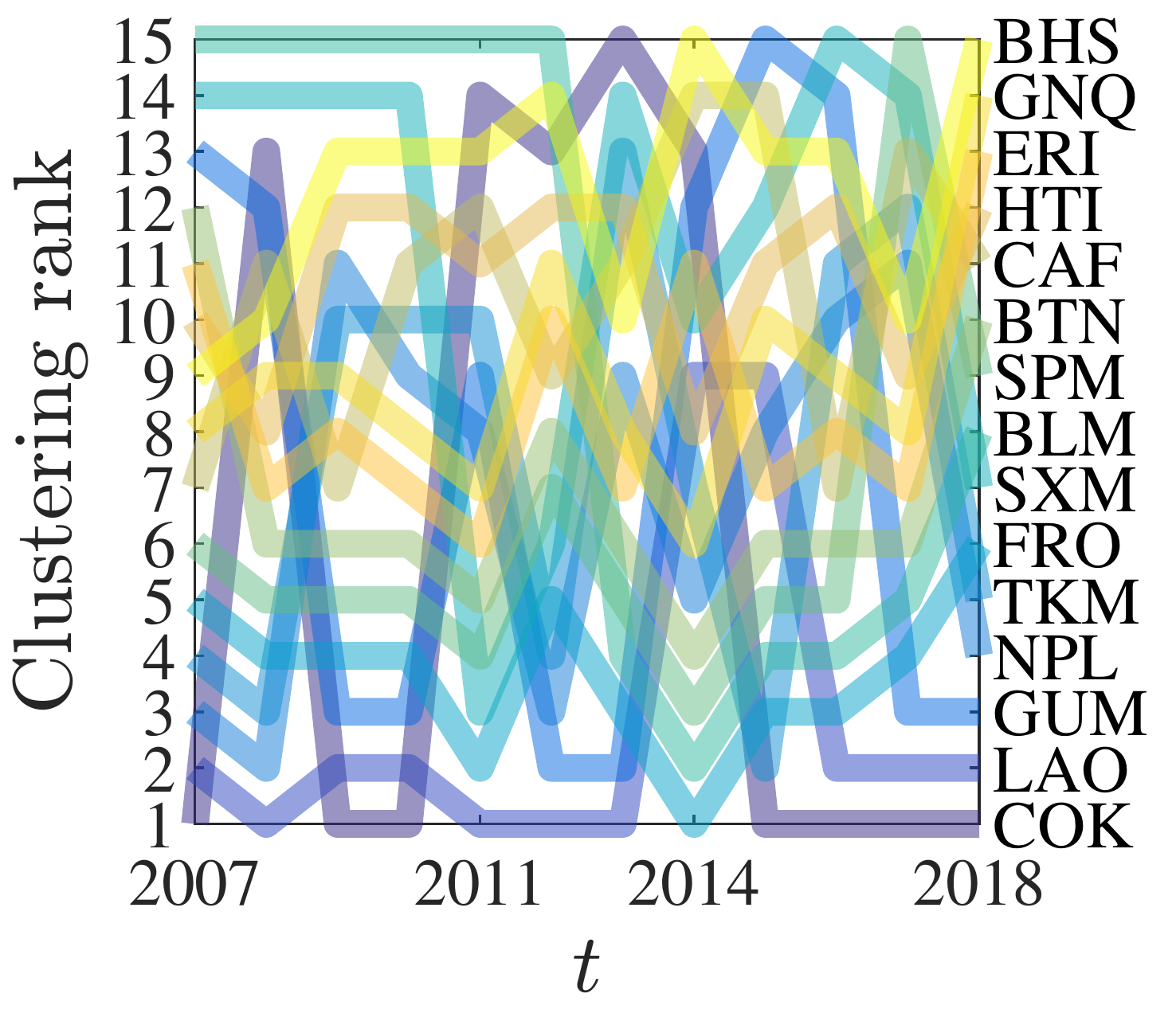}
    \includegraphics[width=0.33\linewidth]{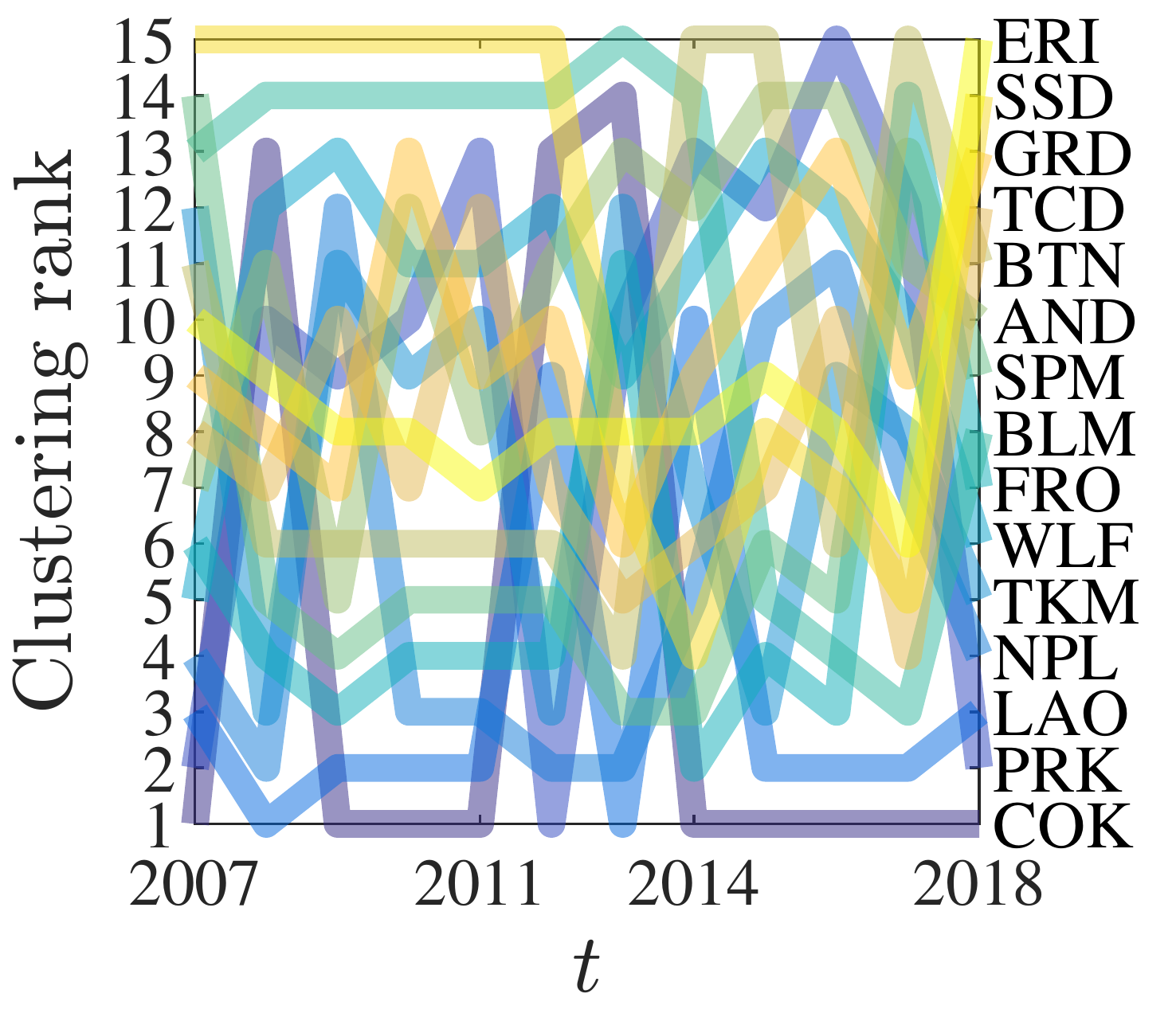}
    \includegraphics[width=0.33\linewidth]{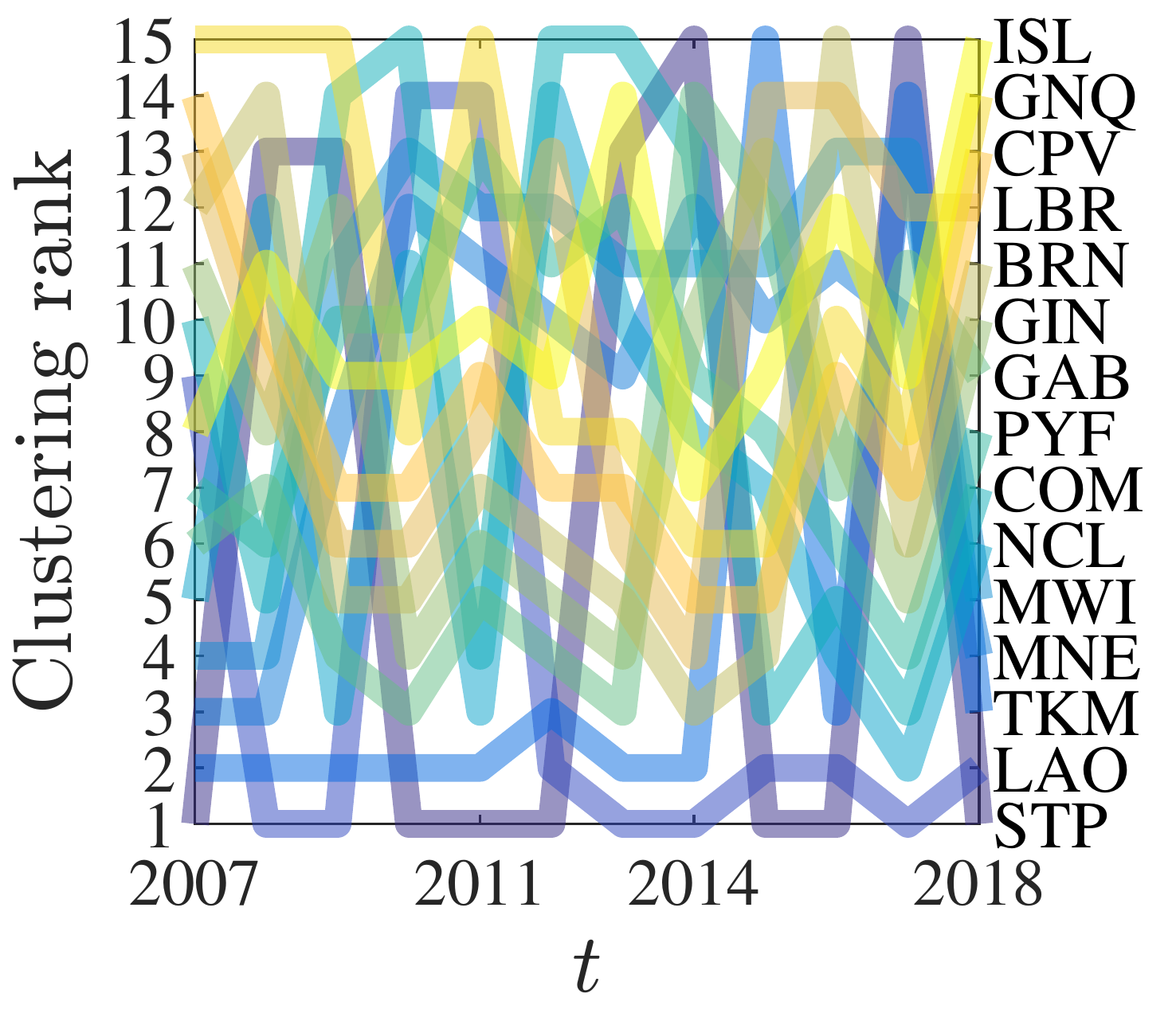}
    \includegraphics[width=0.33\linewidth]{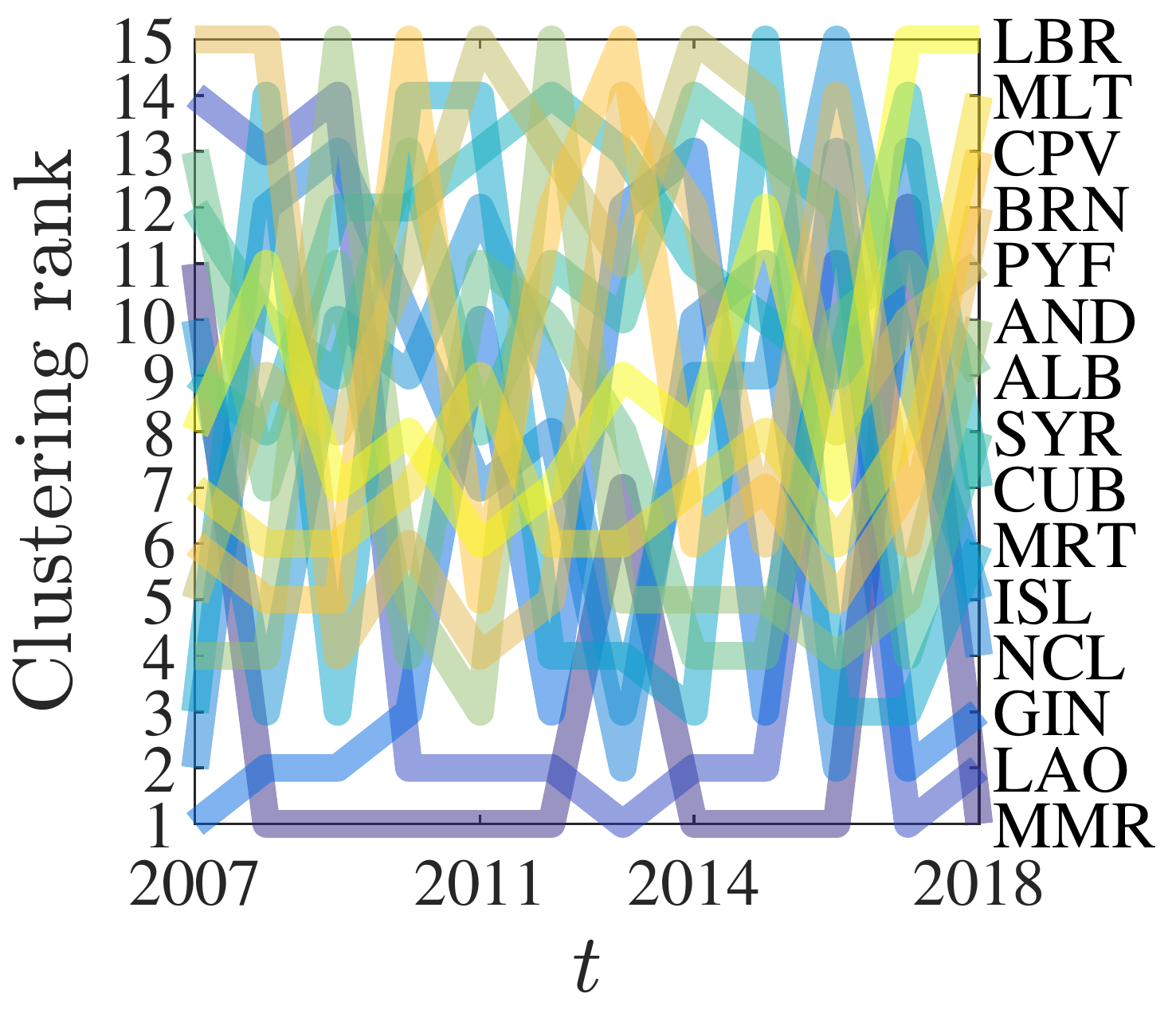}
    \includegraphics[width=0.33\linewidth]{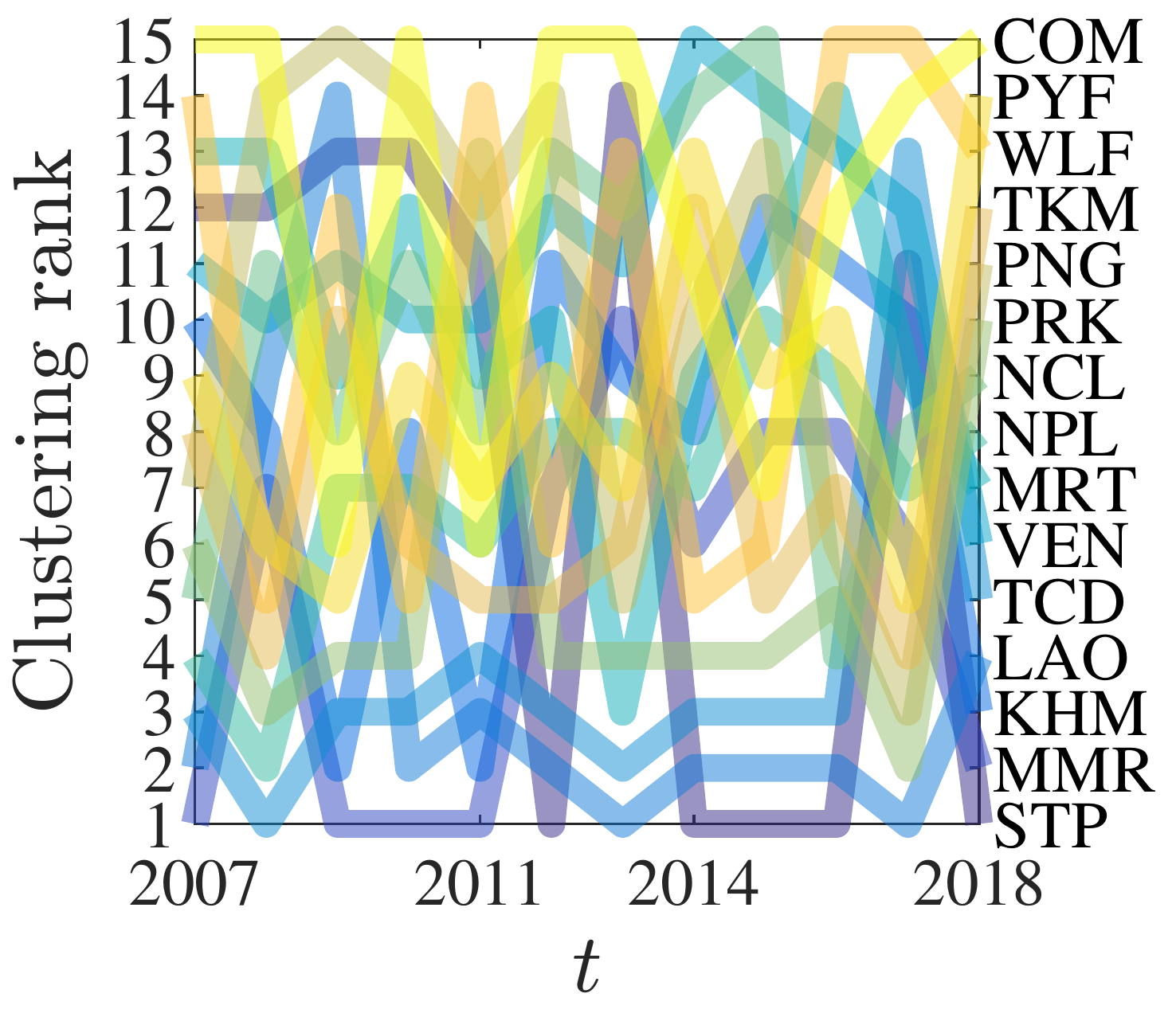}
    \includegraphics[width=0.33\linewidth]{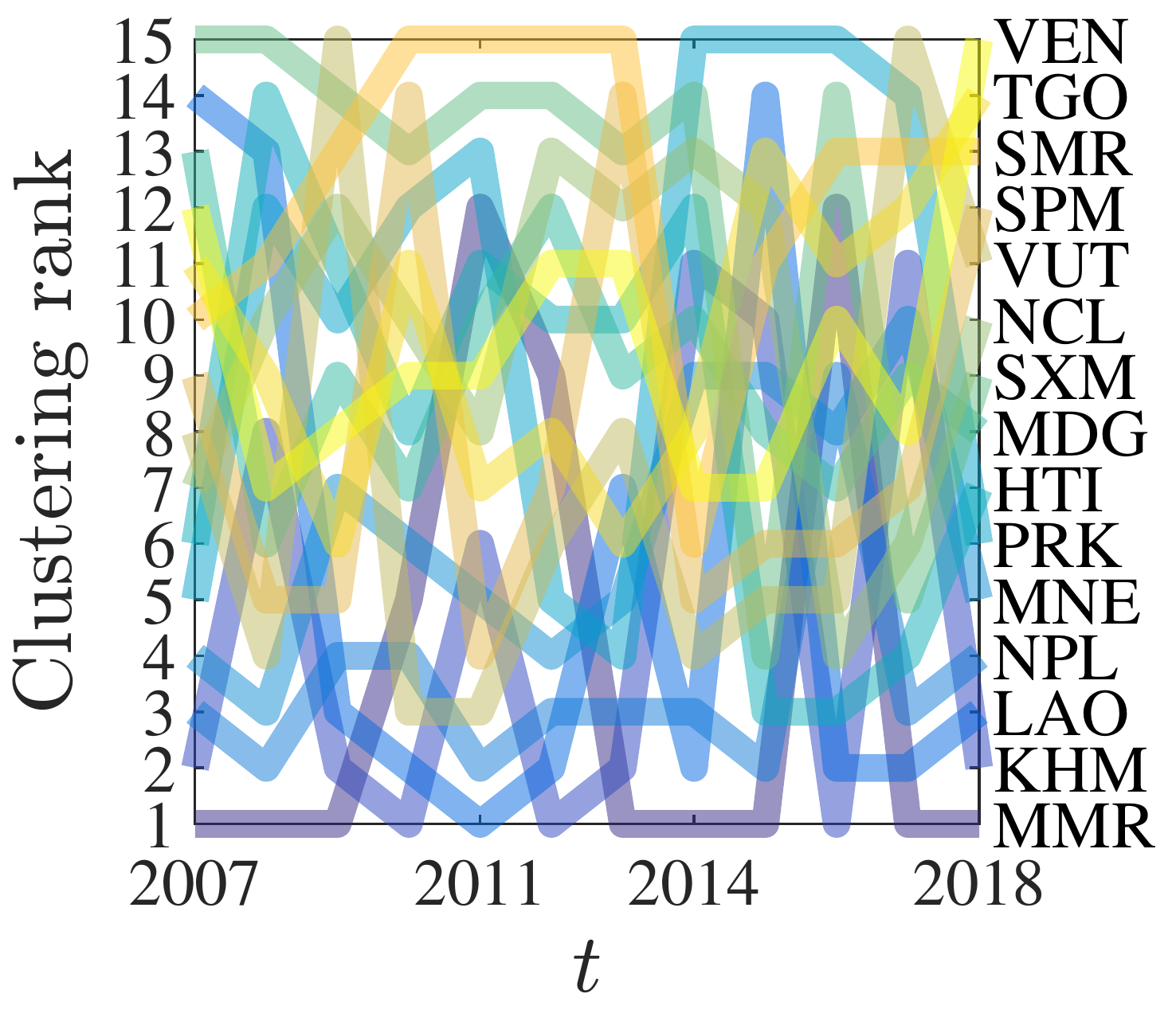}
\vskip    -9.55cm   \hskip   -16.0cm {(a)}
\vskip    -0.43cm   \hskip    -4.9cm {(b)}
\vskip    -0.43cm   \hskip     6.0cm {(c)}
\vskip     4.33cm   \hskip   -16.0cm {(d)}
\vskip    -0.43cm   \hskip    -4.9cm {(e)}
\vskip    -0.43cm   \hskip     6.0cm {(f)}
\vskip 4.05cm
  \caption{Rank evolution of the top-15 economies in the international trade networks for all pesticides (a), insecticides (b), fungicides (c), herbicides (d), disinfectants (e), and rodenticides and other similar products (f) from 2007 to 2018. The ranking is based on nodes' clustering coefficients.}
  \label{Fig:Pesticide:Rank:Evo:Clustering}
\end{figure}

\begin{figure*}[!ht]
\centering
    \includegraphics[width=0.99\linewidth]{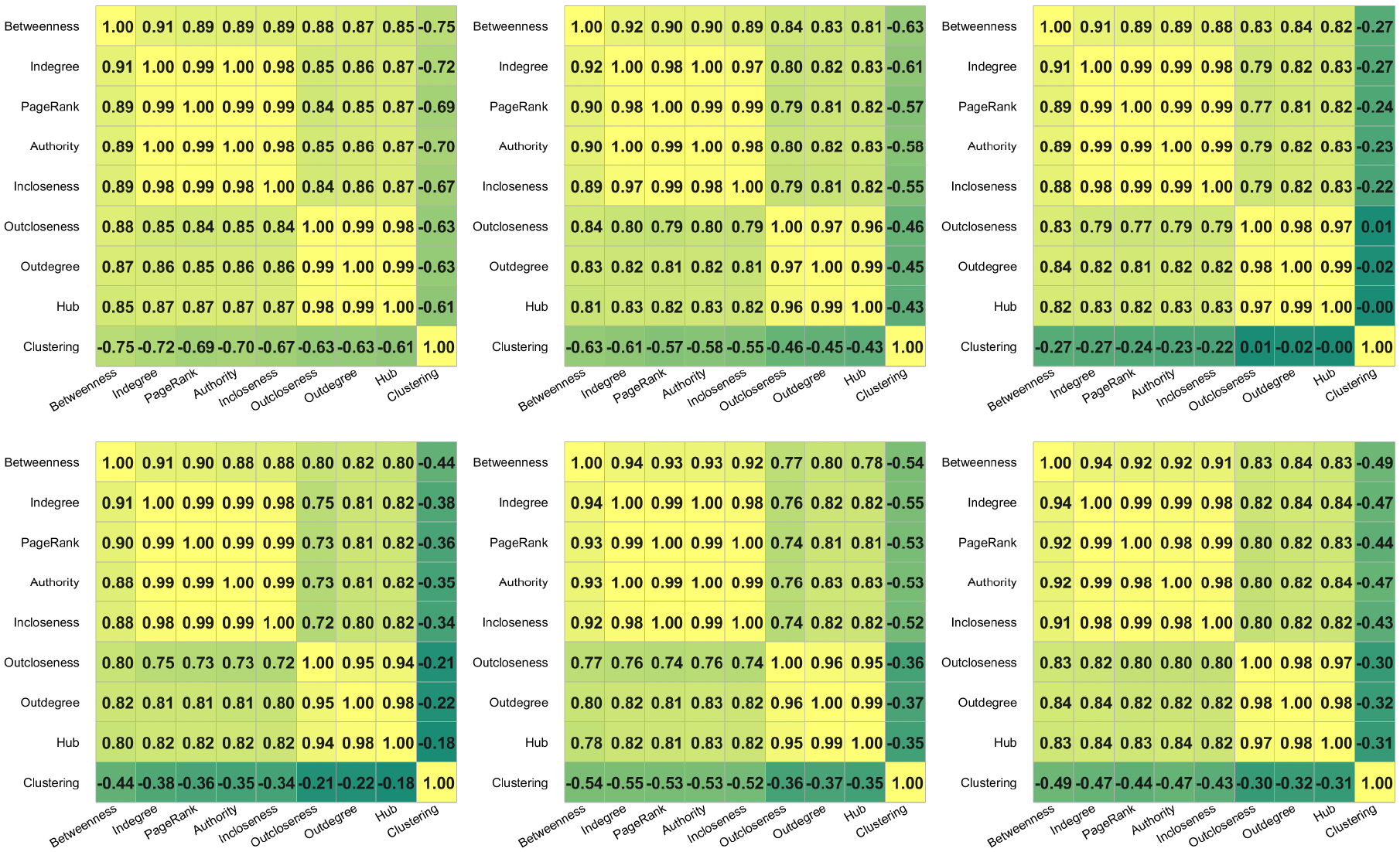}
  \caption{Cross-correlation matrices of the nine node metrics for the aggregated network and the five categorized networks in 2018. There are four blocks: Block I (Betweenness), Block II (In-degree, PageRank, authority, and in-closeness), Block III (Out-closeness, out-degree, and hub), and Block IV (Clustering coefficient).}
  \label{Fig:Pesticide:Corr:888888}
\end{figure*}

We calculate the clustering coefficients of nodes of each international pesticide trade network and rank the economies in each year from 2007 to 2018. The rank evolution of the top-15 economies based on betweenness centrality is illustrated in Fig.~\ref{Fig:Pesticide:Rank:Evo:Clustering}. For the aggregated network in 2018, as shown in Fig.~\ref{Fig:Pesticide:Rank:Evo:Clustering}(a), the top-15 economies are the Cook Islands, the Lao People's Democratic Republic, Guam, Nepal, Turkmenistan, Faeroe Islands, Sint Maarten, Saint Barthelemy, Saint Pierre and Miquelon, Central African Republic, Haiti, Eritrea, Equatorial Guinea, and the Bahamas, which are mainly small developing economies. This list is very different from those in the previous subsections. The situation for the categorized networks is similar.

\subsection{Cross-correlations between node metrics}

As shown in the previous subsections, the rankings of top economies are similar to some extent across different node metrics, except for the clustering coefficient. We calculate the cross-correlation coefficients of the nine node metrics for the aggregated network and the five categorized networks, using Spearman's rank correlation, which is the correlation of the rankings of the node metrics. We find that the cross-correlation matrices of a given type of network (aggregated, insecticides, fungicides, herbicides, disinfectants, and rodenticides) are very similar in different years from 2007 to 2018. We thus present in Fig.~\ref{Fig:Pesticide:Corr:888888} only the cross-correlation matrices of the six networks in 2018. It is intriguing to observe that the six matrices for different networks are also very similar.

To ensure that large correlations are closer to the diagonal, we reorder the node metrics in the axes \citep{Meng-Xie-Jiang-Podobnik-Zhou-Stanley-2014-SciRep}. We observe four evident blocks: Block I (Betweenness), Block II (In-degree, PageRank, authority, and in-closeness), Block III (Out-closeness, out-degree, and hub), and Block IV (Clustering coefficient). This classification can be explained by the following facts: (1) Node metrics in Block II are determined by nodes pointing to them (thus related to ``in-'') and these metrics reflect their trait as targets (or sinks); (2) Node metrics in Block III are determined by nodes pointing from them (thus related to ``out-'') and these metrics reflect their trait as sources; (3) Betweenness in Block I reflects an aspect of global directional traits that does not distinguish targets or sources; and (4) Clustering coefficient is a local trait of a node for undirected networks.

It is found that the clustering coefficients are negatively correlated with all eight other node metrics. Excluding Block IV for the clustering coefficient, the remaining correlation coefficients are significantly positive. The node metrics within the same blocks are strongly correlated, with the correlation coefficients close to 1. The node metrics between different blocks are also strongly correlated with large correlation coefficients, although not close to 1.

These interesting results of the cross-correlations between different node metrics will impact the structural robustness and fragility of the networks when different node metrics are adopted to design attack strategies. There are also interesting open questions for future studies concerning the impacts of external driving factors on the rankings of economies. In particular, changing economic situations (such as of the diminished European Union GDP caused by the global financial crisis of 2008 and the debt crisis of 2012), public health crises (such as Ebola epidemic and COVID-19 pandemic), weather extremes (such as droughts, floods, or the El Ni{\~n}o/La Ni{\~n}a effect), and wars, conflicts and social unrests all have an impact on the node metrics of influenced economies, especially those agricultural intensive economies.

\section{Impact of economy shocks on the size of giant component}
\label{S1:GiantComponent}

\subsection{Robustness curves under different shocks}

To simulate shocks to economies, we adopt three strategies ${\mathcal{S}}=\{{\mathcal{R}}, {\mathcal{D}}, {\mathcal{A}}\}$: random node removal strategy (${\mathcal{S}}={\mathcal{R}}$), intentional node removal strategy in descending order of node metrics (${\mathcal{S}}={\mathcal{D}}$), and intentional node removal strategy in ascending order of node metrics (${\mathcal{S}}={\mathcal{A}}$).  The severity of the aftermath of a shock is quantified by the fraction $p$ of nodes being removed (i.e., economies ceasing import and export). The nine node metrics investigated in Section~\ref{S1:NodeRank} are considered, and the results based on different node metrics are compared.

\begin{figure*}[!ht]
    \centering
    \includegraphics[width=0.95\linewidth]{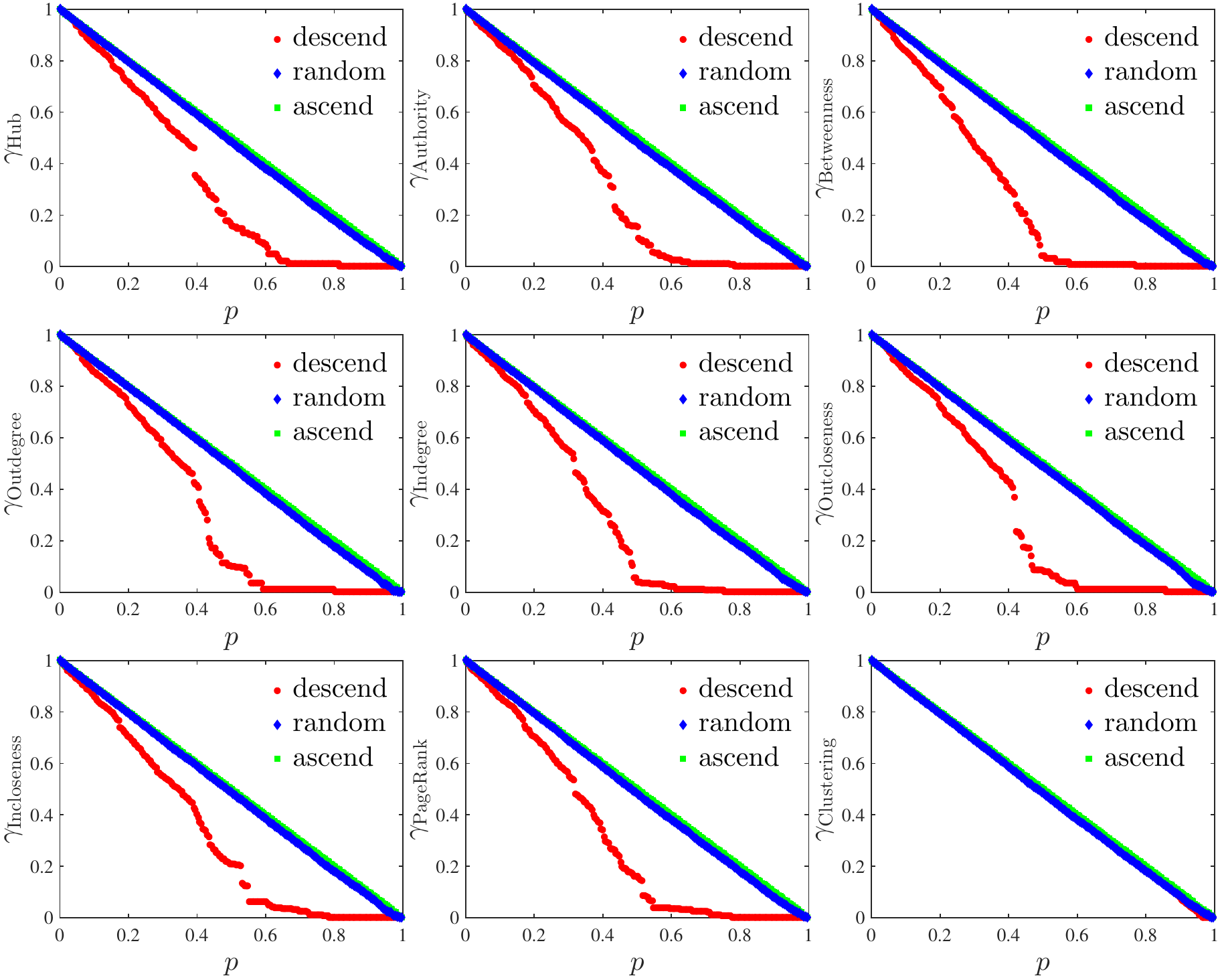}
    \caption{Robust curves $\gamma_I(p, {\mathcal{S}})$ showing the impact of shocks to economies on the giant component size for the aggregated network for all five pesticides in 2018. The ratio $\gamma_I$ of the number of nodes in the giant component formed by the remaining network nodes after deleting the nodes in the network according to node metric $I$. In the figure, ``descend'', ``ascend'', and ``random'' correspond to the cases of preferentially deleting the largest, smallest, and random nodes with node metric $I$, respectively. The robustness curves for the random removal strategy ${\mathcal{S}}={\mathcal{R}}$ are obtained by averaging the results of 20-round simulations.}
    \label{Fig:Pesticide:node:GC}
\end{figure*} 

The structural characteristic we investigate is the size of the giant component \citep{Albert-Jeong-Barabasi-2000-Nature}. Denoting $N(p, {\mathcal{S}})$ the number of nodes left in the giant component after removing a proportion $p$ of nodes based on a given node metric $I$ and strategy ${\mathcal{S}}$, the size of the giant component is defined as the ratio of $N_I(p, {\mathcal{S}})$ to the total number of nodes, $N=N_I(0, {\mathcal{S}})$, in the original network,
\begin{equation}
   \label{Eq:Robustness:Measure:GC}
   \gamma_I(p, {\mathcal{S}})=\frac{N_I(p, {\mathcal{S}})}{N},
\end{equation}
which can be viewed as the robustness curve of the network under certain shocks to nodes. The value of $\gamma_I$ can be regarded as a measure of the globalization of the international pesticide trade.

In Fig.~\ref{Fig:Pesticide:node:GC}, we digest the results showing the impact of shocks to economies on the giant component size for the aggregated network for all five pesticides in 2018. A few intriguing observations can be obtained. We note that the results remain unchanged for other years and for the five categorized networks.

First, the aggregated international pesticide trade network is extremely robust against intentional node removal in ascending order ${\mathcal{S}}={\mathcal{A}}$. We find that, regardless of the node metric $I$, the robustness curve can be expressed as follows,
\begin{equation}
    \gamma_I(p, {\mathcal{A}}) = 1-p,
    \label{Eq:PesticideNet:RobustCurve:Ascend}
\end{equation}
which is independent of $I$. Combining Eq.~(\ref{Eq:Robustness:Measure:GC}) and Eq.~(\ref{Eq:PesticideNet:RobustCurve:Ascend}), we have
\begin{equation}
    N_I(p, {\mathcal{A}}) = N- Np.
    \label{Eq:PesticideNet:RobustCurve:Ascend:Np}
\end{equation}
In other words, after removing $Np$ nodes of least importance (small $I$ values) from the original network, the resulting network does not fall into pieces and remains as a whole, indicating that the shocks are constrained to the removed nodes only and do not impact other nodes.

\begin{figure*}[!ht]
   \centering
   \includegraphics[width=0.95\linewidth]{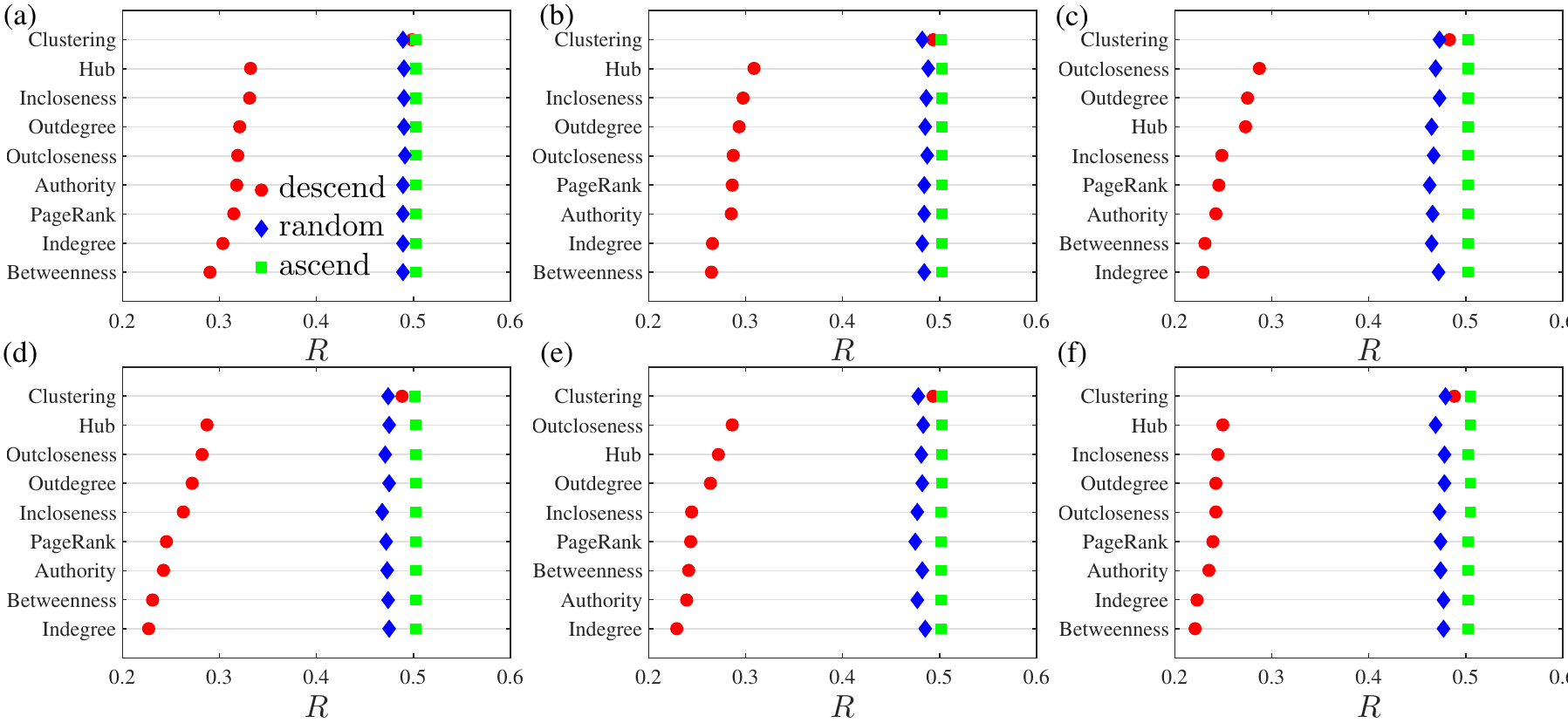}
   \caption{Comparing the influence of different node metrics $I$ on the robustness $R_{I}({\mathcal{S}})$ of the international trade networks for all pesticides (a), insecticides (b), fungicides (c), herbicides (d), disinfectants (e), and rodenticides and other similar products (f) in 2018. }
    \label{Fig:Pesticide:node:R}
\end{figure*}

Second, the aggregated network is very robust against shocks randomly exerted on economies. Regardless of the node metric $I$, the robustness curve can be expressed as follows
\begin{equation}
    \gamma_I(p, {\mathcal{R}}) \approx 1-p - \delta,
    \label{Eq:PesticideNet:RobustCurve:Random}
\end{equation}
where $\delta$ is a very small positive number. Hence, randomly removing nodes has a minor impact on the connectance of the remaining network.

Third, the aggregated network is fragile under intentional attacks to nodes in descending order for eight node metrics except for clustering coefficient, since the robustness curves deviate remarkably from the baseline $\gamma_I(p)=1-p$. These eight robustness curves $\gamma_I(p,{\mathcal{D}})$ descend much faster than $\gamma_I(p,{\mathcal{R}})$ and $\gamma_I(p,{\mathcal{A}})$, especially when $p$ is close to 0.5. We observe the existence of a critical point $p_c$ in each $\gamma_I(p,{\mathcal{D}})$ curve, whose value is different for different node metrics and greater than 0.6. When $p \geq p_c$, the network is fully disconnected.

Fourth, when the node metric is clustering coefficient, the robustness curve is mainly between the other two robustness curves for ${\mathcal{S}}={\mathcal{R}}$ and ${\mathcal{S}}={\mathcal{A}}$. It shows that ranking local node metrics such as clustering coefficient almost never has a global influence on the structural robustness of the network. 

\subsection{Measuring network robustness}

To give a parsimonious description of the structural robustness of complex networks, one calculates the area enclosed by the robustness curve and the two coordinate axes,
\begin{equation}
   \label{Eq:Robustness:Measure:sum}
   R_{I}({\mathcal{S}})=\frac{1}{N+1}\sum_{n=0}^{N}\gamma_I\left(n/N,{\mathcal{S}}\right),
\end{equation}
which is defined as the robustness measure \citep{Schneider-Moreira-Andrade-Havlin-Herrmann-2011-ProcNatlAcadSciUSA}. By construction, the maximum value of $R_{I}({\mathcal{S}})$ is reached if the remaining network is connected when any proportion of nodes are removed. Hence, we have
\begin{equation}
   \label{Eq:Robustness:Measure:sum:Max}
   R_{I}({\mathcal{S}}) \leq 0.5,
\end{equation}
The larger is the value of $R_{I}({\mathcal{S}})$, the more robust is the network.

The values of the robustness measure $R_{I}({\mathcal{S}})$ for different node metrics $I$ and node removal strategies ${\mathcal{S}}$ of the aggregated network in 2018 are calculated from the results in Fig.~\ref{Fig:Pesticide:node:GC} and plotted in Fig.~\ref{Fig:Pesticide:node:R}(a). We also calculated the $R_{I}({\mathcal{S}})$ values for the networks for insecticides, fungicides, herbicides, disinfectants, and rodenticides, which are shown in Fig.~\ref{Fig:Pesticide:node:R}(b-f). It is observed that the node metric clustering coefficient is an outlier and we have
\begin{equation}
   \label{Eq:Robustness:Measure:ADR}
   R_{{\mathrm{Clustering}}}({\mathcal{R}}) 
   < R_{{\mathrm{Clustering}}}({\mathcal{D}}) 
   < R_{{\mathrm{Clustering}}}({\mathcal{A}}),
\end{equation}
suggesting that the clustering coefficient is not a meaningful measure to rank nodes' contributions to the robustness of networks. 
In contrast, for the other eight node metrics, we have
\begin{equation}
   \label{Eq:Robustness:Measure:ARD}
   R_{I}({\mathcal{D}}) 
   < R_{I}({\mathcal{R}}) 
   < R_{I}({\mathcal{A}}) = 0.5,
\end{equation}
showing the ability of these node metrics to rank nodes' contributions to the network's structural robustness.

\begin{figure*}[!ht]
  \centering
  \includegraphics[width=0.95\linewidth]{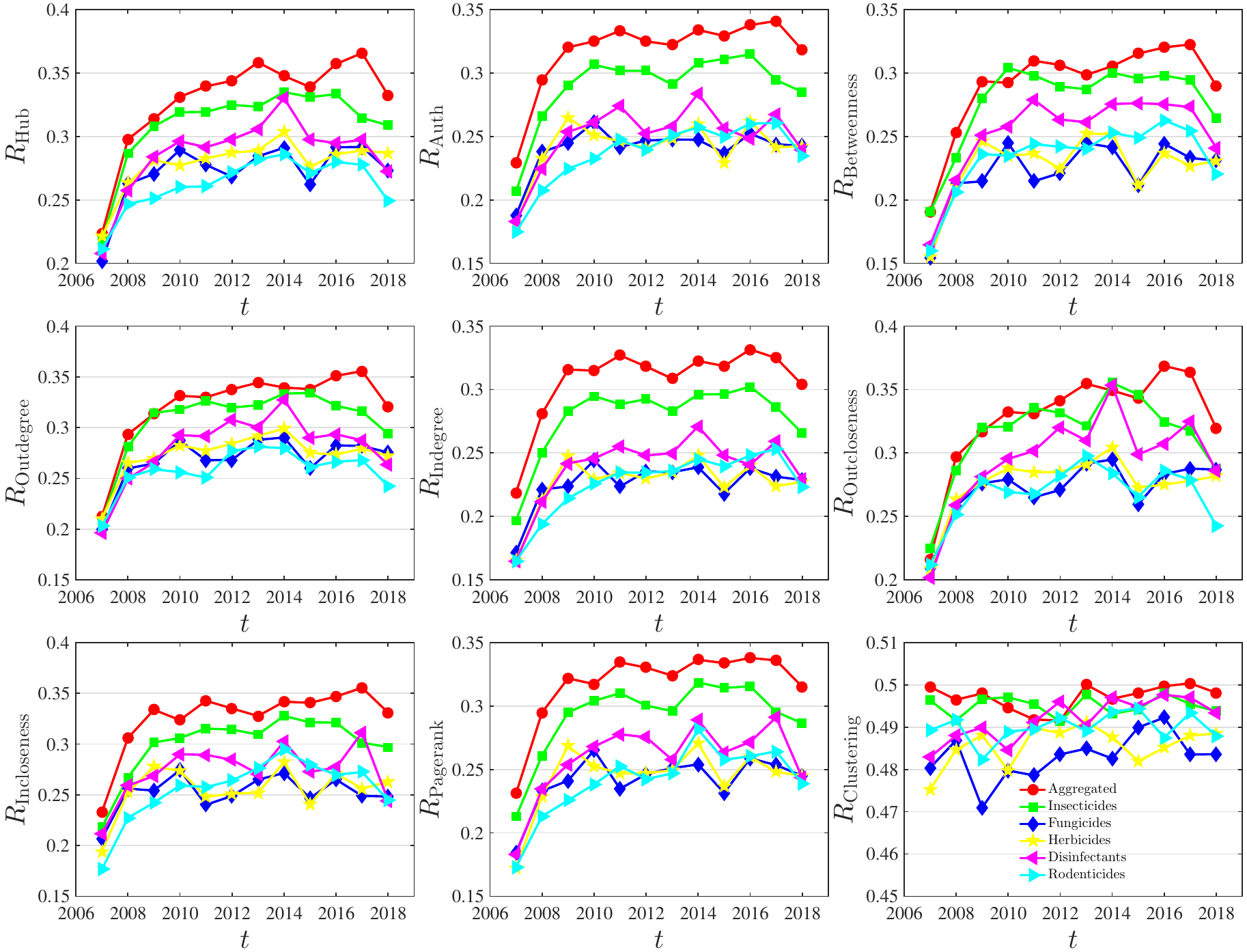}
  \caption{Evolution of network robustness of the international trade networks for all pesticides, insecticides, fungicides, herbicides, disinfectants, and rodenticides and other similar products responding to the strategy of node removal in descending order (${\mathcal{S}}={\mathcal{D}}$) of nine node metrics $I$.}
    \label{Fig:Pesticide:node:R:evo}
\end{figure*}

Economies with higher in-degrees have the largest contributions to the structural robustness of the international trade networks for fungicides, herbicides, and disinfectants, while economies with higher betweenness have the largest contributions to the structural robustness of the international trade networks for insecticides and rodenticides. Together with betweenness, economies having high node metrics calculated based on incoming links (in-degree, authority, PageRank, and in-closeness) are more important in maintaining the structural robustness of the international pesticide trade networks than those with high node metrics calculated based on outgoing links (out-degree, hub, and out-closeness). Speaking differently, it shows that importing economies (or demands) are more important than exporting economies (or supplies) in maintaining the structural robustness of the international pesticide trade networks, suggesting that the international pesticide trade networks are driven by the demands for pesticides. Moreover, the different importance of the metrics is consistent with the results in Fig.~\ref{Fig:Pesticide:Corr:888888}.

\subsection{Evolution of network robustness}

We now turn to investigating the evolutionary behavior of the structural robustness of the yearly international trade networks for all pesticides, insecticides, fungicides, herbicides, disinfectants, and rodenticides and other similar products. Since the random node removal strategy (${\mathcal{S}}={\mathcal{R}}$) and the ascending node removal strategy (${\mathcal{S}}={\mathcal{A}}$) are well and easily understood, we consider only the descending node removal strategy (${\mathcal{S}}={\mathcal{D}}$). We illustrate in Fig.~\ref{Fig:Pesticide:node:R:evo} the evolution of network robustness from 2007 to 2018 for the nine node metrics $I$.

Again, the curves for clustering coefficient are different from the curves for other node metrics. The values of $R_{\mathrm{Clustering}}({\mathcal{D}})$ are all close to 0.5, showing that all the networks are quite robust under targeted node attacks in descending order of the clustering coefficients. The aggregated network is the most robust, while the fungicide network is the least robust. Furthermore, the robustness of networks does not exhibit a clear trend.

For the other eight node metrics, we observe similar patterns with the robustness being at comparable levels between 0.2 and 0.4. The aggregated network is the most robust, the network for insecticides is the second-most robust, and the network for disinfectants is the third-most robust, while the other three networks for fungicides, herbicides, and rodenticides are more fragile. We also find that the robustness of networks has increased since 2007 and decreased after 2017, exhibiting an inverse U-shaped pattern. The decrease in network robustness after 2017 might be caused by the reverse globalization policies of the Trump Administration. We also observe a drop in the robustness of almost all networks in 2015, which corresponds to the 8.5\% drop in global pesticide trade in 2015 \citep{Li-Xie-Zhou-2021-FrontPhysics}. In contrast, the global financial crisis and the global economic recession from 2008 to 2010 did not have obvious impacts on the structural robustness of the international pesticide trade networks.

\section{Summary}
\label{S1:Summary}

In summary, using nine node metrics (in-degree, out-degree, in-closeness, out-closeness, authority score, hub score, PageRank, betweenness, and clustering coefficient), we have measured the importance of economies in the international trade networks for five categories of pesticides. The mutual correlations between the rankings of the nine node metrics were studied. We found that the clustering coefficient correlated negatively with the other eight node metrics, while the other eight node metrics were positively correlated with each other. These node metrics can thus be grouped into four communities: in-degree, PageRank, authority, and in-closeness; out-degree, hub, and out-closeness; betweenness; and clustering coefficient. It is found that developed or large developing economies are more important in both the import and export of the international pesticide trade.

We then investigated the structural robustness of the international pesticide trade networks based on the giant component size under three types of shocks to economies (node removal in descending order, randomly, and in ascending order). We unveiled that, except for clustering coefficient, the international pesticide trade networks are relatively robust under shocks to economies in ascending order and randomly, but fragile under shocks to economies in descending order of node importance. For clustering coefficient, the networks are robust under all three strategies of node removal. We found that economies with large node metrics related to import traits are more important in maintaining the structural robustness of the international pesticide trade networks.

There may be omissions in the pesticide network data, but our results have external validity. The ranking stability of node indicators is related to many factors \citep{Blumm-Ghoshal-Forro-Schich-Bianconi-Bouchaud-Barabasi-2012-PhysRevLett}, and we will not perform detailed analysis in this paper. For instance, the ranking provided by PageRank is sensitive to perturbations in the network topology of random networks. But, there are some super-stable nodes whose ranking is exceptionally stable to perturbations in scale-free networks \citep{Ghoshal-Barabasi-2011-NatCommun}. It has been shown that the top list of nodes is always stable \citep{Iniguez-Pineda-Gershenson-Barabasi-2022-NatCommun}.

We identified two situations that lowered the structural robustness of the international pesticide trade networks. The first one happened in 2015, when the global pesticide trade incurred an 8.5\% drop in international pesticide trade. The second one appeared after 2017, which was very probably triggered by the inverse globalization policies implemented by the Trump Administration. For economies, it is vital to forge more diverse import/export relationships to have higher resistance to external shocks, which can also enhance the structural robustness of the international pesticide trade networks as  whole.

Our work studies the supply networks of pesticides which is at the economy level. Similar analyses can be carried out at the industry or firm level. The microchip shortages we are facing pose a risk to some industries in Europe, and the disruption in the supply chain owing to COVID-19 and trade war effects has affected big countries such as China. There are already many studies on the robustness or resilience of supply chain networks based on simulation data \citep{Hearnshaw-Wilson-2013-IntJOperProdManage,Zhao-Kumar-Yen-2011-IEEETransEngManage,Li-Zobel-2020-IntJProdEcon,Zhao-Yin-Han-Li-2021-PhysicaA} or real data \citep{Zhao-Scheibe-Blackhurst-Kumar-2019-IEEETransEngManage,Li-Zobel-Seref-Chatfield-2020-IntJProdEcon}. However, these studies considered only a very few node metrics. Adopting more node metrics would deepen our understanding of the robustness or resilience of supply chain networks from more dimensions that may mimic different types of shocks on economies. 

There are other interesting topics for future research. One can consider to utilize a Melitz-type structural model  \citep{Melitz-2003-Econometrica,Helpman-Melitz-Yeaple-2004-AmEconRev} and integrate the research framework of network robustness due to shocks (random failures and intentional attacks \citep{Albert-Jeong-Barabasi-2000-Nature}). One can also consider to utilize the gravity model \citep{Isard-1954-QJEcon}, the fitness model \citep{Garlaschelli-Loffredo-2004-PhysRevLett}, and the radiation model \citep{Simini-Gonzalez-Maritan-Barabasi-2012-Nature,Ren-ErcseyRavasz-Wang-Gonzalez-Toroczkai-2014-NatCommun,Yan-Wang-Gao-Lai-2017-NC} for the prediction of trade flows. Such studies will further deepen our understanding of the dynamics of international pesticide trade networks.

\section*{Acknowledgements}

   This work was supported by the National Natural Science Foundation of China (72171083), the Shanghai Outstanding Academic Leaders Plan, and the Fundamental Research Funds for the Central Universities.
   






\end{document}